\documentclass[vecphys]{svmult}

\usepackage{mathptmx}       
\usepackage{helvet}         
\usepackage{courier}        
\usepackage{subeqnar}         
\usepackage{cite}         
\usepackage{graphicx}        
\usepackage[bottom]{footmisc}
\usepackage{bm}

\begin{document}
\title*{Theoretical aspects of Andreev spectroscopy and
tunneling spectroscopy in
non-centrosymmetric superconductors: a topical review}
\author{Matthias Eschrig, Christian Iniotakis, and Yukio Tanaka}
\titlerunning{Surface properties of non-centrosymmetric superconductors} 
\institute{Matthias Eschrig \at 
(a) Fachbereich Physik, Universit\"at Konstanz, D-78464 Konstanz, Germany,
and \\ (b) Institut f\"ur Theoretische Festk\"orperphysik and
DFG-Center for Functional Nanostructures, Karlsruhe Institute of Technology
D-76128 Karlsruhe, Germany, \\
\email{Matthias.Eschrig@kit.de}
\and
Christian Iniotakis \at
Institute for Theoretical Physics, ETH Zurich, 8093 Zurich, Switzerland,\\
\email{iniotaki@phys.ethz.ch}
\and
Yukio Tanaka \at
Department of Applied Physics, Nagoya University, Nagoya, 464-8603, Japan\\
\email{ytanaka@nuap.nagoya-u.ac.jp}
}
\maketitle
\abstract{
Tunneling spectroscopy at surfaces of unconventional superconductors has proven
an invaluable tool for obtaining information about the pairing symmetry.
It is known that mid gap Andreev bound states manifest itself as a
zero bias conductance peak in tunneling spectroscopy.
The zero bias conductance peak
is a signature for a non-trivial pair potential that exhibits different signs on different regions of the Fermi
surface.  Here,  we review recent theoretical results on the spectrum of Andreev bound states near
interfaces and surfaces in non-centrosymmetric superconductors.
We introduce a theoretical scheme to calculate
the energy spectrum of a non-centrosymmetric superconductor.
Then, we discuss the interplay between
the spin orbit vector field on the Fermi surface
and the order parameter symmetry.
The Andreev states carry a spin supercurrent and represent a helical edge mode
along the interface. We study
the topological nature of the resulting edge currents. If the triplet component of the order parameter dominates, then the helical edge mode exists. If, on the other hand, the singlet component dominates, the helical edge mode is absent.
A quantum phase transition occurs for equal spin singlet and triplet order parameter components.  We discuss the tunneling conductance and the
Andreev point contact conductance
between a normal metal and a non-centrosymmetric superconductor.
}

\section{Introduction}
\label{sec:1}
In this chapter, we will discuss the surface and interface 
properties of non-centrosym\-metric superconductors \cite{Bauer}
focusing on the tunneling conductance. 
Since the early sixties tunneling spectroscopy has played an important 
role in gathering information 
about the gap function of conventional superconductors \cite{Giaever}. 
In the context of unconventional superconductivity 
tunneling spectroscopy appeared 
as an important tool to probe the internal phase structure of the 
Cooper pair wave functions \cite{Tanaka95,KashiwayaReport}. 
Surface states with sub-gap energy, known as Andreev bound states (ABS)
\cite{Hu,Buchholtz,MABS,Bruder} 
provide channels for resonant tunneling leading to so-called zero-bias anomalies in $\D I/\D V$.
Zero-bias anomalies observed in high-temperature superconductors showed the presence
of zero-energy bound states at the surface, giving strong evidence for $d$-wave pairing \cite{Tanaka95,KashiwayaReport,Hu,Buchholtz,Bruder}. 
Similarly the tunneling spectrum observed in Sr$_2$RuO$_4$ is consistent with
the existence of chiral surface states as expected for a chiral $p$-wave superconductor  \cite{Honerkamp,Matsumoto,Kuroki}.
Zero bias conductance peaks due to Andreev bound states have been  observed in numerous experiments, e.g. in
high-$T_{\rm c}$ cuprates \cite{Experiments}, 
Sr$_{2}$RuO$_{4}$ \cite{Laube,Mao}, UBe$_{13}$ \cite{Ott}, 
CeCoIn$_{5}$\cite{Wei},  
the two dimensional organic superconductor 
$\kappa$-(BEDT-TTF)$_{2}$Cu[N(CN)$_{2}$]Br \cite{Ichimura} and 
PrOs$_{4}$Sb$_{12}$ \cite{Turel}.
Andreev bound states have also been observed in the Balian-Werthammer phase of
superfluid $^{3}$He \cite{Aoki}. 
The study of Andreev bound states in 
unconventional superconductors and superfluids has emerged as an
important phase sensitive probe.

In section  \ref{andspec} we present the theory for Andreev spectroscopy using
Bogoliubov wave function technique in Andreev approximation. Starting
with superconductors exhibiting $d$-wave or $p$-wave pairing, we
proceed with non-centrosymmetric superconductors.  
In section \ref{qcl}, we develop the theoretical tools for describing
Andreev spectroscopy in non-centrosymmetric superconductors in the framework of Nambu-Gor'kov Green's functions 
within the quasiclassical theory of superconductivity.

\section{Andreev spectroscopy in unconventional superconductors}
\label{andspec}

\subsection{Andreev conductance in $s$- and $d$-wave superconductors}

We discuss first the example of zero-bias resonant states at the interface of
a normal metal/spin-singlet $d$-wave superconductor junction. 
In general, the pair potential can be expressed in terms of
two coordinates, ${\bm x}$ and ${\bm{x'}}$, as $\Delta({\bm x},{\bm x'})$.
In uniform systems it only depends on the relative coordinate $\bm{x-x'}$, 
and a Fourier transform  with respect to it yields $\Delta(\bm{k})$ with relative momentum $\bm{k}$. 
For illustrative purposes, we assume in
the following a cylindrical Fermi surface and concentrate on 
two-dimensional systems. 
The pair potential for spin-singlet $d$-wave pairing is
$\Delta(\theta) = \Delta_{0}\cos(2\theta)$, 
with $\E^{\I \theta} =(k_x+\I k_y)/\mid \bm{k} \mid$, while 
the corresponding spin-singlet $s$-wave one is isotropic,
$\Delta(\theta)=\Delta_{0}$. 
The bulk quasiparticle density of states normalized by its value in the 
normal state is given by 
\begin{equation}
\rho_{\rm B}(E) = \frac{1}{\pi}
\int^{\pi}_{0}\D \theta \rho_{0}(E,\Delta(\theta)), \ \ 
\rho_{0}(E,\Delta(\theta))
= \frac{E}{\sqrt{E^{2} -\Delta_{0}^{2}\cos^{2}(2\theta)}}. 
\label{LDOS}
\end{equation}
For a spin-singlet  $d$-wave superconductor this quantity behaves linearly
at low energies, $\rho(E) \propto |E|$.
As shown below
if the angle between the interface normal
and the lobe direction of the $d$-wave pair potential 
has  a nonzero value $\alpha$ with $0<\alpha<\pi/2$, then the 
resulting tunneling conductance $\sigma_{\rm T}(E)$ has a zero bias 
conductance peak.

\begin{figure}[t]
\begin{center}
\scalebox{0.4}{
\includegraphics[width=10cm,clip]{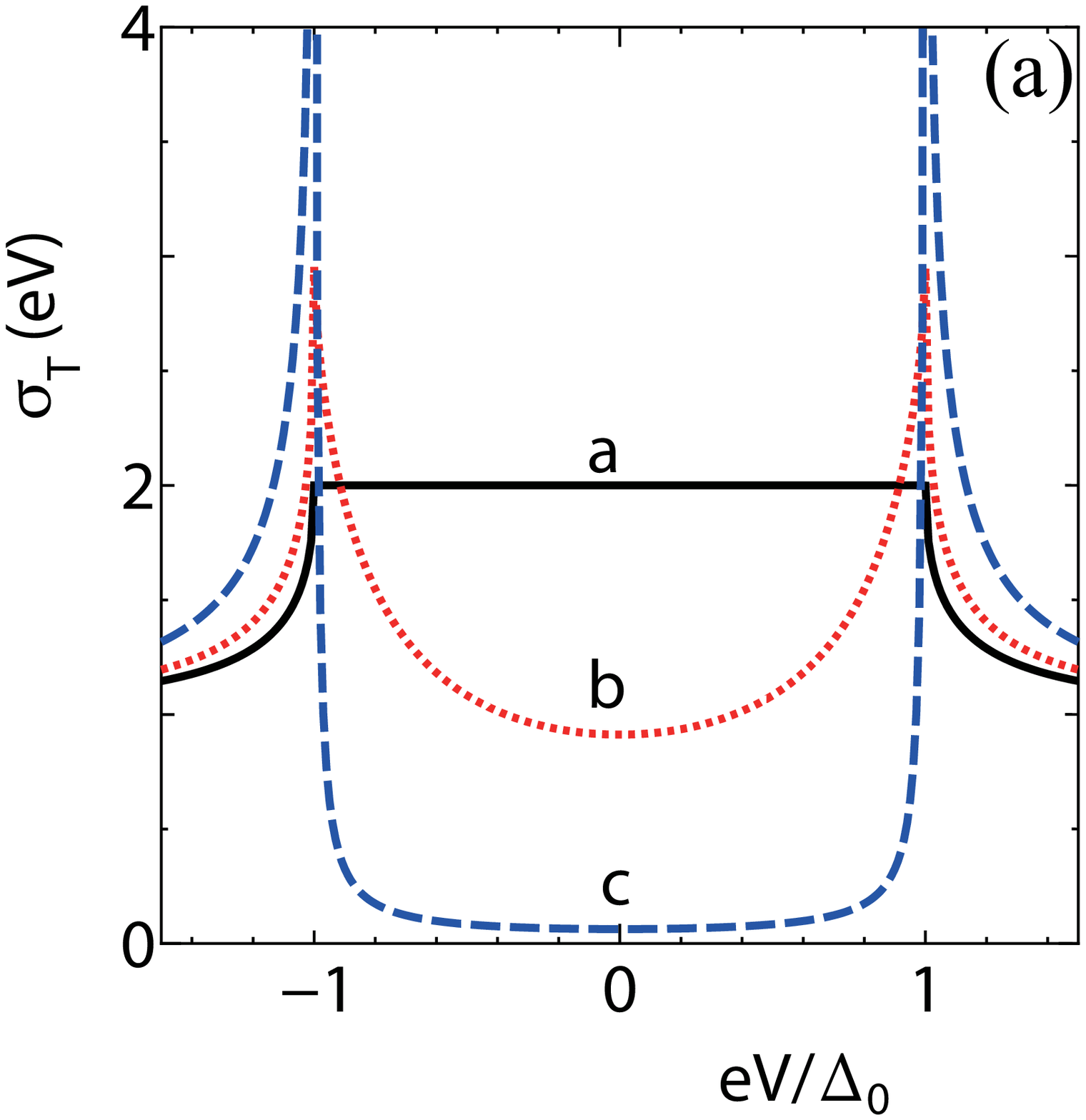} 
\hspace{2cm}
\includegraphics[width=10cm,clip]{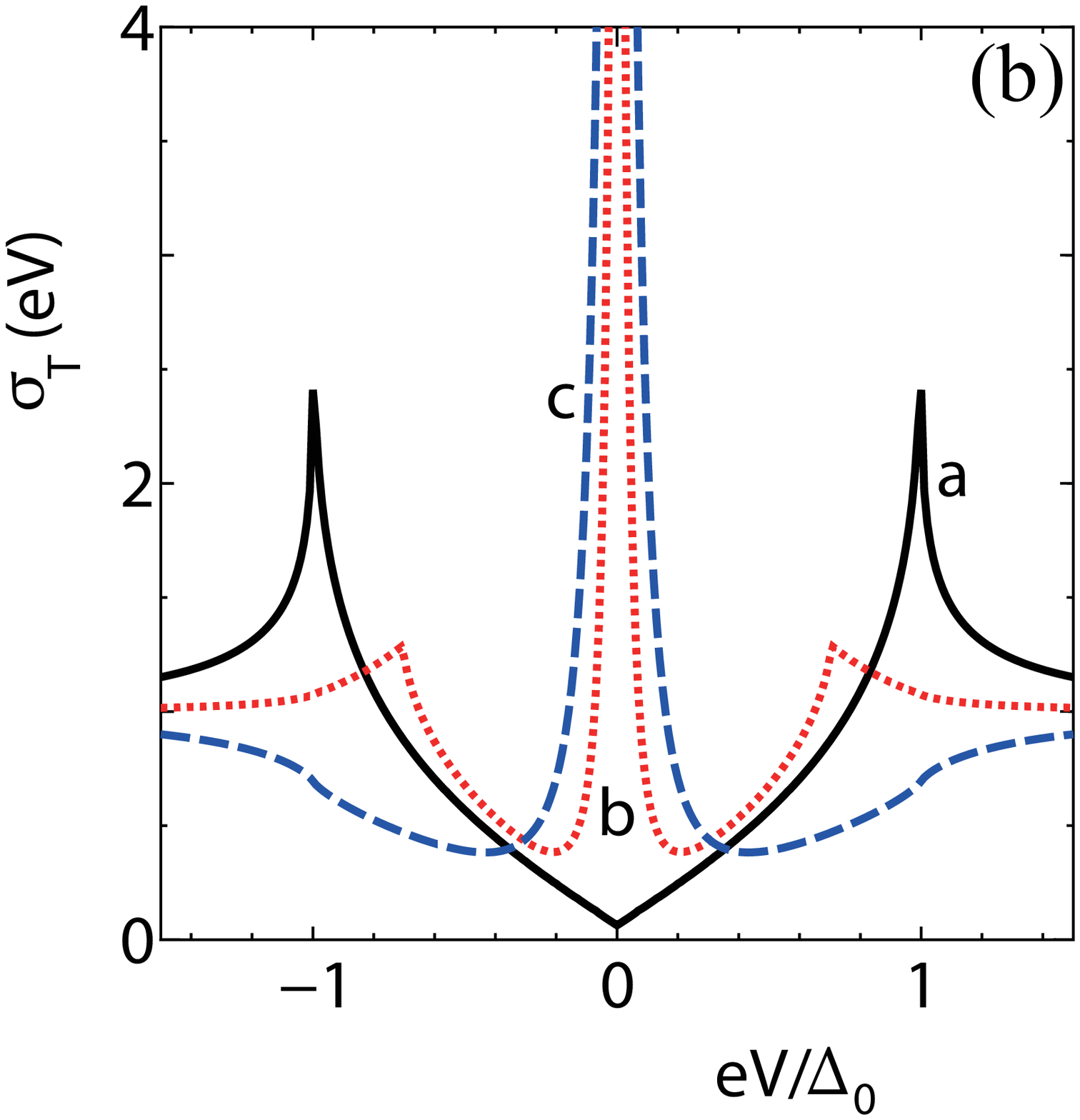}
}
\end{center}
\caption{(Color online) Left: Tunneling conductance for an
$s$-wave superconductor. 
a: $Z=0$, b: $Z=1$ and c: $Z=5$.  
Right:
Tunneling conductance for a
$d$-wave superconductor for $Z=5$. 
a: $\alpha=0$, b: $\alpha=0.125\pi$ and c: $\alpha=0.25\pi$. 
}
\label{fig:1}
\end{figure}

The Andreev conductance
for a normal metal/insulator/spin singlet $s$-wave superconductor 
junction is described by the model of
Blonder, Tinkham, and Klapwijk (BTK) \cite{Blonder}. 
Within this model, $\sigma_{\rm T}(E)$ at zero temperature
is given by 
\begin{equation}
\sigma_{\rm T}(E) \propto \mbox{$\sum_{\theta}$}
(1 + \mid a(E,\theta) \mid^{2} - \mid b(E,\theta) \mid^{2} )
\end{equation}
where $a(E,\theta)$, and $b(E,\theta)$ are probability amplitude
coefficients for Andreev reflection and for normal reflection, 
respectively.  We apply the BTK model in the following to the case of
spin-singlet $d$-wave pairing.
\begin{petit}
We assume that the Fermi energy 
$E_{F}$ is much larger than $\mid \Delta(\theta) \mid $, such
that the Andreev approximation can be applied to the Bogoliubov wave functions. 
For simplicity, we also assume equal effective masses and Fermi momenta in 
the normal metal and in the superconductor.
The spatial dependence of the pair potential is chosen to be 
$\Delta(\theta)\Theta(x)$ (with the Heaviside step function $\Theta $). 
The insulating barrier at the atomically clean interface 
is modeled by a $\delta$-function potential, $V(x)= H\delta(x)$. 
Since the momentum parallel to the interface is conserved,  
the two component Bogoliubov wave function is given in Andreev approximation by
\begin{equation}
\Psi(\theta,x)
=
\left(
\begin{array}{c}
u_{+}(\theta,x) \\
v_{+}(\theta,x)
\end{array}
\right) \exp(\I k_{\rm F}x \cos \theta )
+\left(
\begin{array}{c}
u_{-}(\theta,x) \\
v_{-}(\theta,x)
\end{array}
\right) \exp(-\I  k_{\rm F}x \cos \theta )
\end{equation}
where $u_{j}(\theta,x)$ and $v_{j}(\theta,x)$ with $(j=+,-)$ obey the
Andreev equations
\begin{eqnarray}
Eu_{j}(\theta,x)&=&
-\Big[\frac{\I \hbar^{2}\sigma_j k_{\rm F}\cos\theta}{m}\frac{\D }{\D x}-H\delta(x)\Big]\;
u_{j}(\theta,x) + \Delta(\theta_{j})\Theta(x)v_{j}(\theta,x), 
\nonumber
\\
Ev_{j}(\theta,x)&=&
\Big[\frac{\I  \hbar^{2}\sigma_j k_{\rm F}\cos\theta}{m}\frac{\D  }{\D  x}-H\delta(x)\Big]\;
v_{j}(\theta,x) + \Delta^{*}(\theta_{j})\Theta(x)u_{j}(\theta,x), 
\label{AndEq}
\end{eqnarray}
with $\sigma_+=1$, $\theta_{+}=\theta$, and $\sigma_-=-1$, $\theta_{-}=\pi-\theta$.
For a $d$-wave superconductor the corresponding effective pair 
potentials $\Delta(\theta_{\pm})$ are given by 
\begin{equation}
\Delta(\theta_{+})=\Delta_{0}\cos(2\theta - 2\alpha), \quad
\Delta(\theta_{-})=\Delta_{0}\cos(2\theta + 2\alpha), 
\end{equation}
where the angle between the interface normal and 
the lobe direction of the $d$-wave pair is $\alpha$. 
The wave functions $u_{\pm}(\theta,x)$ and $v_{\pm}(\theta,x)$
resulting from Eqs.~(\ref{AndEq})
are obtained from by ansatz
\begin{eqnarray}
\left(
\begin{array}{c}
u_{+}(\theta,x) \\
v_{+}(\theta,x)
\end{array}
\right) 
&=&\left\{
\begin{array}{ll}
\left(
\begin{array}{c}
1 \\
0
\end{array}
\right) \exp(\I \delta x)
+a(E,\theta)\left(
\begin{array}{c}
0 \\
1
\end{array}
\right) \exp(-\I \delta x) & \quad x<0 \\
c(E,\theta)\left(
\begin{array}{c}
\sqrt{(E + \Omega_{+})/2E} \\
\exp(-\I \phi_{+})
\sqrt{(E - \Omega_{+})/2E}
\end{array}
\right) \exp(\I  \gamma_{+}x) & \quad x>0
\end{array}
\right. ,
\\
\left(
\begin{array}{c}
u_{-}(\theta,x) \\
v_{-}(\theta,x)
\end{array}
\right) 
&=&\left\{
\begin{array}{ll}
b(E,\theta)\left(
\begin{array}{c}
1 \\
0
\end{array}
\right) \exp(-\I  \delta x)
& \quad x<0 \\
d(E,\theta)\left(
\begin{array}{c}
\exp(\I  \phi_{-}) 
\sqrt{(E - \Omega_{-})/2E}
\\
\sqrt{(E + \Omega_{-})/2E}
\end{array}
\right) \exp(\I  \gamma_{-}x) & \quad x>0
\end{array}
\right.
\end{eqnarray}
where we used the abbreviations
\begin{equation}
\delta
=\frac{Em}{\hbar k_{\rm F}\cos \theta}, \; \; 
\gamma_{\pm}
=\frac{\Omega_{\pm}m}{\hbar k_{\rm F}\cos \theta}, \;\;
\Omega_{\pm}=\sqrt{E^{2} -\Delta^{2}(\theta_{\pm})}, \;\;
\exp(\I  \phi_{\pm})=
\frac{\Delta(\theta_{\pm})}{\mid \Delta(\theta_{\pm})\mid}. 
\end{equation}
With the help of appropriate boundary conditions, 
\[
\Psi(\theta,x) \mid_{x=0_{-}}
=\Psi(\theta,x) \mid_{x=0_{+}}
\]
\begin{equation}
\frac{\D }{\D  x} \Psi(\theta,x)\Big|_{x=0_{+}}
-\frac{\D}{\D x} \Psi(\theta,x)\Big|_{x=0_{-}}
=\frac{2mH}{\hbar^{2}}\Psi(\theta,x)\Big|_{x=0_{+}}
\end{equation}
we obtain $a(E,\theta)$, $b(E,\theta)$, $c(E,\theta)$, and $d(E,\theta)$. 
\end{petit}
The resulting conductance is 
\[
\sigma_{\rm T}(E)=
\left(\int^{\pi/2}_{-\pi/2}\D \theta \; 
D(\theta )
\; \sigma_{\rm R}(E,\theta) \; \cos\theta \right)
/\left(
\int^{\pi/2}_{-\pi/2} \D \theta \; 
D(\theta )
\; \cos\theta \right),
\]
\begin{equation}
\sigma_{\rm R}(E,\theta) 
=\frac{ 1 + 
D(\theta )
\mid \Gamma_{+}\mid^{2}
-R(\theta)
\mid\Gamma_{+}\Gamma_{-}\mid^{2} }
{\mid 1  
-R(\theta)
\Gamma_{+}\Gamma_{-}\exp[\I  (\phi_{-}-\phi_{+})] \mid^{2}},
\label{conductance}
\end{equation}
with 
$\Gamma_{\pm}=(E-\Omega_{\pm})/\mid \Delta(\theta_{\pm}) \mid$.
The quantities $D(\theta )$ and $R(\theta )$ above are given by
\[
D
(\theta)=4\cos^{2} \theta/(4\cos^{2}\theta + Z^{2}), 
\qquad R(\theta )=1-D(\theta ),
\]
with injection angle $\theta$ and $Z=2mH/\hbar^{2}k_{\rm F}$ \cite{Tanaka95}.  
Choosing $\Delta(\theta_{\pm})=\Delta_{0}$ reproduces the
BTK formula for an $s$-wave superconductor.
Typical line shapes of $\sigma_{\rm T}(eV)$ with $eV=E$ for 
$s$-wave and $d$-wave superconductors are
shown in Fig. \ref{fig:1}. 
The $d$-wave case is shown in Fig. \ref{fig:1}(b).
As can be seen there, if the angle $\alpha$ deviates from 0, 
the resulting $\D I/\D V$ has a zero bias conduction peak (ZBCP)
(curves $b$ and $c$);
the only exceptional case is $\alpha=0$, as shown in curve $a$.
The width of the ZBCP is proportional to 
$D$, 
while its height is proportional to the inverse of 
$D$. 
The origin of this peak are mid gap Andreev bound states (MABS).  
The condition of the formation for Andreev bound states
at the surface of an isolated $d$-wave superconductor 
($D \rightarrow 0$) 
is expressed by
\begin{equation}
1=\Gamma_{+}\Gamma_{-}\exp[\I  (\phi_{-}-\phi_{+})] .
\end{equation}
At zero energy $\Gamma_{+}\Gamma_{-}=-1$ is satisfied, and consequently
a MABS appears provided $\exp[\I  (\phi_{+}-\phi_{-})]=-1$. 
For this case, on the superconducting side of the interface,
the injected electron and the reflected hole 
experience a different sign of the pair potential. 
For $\alpha=\pi/4$, there is a MABS independent of the 
injection angle. 
In this case, the energy dispersion of the resulting ABS, $E_{\rm b}$, is given by 
\begin{equation}
E_{\rm b}=0 .
\label{ZES}
\end{equation} 

\begin{figure}[t]
\sidecaption[t]
\scalebox{0.7}{
\includegraphics[width=10cm,clip]{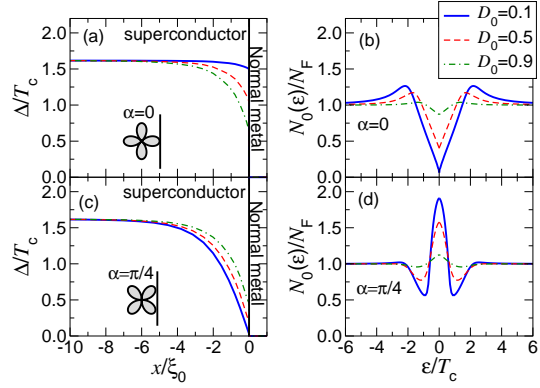}
}
\caption{
(a)+(c):
order parameter amplitude and (b)+(d): local density of states at
the interface for a layered $d$-wave-superconductor/normal-metal junction. 
(a)+(b): $\alpha=0$, (c)+(d): $\alpha=\pi/4$.
The interface is at $x=0$. 
The curves are for the indicated transmission coefficients $D_0$.
The temperature is $T=0.3T_{\rm c}$, and the mean free path $\ell =10 \xi_0$.
After Ref. \cite{eschrig00}.
}
\label{fig:eschrig00}
\end{figure}
Finally, we comment on the effects of order parameter suppression near an
surface or interface in a d-wave superconductor.  
In Fig. \ref{fig:eschrig00} we reproduce a
self-consistent solution for a layered $d$-wave superconductor, showing
that a strong order parameter suppression is always present for
$\alpha =0.25$, whereas for $\alpha=0$ in the tunneling limit the
order parameter suppression can be neglected. The corresponding 
local density of states at the surface is shown in
Fig.~\ref{fig:eschrig00}(b) and (d). 
The interface is modeled by a $\delta $-potential as above, 
with a transmission $D(\theta )=D_0 \cos^2 \theta /(1-D_0\sin^2 \theta)$,
and the parameter $D_0$ is related to $Z$ via $D_0=1/[1+(Z/2)^2]$.

\subsection{Andreev conductance in chiral $p$-wave superconductor}
\label{chiral}

In this section, we discuss the tunneling conductance 
of a normal metal/chiral $p$-wave superconductor junction. 
There is evidence supporting the 
realization of spin-triplet pairing with broken 
time reversal symmetry in the superconducting 
state of Sr$_{2}$RuO$_{4}$ \cite{Maeno,Maeno1,Maeno2,Maeno3,Maeno4,Maeno5}. 
A possible symmetry is given by two-dimensional chiral $p$-wave pairing, 
where the pair potentials are given by 
$\Delta_{\uparrow,\uparrow}=\Delta_{\downarrow,\downarrow}=0$, 
$\Delta_{\uparrow,\downarrow}=
\Delta_{\downarrow,\uparrow}=\Delta_{0}\exp(\I  \theta)$. 
In the following, $\theta$ is measured from the interface normal.
In the actual sample, 
the presence of chirality may produce chiral domain structures. 
A recent experiment is consistent with the presence of chiral domains
\cite{Kambara}. Also, there are several theoretical proposals to detect 
chiral domain structures \cite{Tanuma}. 
Here, for simplicity, we consider a single domain
chiral $p$-wave superconductor. 

Since the $z$-component of the Cooper pair spin is zero, 
we can also use  Eq. (\ref{conductance}) to obtain the
tunneling conductance for normal metal/chiral $p$-wave 
superconductor junctions. 
Before discussing the tunneling conductance, we first 
consider the bulk local density of states (LDOS) of a chiral $p$-wave superconductor.  
In contrast to the spin-singlet 
$d$-wave pairing case, $\rho_{0}(E,\Delta(\theta))$ in 
Eq. (\ref{LDOS}) is given by 
\[
\rho_{0}(E,\Delta(\theta))=E/\sqrt{E^{2} -\Delta_{0}^{2}}. 
\]
It has a fully gapped density of state like in the spin-singlet $s$-wave case.

We now discuss the condition when an ABS is formed at the surface of 
an isolated chiral $p$-wave superconductor. 
The bound state condition is given by \cite{Honerkamp,Matsumoto}
\begin{equation}
E + \sqrt{E^{2}-\Delta_{0}^{2}} 
=-\left( E -\sqrt{E^{2}-\Delta_{0}^{2}} \right)
\exp(-2\I \theta) ,
\end{equation}
showing that the bound state level $E_{\rm b}$ satisfies 
\begin{equation}
\label{bspwave}
E_{\rm b}(\theta ) = \Delta_{0}\sin\theta
\end{equation}
Note that the ABS has a dispersion different from 
that in the $d$-wave case with $\alpha=\pi/4$. 
The presence of the edge state with 
a dispersion induces a spontaneous dissipationless current.

\begin{figure}[b]
\begin{center}
\scalebox{0.4}{
\includegraphics[width=10cm,clip]{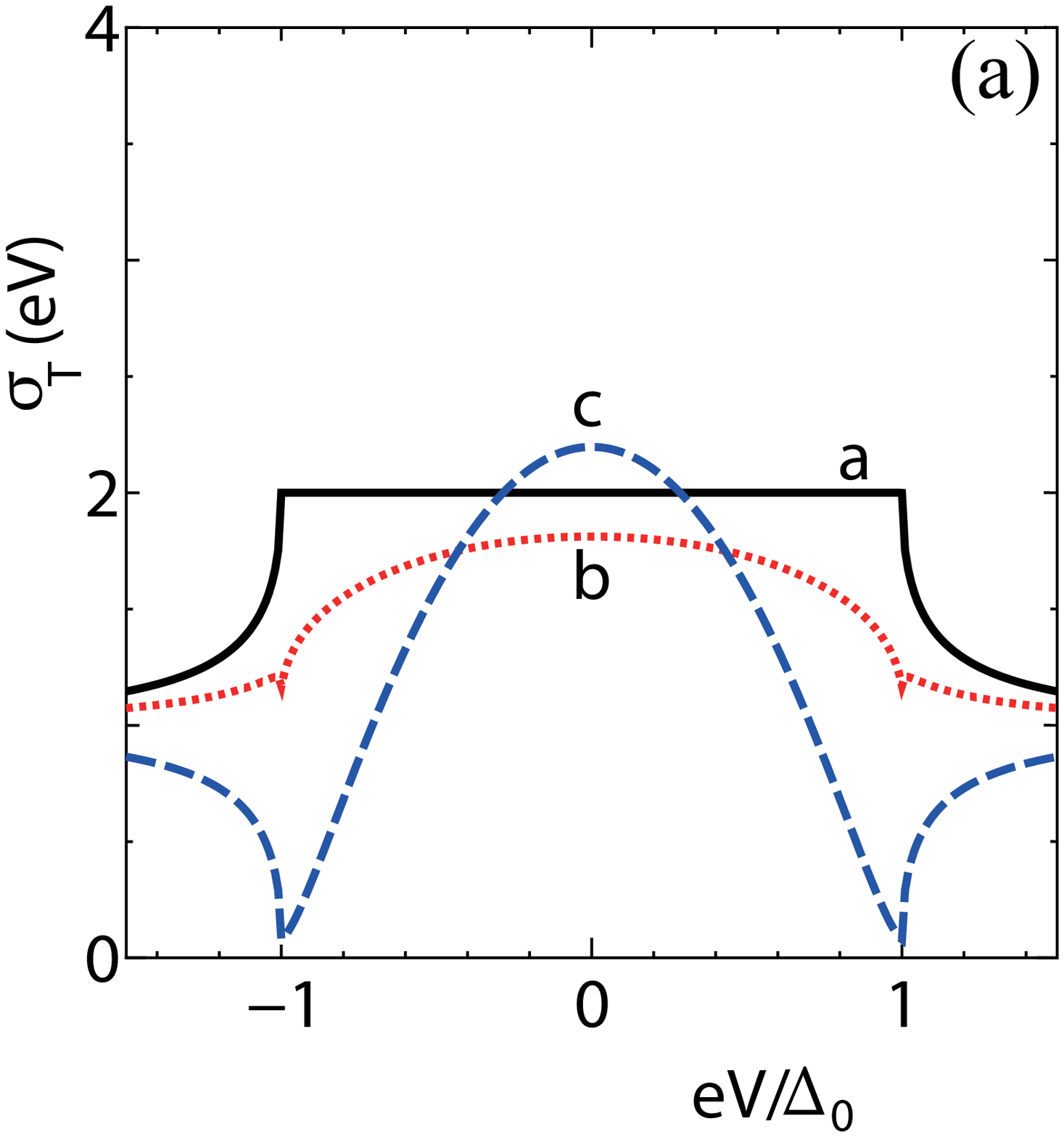}
\hspace{2cm}
\includegraphics[width=10cm,clip]{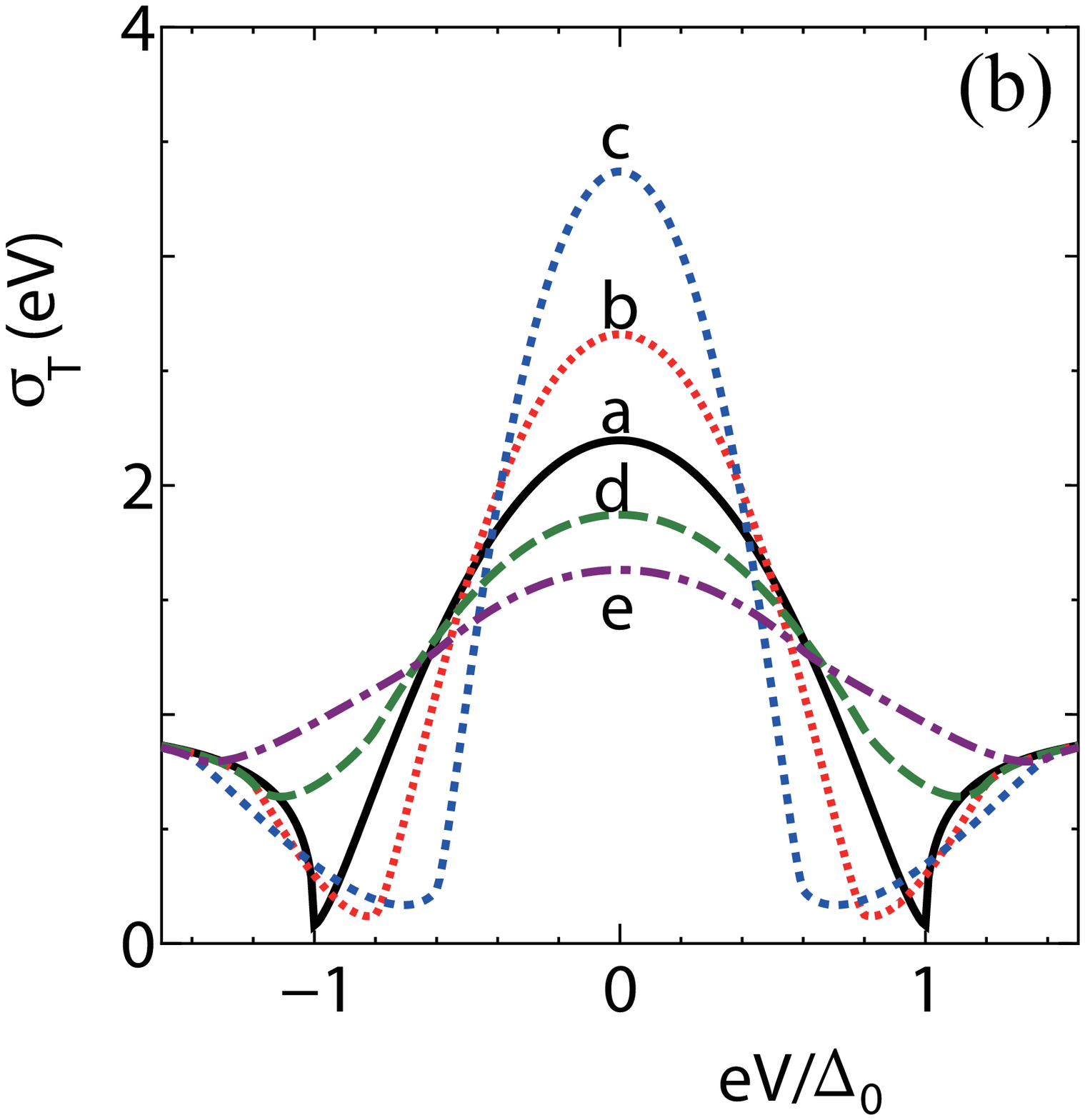}
}
\end{center}
\caption{(Color online) (a): Tunneling conductance for 
a chiral $p$-wave superconductor.  
a: $Z=0$, b: $Z=1$ and c: $Z=5$.  
(b):
Tunneling conductance for a
chiral $p$-wave superconductor in the presence of a magnetic field 
for $Z=0$.  
a: $H=0$, b: $H=0.2H_{0}$ and c: $H=0.4H_{0}$, 
d: $H=-0.2H_{0}$ and e: $H=-0.4H_{0}$. 
}
\label{fig:3}
\label{fig:4}
\end{figure}

As in the previous section we consider the tunneling 
conductance $\sigma_{\rm T}(E)$ in a normal metal/chiral $p$-wave 
superconductor junction, which is shown in
Fig. \ref{fig:3}(a). 
As can be seen,
for $Z=0$, the line shape of conductance is identical to that 
of a spin-singlet $s$-wave superconductor (see curve $a$), whereas
with increasing $Z$  a zero bias conductance peak emerges 
(curves $b$ and $c$).  
The resulting ZBCP is broad in contrast to the 
spin-singlet $d$-wave case due to the fact that 
the position of the ABS depends on the injection angle $\theta$ according
to Eq.~(\ref{bspwave})
\cite{Honerkamp}. 
The presence of the ABS has been confirmed by tunneling experiments 
\cite{Laube,Mao}.

Next we consider the situation where a magnetic field $H$ is applied 
perpendicular to the two-dimensional plane, which induces
a shielding current along the interface.
When the penetration depth for the chiral $p$-wave superconducting material
is much longer than the coherence length, the vector potential
can be approximated as
$\bm{A}(\bm{r})=(0,A_{y}(x),0)$ with
$A_{y}(x) = -\lambda_{\rm m} H \exp(-x/\lambda_{\rm m})$, where
$\lambda_{\rm m}$ is the penetration depth.
In the following we consider the situation where 
Landau level quantization can be neglected.  
Then the quasiclassical approximation can be used.
The applied magnetic field shifts the quasiparticle energy $E$ 
to $E + H\Delta_{0}\sin \phi/H_{0}$ with
$H_{0}=h/(2e\pi^{2}\xi \lambda_{\rm m})$
and $\xi=\hbar^{2} k_{\rm F}/(\pi m \Delta_{0})$ \cite{Doppler}. 
The resulting tunneling conductance for various magnetic fields is
plotted in Fig. \ref{fig:4}(b).
As is seen, $\sigma_{\rm T}(E)$ is enhanced for 
positive $H$, while it is reduced for negative $H$. 
This can be roughly understood by looking at the 
bound state levels. 
In the presence of $H$, the bound state energy can be expressed by 
\begin{equation}
E_{\rm b}(\theta )=\Delta_{0}(1-H/H_{0}) \sin\theta
\sim  \Delta_{0}(1-H/H_{0})k_{y}/k_{\rm F} .
\end{equation}
The contribution of the Andreev bound state to the conductance enters
via a term $\delta (E-E_{\rm b}(\theta ))$, which is proportional to
$1/|\D E_{\rm b}(\theta )/\D \theta |$.
It is clear that the slope of the dispersion around $\theta=0$
is reduced for positive $H$, leading to an enhancement of the numerator
in Eq. \ref{conductance} around $\theta=0$, where the bound states are 
close to zero enery.
On the other hand, for negative $H$, the height of the
ZBCP is reduced since the slope of the curve of $E_{\rm b}$ around zero 
energy becomes steeper \cite{Kuroki}.

\begin{figure}[b]
\sidecaption
\scalebox{0.7}{
\includegraphics[width=10cm,clip]{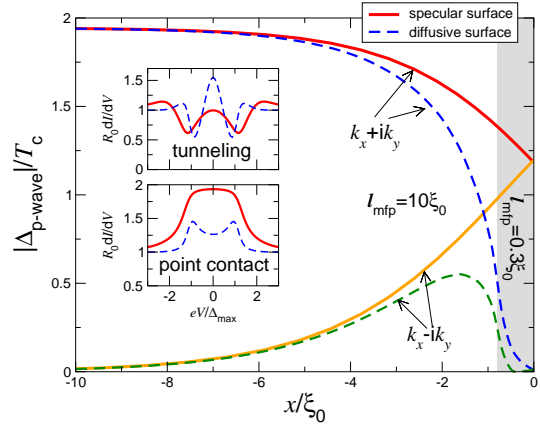}
}
\caption{(Color online) 
Self-consistent order parameter near a surface of a layered $p$-wave
superconductor.  Full lines are for mean free path 
$\ell_{\rm mfp} =10\xi_0$ everywhere;
dashed lines are for a shortened $\ell_{\rm mfp}=0.3 \xi_0$ 
in the gray shaded region.
The calculations are for $T=0.1 T_{\rm c}$.
Insets: point contact spectra for fully transparent interface (bottom) and
tunneling conductace ($D_0=0.05$) (top). After Ref.~\cite{Laube04}.
}
\label{fig:Laube}
\end{figure}

In $p$-wave superconductors self-consistency of the order parameter
and impurity effects can be of importance. In
Fig.~\ref{fig:Laube} we show self-consistent 
order parameters and Andreev spectra at a surface of a 
layered $p$-wave superconductor.
In addition to the bulk
$k_x+\I k_y$ component a subdominant $k_x-\I k_y$ component
is stabilized within a few coherence lengths
($\xi_0=v_{\rm F}/2\pi k_{\rm B} T_{\rm c}$) near the surface.
The full lines are results assuming
a mean free path of $\ell=10 \xi_0$ everywhere.
When replacing the mean free path in
a surface layer (gray shaded region in Fig.~\ref{fig:Laube}) by
$\ell =0.3 \xi_0$, we obtain the results shown as dashed lines.
In contrast to the first case, for the second case
both order parameter components are strongly suppressed near the surface.
The presence of an increased scattering in a surface layer
also modifies the form of point contact spectra and the tunneling conductance
as seen in the insets of Fig.~\ref{fig:Laube}.
In constrast to the surface density of states which for a clean surface
is constant in energy, the tunneling conductance shows
a broad peak similar as in Fig.~\ref{fig:3}, which is however reduced
in height for a self consistent order parameter \cite{Yamashiro99}.
 
Finally, we would like to comment that the above edge state is 
topologically equivalent to that of a quantum Hall system.
In a quantum Hall system it is established that the 
edge channel supports the 
accurate quantization of the Hall conductance $\sigma_{\rm H}$,
which is related to a topological integer \cite{Girvin,Thouless}. 
In the edge state of a chiral $p$-wave superconductor, 
such a topological number can be also defined \cite{Furusaki}. 
For this case, the edge state is topologically protected by the 
bulk energy gap $\Delta_{0}$. 
The topological properties of the electronic states 
have been attracting intensive interest in condensed 
matter physics. In section \ref{topol} we will return to this
question in connection with non-centrosymmetric superconductors.
Before that, we discuss in the following section theoretical
predictions for the Andreev conductance spectra for non-centrosymmetric
superconductors.

\subsection{Andreev conductance in non-centrosymmetric superconductors}
\label{sec:4}
Non-centrosymmetric superconductors such as CePt$_3$Si
are a central topic of current research \cite{Bauer,Frigeri}. 
Two-dimensional non-centrosymmetric superconductors are 
expected e.g. at interfaces and/or surfaces due to a
strong potential gradient. An interesting example is
superconductivity at a LaAlO$_3$/SrTiO$_3$ interface \cite{Interface,Yada}. 
In non-centrosymmetric materials spin-orbit interaction becomes very important.
Frigeri et al. \cite{Frigeri} have shown that the
$(p_{x} \pm \I p_{y})$-pairing 
state has the highest $T_{\rm c}$ within the triplet-channel in CePt$_3$Si. 
It has been shown that singlet ($s$-wave) and triplet ($p$-wave) 
pairing is mixed, and several novel properties related to that mixing,
such as a large upper critical field beyond the 
Pauli limit, have been focused on \cite{Frigeri}. 
On the other hand, a pure $(p_{x} \pm \I p_{y})$-pairing state 
has been studied as a
superconducting analogue of a quantum spin Hall system \cite{Qi}.
Therefore, it is an important and urgent issue to study the 
spin transport properties of the NCS superconductors
from a topological viewpoint.

In this section, we discuss charge and spin transport in 
non-centrosymmetric superconductors \cite{Rapid}. 
We concentrate on
non-centrosymmetric superconductors with time-reversal symmetry,
where a spin-triplet $(p_{x} \pm \I p_{y})$-wave and a spin-singlet $s$-wave 
pair potential can mix with each other, similar
as discussed in the last section.  
We show that when the amplitude of the
$(p_{x} \pm \I p_{y})$-wave component is larger than 
that of the $s$-wave component, then the superconducting 
state belongs to a topologically nontrivial class analogous
to a quantum spin Hall system, and the resulting
helical edge modes are spin current carrying Andreev bound states
that are topologically protected.
Below, we study Andreev reflection \cite{Andreev} at low energy, which is
determined mostly by the helical edge modes, and find the
spin polarized current flowing through an interface as a function of
incident angle.  When a magnetic field is applied, even the
angle-integrated current is spin polarized.

\subsubsection{Andreev bound states}

We start with the Hamiltonian of a non-centrosymmetric superconductor
\begin{eqnarray}
\hat{\cal H}_{\rm S} = \left( 
\begin{array}{*{20}c}
{H\left( {\bf k} \right)} & {\Delta \left( {\bf k} \right)} \\
{ - \Delta ^ * \left( { - {\bf k}} \right)} & { - H^ * \left( 
{ - {\bf k}} \right)}\\
\end{array}\right) 
\label{Hamiltonian}
\nonumber
\end{eqnarray}
with
${H}(\bm{k})=\xi_{\bm{k}} + \bm{g}(\bm{k}) \cdot {\bm{\sigma}}$,
$\bm{g}(\bm{k})=\lambda (\hat{\bm{x}}k_{y}-\hat{\bm{y}}k_{x})$,
$\xi_{\bm{k}}=\hbar^2 {\bm{k}}^{2}/(2m) - \mu$.
Here, $\mu$, $m$,
${\bm{\sigma}}$ and $\lambda$ denote
chemical potential, effective mass,
Pauli matrices and coupling constant of Rashba spin-orbit
interaction, respectively \cite{Frigeri}.
The pair potential ${\Delta}(\bm{k})$
is given by 
\begin{equation}
{\Delta}(\bm{k}) = 
[\bm{d}(\bm{k})\cdot {\bm{\sigma}}
+ \psi(\bm{k})] \I {\sigma}_{y}. 
\end{equation}
We choose $(p_{x} \pm \I p_{y})$ 
with $\bm{d}(\bm{k})=\Delta_{p}(\hat{\bm{x}}k_{y}-\hat{\bm{y}}k_{x})/ \mid {\bm k}\mid$ 
for the spin-triplet component \cite{Frigeri},
and  $\psi(\bm{k})=\Delta_{s}$ with $\Delta_{p} \geq 0$ and 
$\Delta_{s} \geq 0$. 
The superconducting gaps 
$\Delta_{1}=\Delta_{p}+\Delta_{s}$ and 
$\Delta_{2}=\mid \Delta_p-\Delta_s \mid$ open for the 
two spin-split energy bands, respectively, in the homogeneous 
state \cite{Iniotakis07}.
As we will show below, surface states 
are crucially influenced by the relative magnitude 
between $\Delta_{p}$ and $\Delta_{s}$.

\begin{petit}
Let us consider the wave functions,
focusing on those for ABS localized at the surface. 
Consider a two-dimensional semi-infinite superconductor
for $x>0$ where the surface is located at $x=0$. 
The corresponding wave function 
is given by \cite{Yokoyama}
\begin{eqnarray}
&\displaystyle
\Psi_{\rm S}(x)=\E^{\I k_{y}y}
\left[c_{1}\psi_{1}\E^{\I q^{+}_{1x}x} + c_{2}\psi_{2}\E^{-\I q^{-}_{1x}x}
+ c_{3}\psi_{3}\E^{\I q^{+}_{2x}x} + c_{4}\psi_{4}\E^{-\I q^{-}_{2x}x} \right], 
&
\label{wavefunction}
\nonumber
\\
&\displaystyle
q^{\pm}_{jx}
=k^{\pm}_{jx} \pm (k_{j}/k^{\pm}_{jx})
\sqrt{(E^{2}-\Delta_{j}^{2})/
(\lambda^{2} + 2\hbar^{2}\mu /m )}, 
&
\end{eqnarray}
with $j=\{1,2\}$, and
$k^{+}_{jx}=k^{-}_{jx}=k_{jx}$ for $\mid k_{y} \mid \leq k_{j}$ 
and $k^{+}_{jx}=-k^{-}_{jx}=k_{jx}$ for $\mid k_{y} \mid >k_{j}$.  
Here, $k_1$ and $k_2$ with
\begin{equation}
k_{1(2)}=
\mp m\lambda/\hbar^{2} + \sqrt{(m\lambda/\hbar^{2})^{2} + 2m\mu/\hbar^{2} }
\end{equation}
are the Fermi momenta of the small and large
Fermi surface, respectively
(the upper sign holds for $k_1$), and $k_{jx}$ denotes the $x$
component of the Fermi momentum $k_{j}$, with
$k_{jx}=\sqrt{k_{j}^{2} -k_{y}^{2}}$. The wave functions are
given by  
\begin{eqnarray}
\psi_{1} =\left(
\begin{array}{c}
u_{1} \\
-\I \alpha_{1}^{-1}u_{1} \\
\I \alpha_{1}^{-1}v_{1} \\
v_{1} 
\end{array}
\right),  
\psi_{2} =\left( 
\begin{array}{c}
v_{1} \\ 
-\I \tilde{\alpha}_{1}^{-1}v_{1} \\
\I \tilde{\alpha}_{1}^{-1}u_{1} \\
u_{1} 
\end{array}
\right),  
\psi_{3} =\left(
\begin{array}{c}
u_{2} \\
\I \alpha_{2}^{-1}u_{2} \\
\I \gamma \alpha_{2}^{-1}v_{2} \\ 
-\gamma v_{2} 
\end{array}
\right),  
\psi_{4} =
\left( 
\begin{array}{c}
v_{2} \\
\I \tilde{\alpha}_{2}^{-1}v_{2} \\ 
\I \gamma \tilde{\alpha}_{2}^{-1}u_{2} \\ 
-\gamma u_{2} 
\end{array}
\right) 
\nonumber
\end{eqnarray}
with $\gamma = {\rm sgn}(\Delta_{p}-\Delta_{s})$. 
In the above, 
\begin{equation}
u_{j}=
\sqrt {(E + \sqrt{E^2 - \Delta_{j} ^2 }) /2E },\quad
v_{j}=
\sqrt
{(E - \sqrt {E^2 - \Delta_{j} ^2 })/2E }. 
\end{equation}
Here we have introduced
$\alpha_{1}=(k^{+}_{1x}-\I k_{y})/k_{1}$,
$\alpha_{2}=(k^{+}_{2x}-\I k_{y})/k_{2}$,
$\tilde{\alpha}_{1}=(-k^{-}_{1x}-\I k_{y})/k_{1}$, and
$\tilde{\alpha}_{2}=(-k^{-}_{2x}-\I k_{y})/k_{2}$.  
$E$ is the
quasiparticle energy measured from the Fermi energy.
By postulating $\Psi_{\rm S}(x)=0$ at $x=0$, we can determine the ABS. 
\end{petit}

The bound state condition can be expressed by 
\begin{eqnarray}
\sqrt{(\Delta_{1}^{2}-E^{2})(\Delta_{2}^{2}-E^{2})}
=\frac{1 -\zeta}{1 + \zeta}(E^{2} + \gamma \Delta_{1}\Delta_{2}), 
\label{bound}
\\
\displaystyle
\zeta =
\left\{
\begin{array}{ll}
\frac{\sin^{2}[\frac{1}{2}(\theta_{1} + \theta_{2})]}
{\cos^{2}[\frac{1}{2}(\theta_{1}-\theta_{2})]}
& \quad \mbox{for} \quad \mid \theta_{2} \mid \leq \theta{\rm c} \\
1 & \quad \mbox{for} \quad \theta_{\rm c} < \mid \theta_{2} \mid \leq \pi/2, 
\end{array}
\right.
\end{eqnarray}
with $\zeta \leq 1$, 
$\cos\theta_{1}=k_{1x}/k_{1}$ and $\cos\theta_{2}=k_{2x}/k_{2}$.
The critical angle $\theta_{\rm c}$ is defined as
$\arcsin(k_{1}/k_{2})$. 
For $\lambda=0$, Eq. (\ref{bound}) reproduces the 
previous results \cite{Iniotakis07}. 
As seen from Eq. (\ref{bound}), a zero energy ABS 
is only possible 
for $\mid \theta_{2} \mid \leq \theta_{\rm c}$ and 
$\gamma=1$, i.e. $\Delta_{p} > \Delta_{s}$.  
This ABS corresponds to a state in which a
localized quasiparticle can move along the edge.
The energy level of this edge state depends crucially
on the direction of the motion of the quasiparticle.
The inner gap edge modes are absent for large magnitude of $k_{y}$, i.e.
$\theta_{2}$. 
In this case, $k_{1x}$ becomes a purely imaginary number due
to the conservation of the Fermi momentum component parallel to the surface.
The parameter regime where the edge modes survive is reduced with
increasing $\lambda$. 
However, as far as we concentrate on normal injection,
the edge modes survive as midgap ABS \cite{Tanaka95,Hu}  
irrespective of the strength of $\lambda$. 
If we focus on the low energy limit, 
the ABS energy can be written as
\begin{eqnarray}
E=\pm \Delta_{p}\left(1 - \frac{\Delta^{2}_{s}}{\Delta^{2}_{p}}\right)
\frac{k_{1} + k_{2}}{2k_{1}k_{2}}k_{y}, 
\end{eqnarray}
with $\Delta_{s}<\Delta_{p}$
for any $\lambda$ with small magnitude of $k_{y}$. 
For $\Delta_{s} \geq \Delta_{p}$, the ABS 
vanishes since the value of right hand side of Eq.~(\ref{bound}) becomes 
negative, due to the negative sign of $\gamma$ 
for $\mid E\mid<\Delta_{1}$ and $\mid E \mid<\Delta_{2}$. 
It should be remarked that the ABS under consideration does not break
time reversal symmetry, since the edge currents carried by the two partners 
of the Kramers doublet flow in opposite directions.
Thus they can be regarded as helical edge modes, with
the two modes related to each other by a time reversal operation.

\subsubsection{Charge and spin conductance}

Now we turn to transport properties governed by 
the ABS in NCS superconductors \cite{Vorontsov,Linder,Borkje}.
First, we point out that the spin Hall effect, i.e., the 
appearance of the spin Hall voltage perpendicular to 
the superconducting current, is suppressed by the 
compressive nature of the superconducting state 
by the factor of $( k_{\rm F} \lambda_{\rm m})^{-2}$
($k_{\rm F}$: Fermi momentum, $\lambda_{\rm m}$: penetration depth) 
\cite{Furusaki}.  
Instead, we will show below that 
spin transport through the junction between 
a ballistic normal metal at $x<0$ and a NCS superconductor,
i.e., through a N/NSC junction, can be enhanced by the Doppler
effect during Andreev reflection. 
The Hamiltonian $\hat{\cal H}_{\rm N}$ of N is given by putting 
${\Delta}(\bm{k})=0$ and $\lambda=0$ in $\hat {\cal H}_{\rm S}$.
We assume an insulating barrier at $x=0$,
expressed by a delta-function potential $U \delta(x)$. 

The quantities of interest are the angle resolved spin
conductance $f_{\rm S}(\theta)$ and charge conductance $f_{\rm C}(\theta)$
defined by \cite{Kashiwaya99}
\begin{eqnarray}
f_{\rm S}(\theta)&=& \frac{1}{2}
\big[\sum_{\sigma,\rho} s_\rho 
(\mid a_{\sigma,\rho} \mid^{2} -
\mid b_{\sigma,\rho} \mid^{2} ) \big]
\cos \theta ,
\\
f_{\rm C}(\theta) &=& 
\big[1 + \frac{1}{2} \sum_{\sigma,\rho} (\mid a_{\sigma,\rho} \mid^{2} -
\mid b_{\sigma,\rho} \mid^{2} ) \big]
\cos \theta ,
\end{eqnarray}
where $s_\rho = +(-) 1 $ for $\rho=\uparrow (\downarrow) $, and
$\theta$ denotes the injection angle measured from the normal to the interface. 
Here, $b_{\sigma,\rho}$ and $a_{\sigma,\rho}$ with $\sigma,\rho \in \{\uparrow,\downarrow\}$ are spin-dependent reflection and Andreev reflection coefficients, respectively.
\begin{petit}
These coefficients are determined as follows.
The wave function for spin $\sigma$ in
the normal metal $\Psi_{\rm N}(x)$ 
is given by
\begin{eqnarray}
\Psi_{\rm N}(x)\!\!&=&\!\!\exp(\I  k_{{\rm F}y}y)
[(\psi_{i \sigma}+\sum_{\rho=\uparrow,\downarrow}a_{\sigma,\rho}\psi_{a\rho})\exp(\I k_{{\rm F}x}x)
+ \sum_{\rho=\uparrow,\downarrow}
b_{\sigma,\rho}\psi_{b\rho}\exp(-\I  k_{{\rm F}x}x)]
\end{eqnarray}
with
$\,^T\!\psi_{i\uparrow}=
\,^T\!\psi_{b\uparrow}
=\left(1,0,0,0 \right)$, 
$\,^T\!\psi_{i\downarrow}=\,^T\!\psi_{b\downarrow}
=\left(0,1,0,0 \right)$, 
$\,^T\!\psi_{a\uparrow}
=\left(0,0,1,0 \right)$, and 
$\,^T\!\psi_{a\downarrow}
=\left(0,0,0,1 \right)$. 
The corresponding $\Psi_{\rm S}(x)$ is given by Eq. (\ref{wavefunction}). 
The coefficients $a_{\sigma,\rho}$ and $b_{\sigma,\rho}$ are 
determined by postulating the boundary condition 
$\Psi_{\rm N}(0)=\Psi_{\rm S}(0)$, and 
$\hbar \hat{v}_{{\rm S}x}\Psi_{\rm S}(0)-\hbar \hat{v}_{{\rm N}x}\Psi_{\rm N}(0)
=-2iU\hat{\tau}_{3}\Psi_{\rm S}(0)$ 
with $\hbar \hat{v}_{{\rm S(N)}x}=\partial \hat{H}_{\rm S(N)}/\partial k_{x}$, 
and the diagonal matrix $\hat{\tau}_{3}$ given by 
$\hat{\tau}_{3}={\rm diag}(1,1,-1,-1)$. 
\end{petit}

\begin{figure}[b]
\sidecaption[b]
\scalebox{0.5}{
\includegraphics[width=10cm,clip]{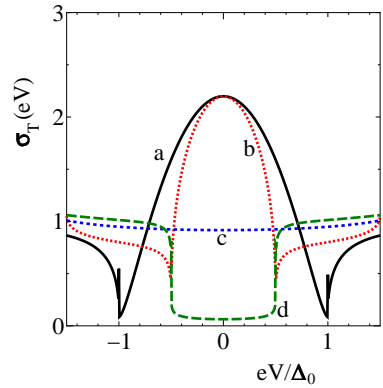}
}
\caption{(Color online) $\sigma_{\rm T}$  for NCS superconductor with 
$2m \lambda /k_{\rm F}\hbar^{2}=0.1$ and  $Z=5$.
a: $\Delta_{s}=0$, b: $\Delta_{s}=0.5\Delta_{p}$, 
c: $\Delta_{s}=\Delta_{p}$ 
and 
d: $\Delta_{s}=1.5\Delta_{p}$. 
 }
\label{fig:5}
\end{figure}

The resulting angle averaged charge conductance (tunneling conductance) 
is given by 
\begin{equation}
\sigma_{\rm C}\equiv \sigma_{\rm T}=
\left(\int^{\pi/2}_{-\pi/2} f_{\rm C}(\theta ) d\theta \right)/\left(
\int^{\pi/2}_{-\pi/2} f_{\rm NC}(\theta ) d\theta \right). 
\end{equation}
We plot in Fig.~\ref{fig:5} the charge conductance by changing the 
ratio of $\Delta_{s}/\Delta_{p}$ in the presence of the 
splitting of the Fermi surface \cite{Rapid}. 
For $\Delta_{s}<\Delta_{p}$, $\sigma_{\rm T}(eV)$ has a ZBCP due to the 
presence of the helical edge modes (curves $a$ and $b$ in Fig. \ref{fig:5}). 
For $\Delta_{s}=\Delta_{p}$, due to the closing of the bulk energy gap, the 
resulting $\sigma_{\rm T}(eV)$ is almost constant. 
For $\Delta_{s}>\Delta_{p}$,  $\sigma_{\rm T}(eV)$ has a gap like structure 
similar to spin-singlet $s$-wave superconductor.

\begin{figure}[b]
\begin{center}
\scalebox{1.0}{
\includegraphics[width=10cm,clip]{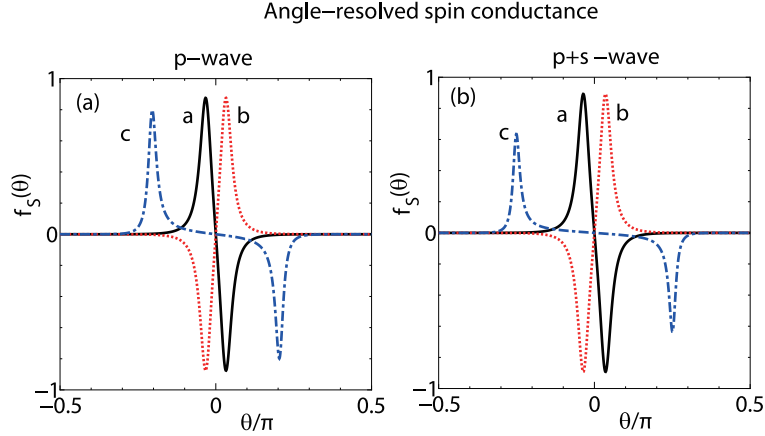}
}
\end{center}
\caption{(Color online) Angle resolved spin conductance for $Z=5$.
a: $eV=0.1\Delta_{p}$, b: $eV=-0.1\Delta_{p}$
and
c: $eV=0.6\Delta_{p}$ with $\lambda k_{\rm F}=0.1\mu$. 
(a) pure $(p_{x}\pm \I p_{y})$-wave case with $\Delta_{s}=0$;
(b) $\Delta_{s}=0.3\Delta_{p}$. 
[from Fig. 2 Phys. Rev. B \textbf{79}, 060505(R) (2009).]  
}
\label{fig:06}
\end{figure}

Next, we focus on the spin conductance. 
First we consider a pure $(p_{x} \pm \I p_{y})$-wave state. 
In Fig. \ref{fig:06}, the
angle resolved spin conductance is plotted as a function of
injection angle $\theta $ and bias voltage $V$ with $E=eV$.  
Note here that the $k_y$ is related to 
$\theta $ as $k_{y} = k_{\rm F} \sin \theta $. 
It is remarkable that the spin conductance has a non zero value although the
NCS superconductor does not break time reversal symmetry. 
The quantity
$f_{\rm S}(\theta )$ has a peak when the angle $\theta $ or $k_y$  corresponds to
the Andreev bound state energy $E$ in the energy dispersion.
With this condition, the spin-dependent Andreev reflection 
occurs to result in a spin current. 
Besides this property, we can show that
$f_{\rm S}(\theta )=-f_{\rm S}(-\theta )$ is satisfied. By changing the sign of
$eV$, $f_{\rm S}(\theta )$ changes sign as seen in Fig. \ref{fig:06}(a). 
Next, we look at the case where an $s$-wave component coexists. 
We calculate the spin conductance similar to that for the pure 
$(p_{x} \pm \I p_{y})$-wave case. 
For $\Delta_{s} < \Delta_{p}$, where helical edge modes exist, 
$f_{\rm S}(\theta )$ shows a sharp peak and 
$f_{\rm S}(\theta )=-f_{\rm S}(-\theta )$ is satisfied [see Fig. \ref{fig:06}(b)]. 
These features are similar to those of the pure
$(p_{x} \pm \I p_{y})$-wave case. 
On the other hand, for  
$\Delta_{s} > \Delta_{p}$, where the helical edge modes are absent, 
sharp peaks of $f_{\rm S}(\theta )$ as shown in Fig. \ref{fig:06} are absent. 

We have checked that there is negligible quantitative change, i.e., less 
than 0.5\% change of the peak height, 
by taking the $\lambda=0$ limit compared to Fig.~\ref{fig:06}.
In this limit, for the pure $(p_{x} \pm \I p_{y})$-wave state,
$f_{\rm S}(\theta )$ is given simply as follows 
\[
\frac{-8RD^{2}\sin2\theta  \sin2\varphi \cos\theta }
{\mid 4(\sin^{2}\theta  - \sin^{2}\varphi) 
+ D[2 \cos2\theta  -(1+R)\exp(-2\I  \varphi) ]\mid^{2}}
\label{spincurrent}
\]
for $ \mid E \mid< \Delta_{p}$ and $f_{\rm S}(\theta )=0$ for 
$ \mid E \mid > \Delta_{p}$ with 
$\sin \varphi = E/\Delta_{p}$
The transparency of the interface 
$D$ is given as before by $4\cos^{2} \theta /(4\cos^{2} \theta + Z^{2})$,
with the dimensionless constant $Z=2mU/\hbar^{2}k_{\rm F}$.
The magnitude of $f_{\rm S}(\theta )$ is largely enhanced at $E= \pm \Delta_{p}\sin\theta $ corresponding to the energy dispersion of the ABS.
The origin of the nonzero $f_{\rm S}(\theta )$ even for  $\lambda=0$ is due to
spin-dependent Andreev bound states.
We have checked that even if  we take into account the 
spatial dependence of the $(p_{x} \pm \I p_{y})$-wave 
pair potential explicitly, 
the resulting $f_{\rm S}(\theta )$ does not qualitatively change \cite{Vorontsov}. 

Summarizing these features, we can conclude that the presence of the helical
edge modes in NCS superconductors is the
origin of the large angle resolved spin current through
normal-metal/NCS superconductor junctions. However, the angle
averaged normalized spin conductance becomes zero since
$f_S(\theta )=-f_S(-\theta )$ is satisfied. 

Magnetic field offers an opportunity to observe the spin current in a 
more accessible way, where the time reversal ($\tens{T}$) symmetry is broken by the 
shielding current at the interface. Here
we consider the angle averaged normalized spin conductance 
$\sigma_{\rm S}$ and charge conductance $\sigma_{\rm C}$ 
as a function of magnetic field. 
The spin conductance is given by \cite{Kashiwaya99}
\begin{equation}
\sigma_{\rm S}=\left(
\int^{\pi/2}_{-\pi/2} f_{\rm S}(\theta ) d\theta  \right) /
\left(\int^{\pi/2}_{-\pi/2} f_{\rm NC}(\theta ) d\theta \right),
\end{equation}
where $f_{\rm NC}(\theta )$ denotes the angle resolved 
charge conductance in the normal state with 
$\Delta_{p}=\Delta_{s}=0$. 
We consider a magnetic field $H$ applied 
perpendicular to the two-dimensional plane, which induces
a shielding current along the normal-metal/NCS superconductor interface.
When the penetration depth of the NCS superconductor
is much longer than the coherence length, the vector potential
can be approximated as described in section ~\ref{chiral}.
As in the case of a chiral $p$-wave superconductor,
the applied magnetic field shifts the quasiparticle energy 
$E$ 
to $E + H\Delta_{p}\sin \theta /H_{0}$.
For typical values of $\xi \sim 10$ nm, $\lambda_{\rm m} \sim 100$ nm,
the magnitude of $H_0$ is of the order of 0.2 Tesla.
The order of magnitude of the 
Doppler shift is given by  $H \Delta_{p}/H_{0}$. 
Since the Zeeman energy is given by 
$\mu_{\rm B}H$, the energy shift due to the Doppler effect is 
by a factor $k_{\rm F}\lambda_{\rm m}$ larger than that due to the Zeeman effect.  
Thus, we can neglect the Zeeman effect in the present analysis.
This is in sharp contrast to quantum spin Hall systems where the 
Zeeman effect is the main effect of a magnetic field, 
which opens a gap in the helical edge modes and modulates the transport 
properties \cite{Konig}. The enhanced spin current due to Doppler 
shifts is specific to the superconducting state, and is
not realized in quanum spin Hall systems. 

\subsubsection{Topological aspects}
\label{topol}
We now focus on the topological 
aspect of non-centrosymmetric superconductors. 
Recently, the concept of the quantum Hall system has been generalized to 
time-reversal ($\tens{T}$) symmetric systems, i.e., quantum spin Hall systems
\cite{Mele,Bernevig,Fu}. A quantum spin Hall system could be regarded as  
two copies of a quantum Hall system, for up and down spins, that are
characterized by opposite chiralities. 
In the generic case, however, a mixture of up and 
down spins occurs due to spin-orbit interaction,
which necessitates a new topological number to characterize
a quantum spin Hall system \cite{Mele,Fu}. 
In quantum spin Hall systems, there exist helical edge modes, i.e.,
time-reversed partners of right- and left-going 
one-dimensional modes. This has been
experimentally demonstrated for the quantum well of the HgTe system by
measurements of the charge conductance \cite{Konig}.

As shown in Fig. \ref{fig:06}, to discuss the topological nature of the 
helical edge modes, 
it is sufficient to consider the pure $(p_{x} \pm \I p_{y})$-wave state. 
Here, we give an argument 
from the viewpoint of the $Z_2$ (topological) class \cite{Mele},
why the superconducting state with $\Delta_{p}>\Delta_{s}$ has an 
Andreev bound state.
We commence with a pure $(p_{x} \pm \I p_{y})$-wave state 
without spin-orbit interaction, i.e. $\lambda=0$. 
The spin Chern number \cite{Fu} for the corresponding
Bogoliubov-de Gennes Hamiltonian is 2. 
Turning on $\lambda$ adiabatically leaves the time reversal
$\tens{T}$-symmetry intact and keeps the gap open. 
Upon this adiabatic change of $\lambda$, the number of 
the helical edge mode pairs
does not change. 
The reason is that this number is a topological number and
consequently can only change by integer values.
We now increase the magnitude of $\Delta_{s}$ from zero.
As far as $\Delta_{p} > \Delta_{s}$ is satisfied, 
the number of helical edge modes does not change. 
However, if $\Delta_{s}$ exceeds $\Delta_{p}$, the helical mode 
disappears. In this regime, the topological nature of 
the superconducting state belongs 
to a pure $s$-wave state with $\lambda=0$.  
It is remarkable that just at $\Delta_{s}=\Delta_{p}$ one of the 
two energy gaps for quasiparticles in the bulk closes.
At precisely this point a quantum phase transition occurs. 

\begin{figure}[b]
\begin{center}
\scalebox{1.0}{
\includegraphics[width=10.0cm,clip]{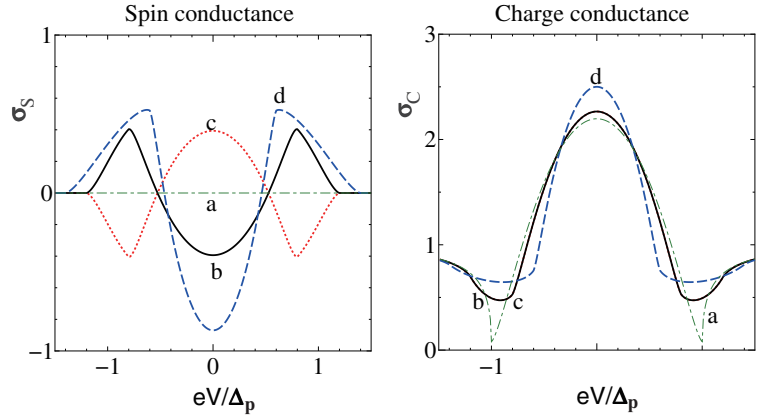}
}
\end{center}
\caption{(Color online) Angle averaged
spin conductance and charge conductance
as a function of $eV$ with bias voltage $V$ with 
$\lambda k_{\rm F}=0.1\mu$.
a: $H=0$, b: $H=-0.2H_{0}$, c:
$H=0.2H_{0}$, and d: $H=-0.4H_{0}$. Curves $b$ and $c$ of the right panel 
are identical.
[from Fig. 2 Phys. Rev. B \textbf{79}, 060505(R) (2009).]  
}
\label{fig:07}
\end{figure}

In the following, we discuss the pure $(p_{x} \pm \I p_{y})$-wave case in
more detail.
In Fig. \ref{fig:07}, the spin conductance $\sigma_{\rm  S}$ and charge conductance
$\sigma_{\rm C}$ normalized by the charge conductance in
the normal state are plotted. 
It should be noted that
$\sigma_{\rm S}$ becomes nonzero in the presence of a magnetic field
$H$ (see curves $b$, $c$ and $d$), since  $f_{\rm S}(\theta )$ 
is no more an odd function of 
$\theta $ due to the imbalance of the helical edge modes. 
For $\lambda=0$, the corresponding 
helical edge modes are given  by 
$E= \Delta_{p}(1 - H/H_{0})\sin\theta $ and $E= -\Delta_{p}(1 + H/H_{0})\sin\theta $. As seen from the curves $b$ and $c$, the
sign of $\sigma_{\rm S}$ is reversed when changing the direction of the
applied magnetic field. 
On the other hand, the corresponding charge
conductance has different features. For $H=0$, the resulting line
shape of $\sigma_{\rm C}$ is the same as that for a chiral $p$-wave
superconductor (see curve $a$ of right panel)
\cite{Yokoyama,Iniotakis07,Linder}. 
As seen
from curves $b$ and $c$ in the right panel, $\sigma_{\rm C}$ does not change
with the direction of the magnetic field $H$. 

In summary, we have clarified the charge and
spin transport properties of non-centrosymmetric
superconductors from the viewpoint of topology and 
Andreev bound state. We have
found an incident angle dependent spin polarized current flowing
through the interface. When a weak magnetic field is applied, even the
angle-integrated current is largely spin polarized. 
In analogy to quantum spin Hall systems,
the Andreev bound states in non-centrosymmetric superconductors 
corresponds to helical edge modes.
Andreev reflection via helical edge modes 
produces the enhanced spin current specific to non-centrosymmetric superconductors.

\section{Quasiclassical Theory of Superconductivity for 
Non-Centrosymmetric Superconductors} 
\label{qcl}

\newcommand{\be}{\begin{equation}}
\newcommand{\ee}{\end{equation}}
\newcommand{\bea}{\begin{eqnarray}}
\newcommand{\eea}{\end{eqnarray}}
\def\ul#1{\underline{#1}}
\def\sm#1{{\mbox{\tiny #1}}}
\newcommand{\avec}[1]{\stackrel{\to }{#1}}
\renewcommand{\vec}[1]{\bm{#1}}

\def\cH{{\cal H}}
\def\cY{{\cal Y}}
\def\cM{{\cal M}}
\def\grad{{\bf \nabla }}
\newcommand{\vsigma}{\mbox{$\bm{\sigma}$}}
\newcommand{\vgamma}{\mbox{$\bm{\gamma}$}}
\newcommand{\vPi}{\mbox{$\bm{\Pi}$}}
\renewcommand{\phi}{\varphi}
\renewcommand{\epsilon}{\varepsilon}
\newcommand{\ud}{\uparrow,\downarrow}
\renewcommand{\u}{\uparrow}
\renewcommand{\d}{\downarrow}
\newcommand{\ket}[1]{| {#1}\rangle}
\newcommand{\bra}[1]{\langle {#1}|}

\def\vk{{\vec k}}
\def\vp{{\vec p}}
\def\vx{{\vec x}}
\def\vy{{\vec y}}
\def\vv{{\vec v}}
\def\vR{{\vec R}}
\def\vq{{\vec q}}
\def\vg{{\vec g}}
\def\vz{{\vec z}}
\def\vn{{\vec n}}
\def\vJ{{\vec J}}
\newcommand{\vare}{\varepsilon}
\newcommand{\hg}{\hat{g}}
\newcommand{\ha}{\hat{a}}
\newcommand{\hf}{\hat{f}}
\newcommand{\hp}{\hat{p}}
\newcommand{\hk}{\hat{k}}

\subsection{Quasiparticle Propagator}

Electronic quasiparticles in normal Landau Fermi liquids are restricted
in phase space to a region that comprises only a small part of
the entire electronic phase space \cite{landau57,landau59}. 
It consists of a narrow
(compared to the Fermi momentum $\vp_{\rm F}$) shell around
the Fermi surface, and a small (compared to the Fermi energy
$E_{\rm F}$)
region around the chemical potential.
Quasiparticles are characterized by their spin and charge, and
their group velocity is the Fermi velocity, $\vv_{\rm F} (\vp_{\rm F})$.
Quasiclassical theory is the appropriate framework to describe such a system.
It consists of a {\it systematic} classification of all
interaction processes according to their relevance, i.e. their smallness
with respect to an expansion parameter {\sc small} \cite{Serene,rai86,eschrig94,rai95,esc99fluk}. 
This expansion parameter
assumes the existence of a well defined scale separation between a {\it low-energy scale}
and a {\it high-energy scale}. 

Superconducting phenomena are governed by the low-energy scale. That means that
the energy scales determined by
the energy gap $\Delta $ and the transition temperature $T_{\rm c}$ are small.
In contrast the energy scales determined by the Fermi energy $E_{\rm F}$ or
the Coulomb repulsion $U_{\rm C}$ are large energies. 
Disorder can be described within quasiclassical approximation as long as
the energy associated with the scattering rate, $\hbar/\tau $, 
is classified as a small energy. 
A systematic classification shows that a consistent 
treatment of disorder requires the $t$-matrix approximation.
Localization effects due to disorder are beyond the leading order precision
of quasiclassical theory.
Associated with the energy scales are small and large length scales. For example
the superconducting coherence length $\xi_0=\hbar v_{\rm F}/2\pi k_{\rm B}T_{\rm c}$,
and the elastic mean free path $\ell = v_{\rm F} \tau $ are large compared to
the lattice constant $a$ and the Fermi wave length $\lambda_{\rm F}=\hbar /p_{\rm F}$.

This separation in energy and length scales is associated with a low-energy region in 
phase space, that includes low quasiparticle energies $\sim \Delta $,
$k_{\rm B} T$,
and a momentum shell around the quasiparticle Fermi momentum 
$\vp_{\rm F}$ of extend
$\delta p \sim \Delta/|\vv_{\rm F}(\vp_{\rm F})|$.  The phase space volume of this
low-energy region, divided by the entire phase space volume, is employed for 
a systematic diagrammatic expansion of a Dyson series within a path-ordered
Green's function technique (e.g. Matsubara technique for the Matsubara path,
Keldysh-Nambu-Gor'kov technique for the Schwinger-Keldysh path).
Within the framework of Green's function technique, all diagrams in a
Feynman diagrammatic expansion can be classified according to their
order in this expansion parameter, which is denoted as {\sc small}.
The leading order theory in this expansion parameter is called
the ``Quasiclassical Theory of Metals and Superconductors'' \cite{larkin68, eilen,Serene}.

The possibility to define a quasiparticle Fermi surface around which all
quasiparticle excitations reside is a requirement for the quasiclassical theory
to work. 
Its presence ensures that the Pauli principle is still effective 
in placing stringent kinetic restrictions on the possible scattering events.
It is essential to note that such a definition need not be
sharp, i.e. the theory is not restricted to normal Fermi liquids with 
a jump in the momentum distribution at zero temperature. Thus, the theory
includes superconducting phenomena as well as strong coupling metals.
It is convenient to introduce a local coordinate system at each momentum 
point of the Fermi surface $\vp_{\rm F}$,
with a variation along the surface normal, i.e. 
in direction of the Fermi velocity $\vv_{\rm F}(\vp_{\rm F})$,
that is determined by a variable $\xi_\vp $ (this variable is zero at the
Fermi momentum), and a
tangential variation along the Fermi surface at constant $\xi_\vp $.
A consistent approximation requires to consider the Fermi velocity constant
across the low-energy momentum shell, and thus the local coordinate system
stays an orthogonal system as long as $\xi_\vp $ varies within this momentum
shell, and furthermore, $\xi_\vp$ stays small within this momentum shell.
The coordinate $\xi_\vp $ around each Fermi surface point $\vp_{\rm F}$
varies then approximately as $\xi_\vp \approx \vv_{\rm F} (\vp -\vp_{\rm F})$.

The quasiclassical theory is obtained by defining quasiparticle
propagators for the low-energy regions of the phase space, and in combining
all diagrams involving Green's functions with their variables residing in
the high-energy regions into new effective {\it high-energy interaction vertices}.
This process of integrating out high-energy degrees of freedom is 
highly non-trivial and must be solved by microscopic theories. In the spirit of
Fermi liquid theory it is, however, possible to regard all high-energy 
interaction vertices as phenomenological parameters of the theory. 
In quasiclassical approximation they do not depend on any low-energy variables
as temperature or superconducting gap, and they do not vary as function of
$\xi_\vp$ as long as $\xi_\vp $ stays within the momentum shell that harbors the
quasiparticle excitations.  However they do depend in general on
the position of the Fermi momentum on the Fermi surface.

In addition to introducing new effective interaction vertices the above procedure
also introduces a quasiparticle renormalization factor $a^2(\vp)\sim 1/Z(\vp) $, 
that is due to the self energies of the low-energy
quasiparticles moving in the background of
the high energy electrons. This renormalization leads to a modification
of the quasiparticle Fermi velocity compared to the bare Fermi velocity of the
system, and to a deformation of the quasiparticle Fermi surface compared to the
bare Fermi surface. It also determines quasiparticle weight as the
residua of the quasiparticle poles in the complex energy plane.

One has to keep these remarks in consideration when including 
additional interaction, like spin-orbit interaction or exchange interaction, 
in a quasiclassical theory.
First, it is important to decide if this interaction is going to be
treated among the low-energy terms or among the high-energy terms.
Depending on that one obtains two different quasiclassical theories, that
cannot in general adiabatically be connected with each other.
Going from one limit to the other includes the un-dressing of all effective
interaction vertices and of the quasiparticles, and re-dressing with
new types of effective interaction vertices and self energies.
Importantly, this dressing leads to strongly spin dependent effective
interactions and quasiparticle renormalizations in one limit, 
and to leading order spin-symmetric interactions and quasiparticle renormalizations
in the other limit.
The former case, when spin-dependent interactions are included in 
the high-energy scale, leads to a complete reorganization of the Fermi surface
geometry, with in general new spin-dependent quasiparticle energy bands. In this
case, it is not sensible anymore to keep the spin as a good quantum number, 
but it is necessary to deal directly with the representation that diagonalizes
the energy bands including the spin-dependent interaction. In the case of
a strong exchange energy this leads to exchange split energy bands, and in the
case of strong spin-orbit interaction this leads to helicity bands.

The basic quantities in the theory are the quasiparticle Fermi surface,
the quasiparticle velocity, and quasiparticle interactions. 
Here we give a short sketch of how they enter the theory.
The bare propagator (without inclusion of exchange interaction or spin-orbit
coupling) in the quasiparticle region of the phase space has the
general structure
\begin{equation}
\label{bareG}
G^{(0)}_{\alpha\beta }(\vec{p},\epsilon ) =
\frac{\delta_{\alpha \beta }}{\epsilon - \xi^{(0)}_{}(\vec{p})}
\end{equation}
where $\xi^{(0)} (\vec{p})$ is the bare energy dispersion of the energy band
(measured from the electrochemical potential of the electrons).
It does not include electron-electron interaction effects yet, and thus
determines a {\it bare } Fermi surface that does not coincide with the
quasiparticle Fermi surface defined below.
The quantum number $\alpha $ labels the spin.
The leading order self energy is solely due to coupling of
low-energy electrons (superscript L) to high-energy electrons (superscript
H), and consequently the corresponding self energy, $\Sigma^{\rm (H)}$,
must be classified as a pure high-energy quantity. 
In general, when either exchange interaction or spin-orbit coupling are
large energy scales, this self energy contribution will be spin-dependent
(and will ultimately lead to new, spin-split energy bands as explained below).
The self energy $\Sigma^{\rm (H)}$ is,
however, slowly varying in energy on the low-energy scale 
and thus can be expanded around the chemical potential,
\begin{equation}
\Sigma^{\rm (H)}_{\alpha \beta}(\vec{p},\epsilon )=
\Sigma^{\rm (H)}_{\alpha \beta}(\vec{p},0 ) + \epsilon \partial_\epsilon
\Sigma^{\rm (H)}_{\alpha \beta}(\vec{p},\epsilon )\Big|_{\epsilon =0}
+ {\cal O}(\epsilon^2 ).
\end{equation}
The second term can be combined with energy $\epsilon $ 
into a renormalization function
\be
\epsilon - \epsilon \partial_\epsilon
\Sigma^{\rm (H)}_{\alpha \beta}(\vec{p},0)= Z^{\rm (H)}_{\alpha \beta}(\vec{p})\epsilon ,
\ee
such that to leading order the Dyson equation for the low-energy propagator
$G^{\rm (L)}_{\beta \gamma }$ reads 
\be
\left\{ Z^{\rm (H)}_{\alpha \beta}(\vec{p})\epsilon 
- [\xi^{(0)}_{}(\vec{p})\delta_{\alpha \beta }+\Sigma^{\rm (H)}_{\alpha \beta}(\vec{p},0 ] 
- \Sigma^{\rm (L)}_{\alpha \beta} (\vec{p}, \epsilon ) \right\}
\otimes G^{\rm (L)}_{\beta \gamma }(\vec{p},\epsilon ) =
\delta_{\alpha \gamma },
\end{equation}
where $\Sigma^{\rm (L)}$ includes all self energy terms of
order {\sc small }.
Here, and in the following, summation over repeated indices is implied.
The $\otimes$ sign accounts for possible spatial or temporal inhomogeneities,
in which case it has the form of a convolution product in
Wigner representation (see Ref.~\cite{Serene}
for details).
The equation holds in this form either in Matsubara 
or in Keldysh representation (in which case all quantities are 2$\times $2
matrices in Keldysh space \cite{keldysh64}).
Low-energy excitations reside in momentum regions 
differing considerably from that for the bare propagators Eq.~(\ref{bareG}).

The quantities $Z^{\rm (H)}_{\alpha \beta }(\vec{p})$ and
$\Sigma^{(H)}_{\alpha \beta}(\vec{p},0 )$ can be defined such that they
have real eigenvalues.
The next step is to eliminate the high-energy renormalization factor
$Z^{\rm (H)}_{\alpha \beta }(\vec{p})$ from the low-energy theory.
This is done with the help of quasiparticle weight factors $a_{\alpha \beta}(\vp)$,
that are the solution of
\begin{equation}
a_{\alpha \gamma} (\vec{p})
Z^{\rm (H)}_{\gamma \gamma'}(\vec{p})
a_{\gamma' \beta}(\vec{p})=\delta_{\alpha \beta}.
\end{equation}
They exist as long as $Z^{\rm (H)}_{\gamma \gamma'}$ has non-zero
eigenvalues.
Then we can define the {\it quasiparticle Green's function} $G^{\rm (QP)}_{\alpha \beta}$
as the solution of 
\begin{equation}
a_{\alpha \gamma} (\vec{p}) G^{\rm (QP)}_{\gamma\gamma'} a_{\gamma' \beta} (\vec{p}) =
G^{\rm (L)}_{\alpha \beta }(\vec{p},\epsilon) ,
\end{equation}
which exists under the condition that $ a_{\alpha \gamma}$ has non-zero 
eigenvalues (i.e. the quasiparticle weights are non-zero; otherwise
the quasiparticle approximation breaks down).
It fulfills the Dyson equation
\begin{equation}
\left[\epsilon - \xi^{QP}(\vec{p})
- \Sigma^{QP}(\vec{p},\epsilon )\right]_{\alpha \beta}
\otimes G^{(QP)}_{\beta \gamma }(\vec{p},\epsilon ) = \delta_{\alpha \gamma }
\end{equation}
with the quasiparticle dispersion
\begin{equation}
\xi^{(QP)}_{\alpha \beta }(\vec{p})=
a_{\alpha \gamma} (\vec{p})
\left(\xi^{(0)}_{}(\vec{p}) \delta_{\gamma \gamma'} + \Sigma^{(H)}_{\gamma \gamma' }(\vec{p},0 )\right)
a_{\gamma' \beta} (\vec{p})
\end{equation}
and the quasiparticle self energies
\begin{equation}
\Sigma^{(QP)}_{\alpha \beta }(\vec{p},\epsilon)=
a_{\alpha \gamma} (\vec{p})
\Sigma^{(L)}_{\gamma \gamma' }(\vec{p},\epsilon )
a_{\gamma' \beta} (\vec{p}) .
\end{equation}
The effective (renormalized by high-energy processes)
interactions vertices for the low-energy propagators,
$G^{\rm (L)}_{\alpha \beta }$,
which enter the diagrammatic expressions for the quasiparticle self energy,
have the general structure
$V_{\beta_1 \ldots \beta_n }(\epsilon_1,\vec{p}_1;\ldots ;\epsilon_n,\vec{p}_n)$.
In leading order the energy dependence of these vertices can be neglected near
the chemical potential, i.e. the
arguments can be restricted to the chemical potential.
Furthermore,
instead of working with $G^{\rm (L)}_{\alpha \beta }$ and $V_{\beta_1 \ldots \beta_n }$ 
the common and completely equivalent description in terms of the above defined
quasiparticle propagators,
$G^{\rm (QP)}_{\alpha \beta }$, and renormalized quasiparticle interactions,
$V^{\rm (QP)}_{\beta_1\ldots \beta_n}$, given by
\begin{eqnarray}
&&
V^{\rm (QP)}_{\beta_1\ldots \beta_n}(\vec{p}_{1}\ldots \vec{p}_{n})=
a_{\beta_1\beta'_1}(\vec{p}_{1}) \ldots
a_{\beta_n \beta'_n}(\vec{p}_{n})
V_{\beta'_1 \ldots \beta'_n }(0,\vec{p}_{1};\ldots ;0,\vec{p}_{n}) 
\end{eqnarray}
can be used.

It is important to note that the quasiparticle self energies can be written down
as functionals of the quasiparticle Green's functions only in 
leading order in the expansion in {\sc small }, which is the order relevant for the
quasiclassical approximation. 
In this case, the quasiparticle weights have disappeared from the 
theory and cannot
in principle be determined from low-energy processes that only involve 
quasiparticle dynamics. They must be obtained from a microscopic theory
by considering high energy scattering processes,
which is beyond the quasiclassical approximation.

It is obvious, that the appearance of the quasiparticle renormalization factors
renders all self energies and interactions non-diagonal in spin unless
spin-dependent interactions are small enough the be omitted from the
high-energy quantities.
From the above expressions one obtains
the quasiparticle Fermi surfaces by diagonalizing the quasiparticle dispersion
\begin{equation}
U_{\vp \lambda \alpha} \xi^{(QP)}_{\alpha \beta}(\vec{p}) =
\xi^{(QP)}_{\lambda }(\vec{p}) U_{\vp \lambda \beta} 
\end{equation}
with band index $\lambda $, and solving the equation
\begin{equation}
\xi^{(QP)}_\lambda (\vp) =0 \to
\vp=\vec{p}_{F}^{\lambda }.
\end{equation}
The corresponding quasiparticle Fermi velocity is then given by
\begin{equation}
\vec{v}_{F}^{\lambda }=
\frac{\partial}{\partial {\vec{p}} }
\xi^{(QP)}_{\lambda }(\vec{p}) 
\Big|_{\vec{p}=\vec{p}_F^{\lambda }}  .
\end{equation}
In the band diagonal frame, the quasiparticle propagator is given by,
\begin{equation}
\left\{ [\epsilon - \xi_\lambda^{QP}(\vec{p})] \delta_{\lambda \lambda_1 }
- \Sigma^{QP}_{\lambda \lambda_1 } (\vec{p},\epsilon ) \right\}
\otimes G^{(QP)}_{\lambda_1 \lambda' }(\vec{p},\epsilon ) = \delta_{\lambda \lambda' },
\end{equation}
where the self energy (and all interactions in the self energy expressions)
must be transformed accordingly, e.g.
\begin{equation}
\Sigma^{(QP)}_{\lambda \lambda'}(\vec{p},\epsilon ) 
=
U^{\; }_{\vp \lambda \alpha} \Sigma^{(QP)}_{\alpha \beta}(\vec{p},\epsilon)
U^\ast_{\vp \beta \lambda' }  .
\end{equation}
In the next section this procedure is carried out for the case of
a strong spin-orbit interaction, e.g. appropriate for some
non-centrosymmetric materials.

\subsection{Spin-orbit interaction and Helicity representation}

As discussed in the introductory chapter of this book, 
for treating a non-centrosym\-metric material it is convenient to perform a
canonical transformation from a spin basis with fermion
annihilation operators $a_{\vk \alpha }$ for spin
$\alpha=\uparrow,\downarrow $ to the so-called helicity basis with
fermion annihilation operators
$c_{\vk \lambda }$ for helicity $\lambda =\pm $. This canonical
transformation diagonalizes the kinetic part of the Hamiltonian,
    \be
\cH_{kin} = \sum_{\vk} \sum_{ \alpha \beta=\uparrow,\downarrow } 
\left[\xi^{\; }(\vk ) + \vg^{\; }(\vk ) \cdot \vsigma)_{\alpha \beta } \right]
a^\dag_{\vk \alpha }
a^{\; }_{\vk \beta}
 = \sum_{\vk } \sum_{\lambda =\pm }\xi^{\; }_{ \lambda }(\vk )  c^\dag_{\vk \lambda } c^{\; }_{\vk \lambda } \,.
\label{eq:HN}
    \ee
Here, $\xi(\vk )$ is the band dispersion
relative to the chemical potential in the absence of spin-orbit
interaction, 
$\vg(\vk )$ is the spin-orbit pseudovector, which is
odd in momentum, $\vg({-\vk}) =-\vg(\vk )$,
and $\vsigma $ is the vector of Pauli matrices.
The resulting helicity band dispersion is 
\be
\xi_{\pm}(\vk ) = \xi(\vk )\pm |\vg(\vk )| . 
\ee
As is easily seen,
spin-orbit interaction locks the orientation of the
quasiparticle spin with respect to its momentum in each helicity
band.
The Hamiltonian, Eq.~(\ref{eq:HN}), is time reversal invariant, however
lifts the spin degeneracy. 

It is convenient to introduce polar and azimuthal angles for the vector
$\vg$, defined by $\{g_x,g_y,g_z\}=|\vg| \{
\sin(\theta_\vg) \cos(\phi_\vg),
\sin(\theta_\vg) \sin(\phi_\vg),
\cos(\theta_\vg) \}$ (where $0\le \theta_\vg \le \pi $).
In terms of those,
the transformation from spin to helicity basis, $U_{\vk \lambda \alpha }$, is defined
by \cite{Frigeri}
\be
U_{\vk  \lambda \alpha } = \left(
\begin{array}{cc}
\cos(\theta_\vg/2) & \sin(\theta_\vg/2) \E^{-i \phi_\vg} \\
-\sin(\theta_\vg/2) \E^{i \phi_\vg} & \cos(\theta_\vg/2)
\end{array} \right) \,, \qquad
c_{\vk \lambda } = \sum_{\alpha } U_{\vk \lambda \alpha} a_{\vk \alpha } .
\label{eq:U}
\ee
Obviously,
$\sum_{\alpha \beta }U^{\; }_{\vk  \lambda \alpha }
[\vg^{\; }(\vk )\cdot \vsigma_{\alpha \beta }] U^\ast_{\vk  \lambda' \beta } = 
|\vg(\vk) | \sigma^{(3)}_{\lambda \lambda' }$. 

For the superconducting state the Nambu-Gor'kov formalism is appropriate \cite{gorkov58}.
The Nambu spinor, $\hat A_{\vk } = (a_{\vk \uparrow},
a_{\vk \downarrow}, a^{\dagger }_{-\vk \uparrow},
a^{\dagger }_{-\vk \downarrow})^T$ transforms under the above canonical transformation into
the helical object
$\hat C_{\vk } = (c_{\vk +},
c_{\vk -}, c^{\dagger }_{-\vk +},
c^{\dagger }_{-\vk -})^T$, where
\be
\hat{C}_{\vk } = \hat U_{\vk } \hat{A}_{\vk }, \qquad \hat{U}_\vk  =  \left(
\begin{array}{cc} U_\vk & 0 \\ 0 & U_{-\vk}^\ast \end{array} \right) .
\ee
Correspondingly, one can 
construct $4\times 4$ retarded Green's functions in 
spin basis,
\be 
\hat G^{(s)}_{\vk_1 \vk_2}(t_1,t_2)=
-i\theta(t_1-t_2) \langle
\big\{ \hat A_{\vk_1}(t_1), \hat A^\dagger_{\vk_2} (t_2) \big\} \rangle_{\cal H},
\ee
and in helicity basis,
\be 
\hat G_{\vk_1 \vk_2}(t_1,t_2)=
-i\theta(t_1-t_2) \langle
\big\{ \hat C_{\vk_1}(t_1), \hat C^\dagger_{\vk_2} (t_2) \big\} \rangle_{\cal H},
=
\hat U_{\vk_1} 
\hat G^{(s)}_{\vk_1 \vk_2}(t_1,t_2)
\hat U_{\vk_2}^\dagger ,
\ee
where 
$\hat A(t)$ and $\hat C(t)$
are Heisenberg operators, the braces denote an anticommutator,
$\langle \ldots \rangle_{\cal H}$ is a grand canonical average,
and $\theta $ is the usual Heaviside step function.
Analogously, advanced, Keldysh, and Matsubara propagators can be
defined in helicity representation.
For dealing with superconducting phenomena it is often convenient
to introduce Wigner coordinates,
\be
\hat G(\vk ,\vR, \epsilon,t)= \int
(d\vq )(d\tau ) \E^{i(\vq \vR+\epsilon \tau )} \hat 
G_{\vk+\frac{\vq}{2},\vk-\frac{\vq}{2}} (t+\frac{\tau
}{2},t-\frac{\tau }{2}) .
\ee
From here, one can proceed along different lines. Either, 
the Dyson equation for the full Gor'kov Green's functions is
solved, which is equivalent to the Bogoliubov-de Gennes
description in wave function techniques. Or, the quasiclassical
approximation is employed, that is equivalent to the
Andreev approximation in wave function language. In the following section
we will adopt the second line.

\subsection{Quasiclassical Propagator}
In the following,
the quasiclassical theory of 
superconductivity \cite{larkin68,eilen,Serene,schmid75,schmid81,rammer86,Larkin86,FLT} will be employed
to calculate electronic transport properties across interfaces with
non-centrosymmetric superconductors.
This method is based on the observation that, in most situations, the superconducting state
varies on the length scale of the superconducting coherence length $\xi_0=\hbar v_{\rm{F}} /2\pi k_{\rm B} T_{\rm c}$. The appropriate many-body Green's function for describing
the superconducting state has been introduced by Gor'kov \cite{gorkov58}, and
the Gor'kov Green's function
can then be decomposed in a fast oscillating component, varying on the scale of $1/k_{\rm{F}}$, and an envelop function varying on the scale of $\xi_0$.
The quasiclassical approximation consists of integrating out the fast oscillating component for each quasiparticle band separately:
\begin{equation} 
\check{g}(\vec{p}^{\lambda }_{\rm{F}} , \vec{R}, \epsilon, t)=
\int d\xi_\vp^{\lambda } \;\hat{\tau}_3 \; \check{G}^{\rm (QP)}(\vec{p}, \vec{R}, \epsilon, t)
\end{equation}
where a ``check'' 
denotes a matrix in Keldysh-Nambu-Gor'kov space,\cite{keldysh64} 
a ``hat'' denotes
a matrix in Nambu-Gor'kov particle-hole space, $\xi_\vp^{\lambda }=v^{\lambda }_{\rm{F}} (\vec{p}-\vec{p}^{\lambda }_{\rm{F}} )$, and 
$\hat{\tau}_3$ is the third Pauli matrix in particle-hole space. 

The quasiclassical Green's function obeys the transport equation\cite{larkin68,eilen}
\begin{equation} 
\label{eilen1} 
i \hbar \vec{v}_{\rm{F}} \cdot \nabla_{\vec{R}}\check{g}+[\epsilon\hat{\tau}_3-\check{\Delta}-\check{h}, \check{g}]_{\circ }=\check{0}.
\end{equation}
Here, $\epsilon$ is the quasiparticle energy, $\check{\Delta}$ is the superconducting order parameter and $\check{h}$ contains all other self-energies and
external perturbations, related to external fields, impurities etc. 
The notation $\circ $ combines a time convolution with matrix multiplication, and 
$[\bullet ,\bullet ]_{\circ }$ denotes the commutator with respect to the
$\circ $-product. 
Equation (\ref{eilen1}) must be supplemented by a 
normalization condition that must be obtained from an explicite calculations
in the normal state \cite{larkin68,shelankov85},
\be
\check{g}\circ \check{g}=-\check{1}\pi^2.
\ee
From the knowledge of $\check g$ one can calculate measurable quantities, e.g.
the current density is related to the Keldysh component of the Green's function via
\begin{equation}
\label{curr}
\vec{j}(\vR,t)=q N_{\rm F}\int \frac{d\epsilon}{8\pi i} 
\mathrm{Tr}
\langle \vec{v}_{\rm{F}} 
\hat{\tau}_3\hat{g}^{\rm K} (\vec{p}^\lambda_{\rm{F}},\vR, \epsilon ,t)\rangle, \end{equation}
where $q=-|e|$ is the electron charge, and
$\langle \cdots \rangle$ denotes a Fermi surface average, which is defined by
\be
\langle \cdots \rangle = \frac{1}{N_{\rm F}} \sum_{\lambda }\int \frac{d^3p^\lambda _{\rm F} }{(2\pi\hbar)^3|\vv^\lambda_{\rm F} |}
\cdots \qquad 
N_{\rm F} = \sum_{\lambda } \int \frac{d^3p^\lambda_{\rm F} }{(2\pi\hbar)^3|\vv^\lambda_{\rm F} |} ,
\ee
and Tr denotes a trace over the Nambu-Gor'kov matrix.

\subsubsection{Case of weak spin-orbit splitting}

In the case of weak spin-orbit splitting the quasiclassical propagator
can be obtained in either spin or helicity representation. 
It is possible then to define a common Fermi surface $\vp_{\rm F}$ for both
spin bands or, equivalently, both helicity bands.
This case applies when
$|\vg (\vp_{\rm F})| \ll E_{\rm F} $ for any Fermi momentum
$\vp_{\rm F}$, where $E_{\rm F}$ is the Fermi energy 
(in addition to the condition that the
superconducting energy scales ($k_{\rm B}T_{\rm c}$ and
the gap $\Delta $ are much smaller than $E_{\rm F}$). 
Under these
circumstances quasiparticles with different helicity but with the
same $\hat \vk \equiv \vk /|\vk |$ propagate 
coherently along a common classical trajectory 
over distances much longer than the Fermi wavelength.
The transport equation is the usual Eilenberger equation modified by
a spin-orbit interaction term
\cite{Serene,alexander85} 
\begin{equation} 
\label{eilen2} 
i \hbar \vec{v}_{\rm{F}} \cdot \nabla_{\vec{R}}\check{g}+[\epsilon \hat{\tau}_3-\check{\Delta}-\check{v}_{\rm SO}, \check{g}]_{\circ }=\check{0}
\end{equation}
with normalization $\hg\circ \hg = - \pi^2 \hat{1}$. 
Here,  in helicity basis
$\hat v_{SO}= |\vg_{\vk_{\rm F} }|\, \sigma^{(3)} $, and
in spin basis
$\hat v_{SO}= \vg_{\vk_{\rm F} }\cdot  \hat{\vsigma} \hat\tau_3 $,
with 
\be
\hat{\vsigma} = 
\left( \begin{array}{cc} \vsigma & 0 \\ 0 & \vsigma^\ast \end{array} \right) 
=
\left( \begin{array}{cc} \vsigma & 0 \\ 0 & -\sigma^{(2)}\vsigma \sigma^{(2)} \end{array} \right) 
.
\ee
The velocity renormalization of order $|\vg| /E_{\rm F}\ll 1$ can safely be neglected.
The quasiparticle trajectories are doubly degenerate in either spin or
helicity space, and coherent mixing between spin states or between 
helicity states can take place.

\subsubsection{Case of strong spin-orbit splitting}

In the case of strong spin-orbit splitting the only possible 
representation for quasiclassical theory is the helicity representation.
In this case, the spin-orbit interaction does not appear anymore as a source
term in the transport equations, however explicitely as the presence of
well defined helicity bands. The transport equation takes the form
\begin{equation} 
\label{eilen3} 
i \hbar \vec{v}^\lambda_{\rm{F}} \cdot \nabla_{\vec{R}}\check{g}+[\epsilon \hat{\tau}_3-\check{\Delta}^\lambda , \check{g}]_{\circ}=\check{0}
\end{equation}
with normalization condition $\hg\circ \hg = - \pi^2 \hat{1}$. 
Here, the velocity is strongly renormalized due to spin-orbit interaction.
The quasiparticle trajectories are different for different helicity,
and no coherence exists between the different helicity states.
The matrix dimension can be reduced by a factor 2 compared to the
case of weak spin-orbit splitting, and instead the number of Fermi
surface sheets is increased by a factor of 2.
If measurements are made that are spin-selective, the corresponding
vector of Pauli spin matrices must be transformed according to
\be
\vsigma_{\lambda \lambda'}=
(U_\vk \vsigma U_\vk^\dagger )_{\lambda \lambda'}=
U_{\vk\lambda \alpha }^{\, } \vsigma_{\alpha \beta } 
U_{\vk \lambda ' \beta }^\ast
.
\ee

\subsection{Riccati parameterization}

One of the main obstacles of quasiclassical theory has been the 
non-linearity that is introduced by the normalization condition.
A powerful way to deal with this problem is the choice of a parameter representation
that ensures the normalization condition by definition.
In this representation, the Keldysh quasiclassical Green's function 
is determined by six parameters in 
particle-hole space, $\gamma^{\rm R,A}, \tilde{\gamma}^{\rm R,A}, x^{\rm K}, \tilde{x}^{\rm K}$, 
of which $\gamma^{\rm R,A}, \tilde{\gamma}^{\rm R,A}$ are the coherence functions, 
describing the coherence between particle-like and hole-like states, 
whereas $x^{\rm K}, \tilde{x}^{\rm K}$ are distribution functions, describing the 
occupation of quasiparticle states \cite{eschrig99,eschrig00}.
The coherence functions are a generalization of the so-called
Riccati amplitudes \cite{nagato93,schopohl95,eschrig99,eschrig04,cuevas06} to non-equilibrium situations.
All six parameters are matrix functions with the dimension determined by the degeneracy
of the quasiparticle trajectories, and depend on
Fermi momentum, position, energy, and time.
The parameterization is simplified by the fact that, due to symmetry relations,
only two functions of the six are independent. The particle-hole symmetry is
expressed by the operation $\tilde{X}$ which is defined for any
function $X$ of the phase space variables by 
\begin{equation}
\label{tilde1}
\tilde Q(\vec{p}_{\rm F},\vec{R},z,t)=Q(-\vec{p}_{\rm F},\vec{R},-z^\ast,t)^{\ast }.
\end{equation}
Here, $z=\epsilon $ is real for the Keldysh components  and $z$ is situated in
the upper (lower) complex energy half plane for retarded (advanced) quantities.
Furthermore, the symmetry relations 
\begin{equation}
\gamma^{\rm A}=(\tilde \gamma^{\rm R})^\dagger , \quad
\tilde \gamma^{\rm A}=(\gamma^{\rm R})^\dagger , \quad
x^{\rm K}=(x^{\rm K})^\dagger 
\end{equation}
hold. As a consequence, it suffices to
determine fully the parameters $\gamma^{\rm R}$ and $x^{\rm K}$.

The quasiclassical Green's function is related to these amplitudes in the following way
[here the upper (lower) sign corresponds to retarded (advanced)]:
\newcommand{\plus}{\;\;\,}
\newcommand{\mat}{\left( \begin{array}{cc} }
\newcommand{\matend}{\end{array}\right)}
\begin{equation}
\label{cgretav0}
\hat g^{\rm R,A} =
\mp i\pi
\mat
(1-\gamma \circ {\tilde \gamma})^{-1}  \circ
(1+\gamma \circ {\tilde \gamma}) & 
2(1-\gamma \circ {\tilde \gamma})^{-1}  \circ
\gamma  
\\ 
-2 
(1-{\tilde \gamma} \circ \gamma )^{-1} \circ
{\tilde  \gamma}
& 
-(1-{\tilde \gamma} \circ \gamma )^{-1} \circ
(1+{\tilde \gamma} \circ \gamma  )
\matend^{\!\!\! \rm R,A }
,
\end{equation}
which can be written in more compact form as \cite{eschrig09}
\begin{equation}
\label{cgretav}
\hat g^{\rm R,A} =
\mp \, 2\pi i\,
\mat \plus {\cal G} & \plus {\cal F} \\ -\tilde{\cal F} &
-\tilde{\cal G} \matend^{\!\!\! \rm R,A }
\pm i\pi \hat \tau_3
 ,
\end{equation}
with the abbreviations 
${\cal G}=({\it 1}-\gamma \circ \tilde\gamma )^{-1}$ and ${\cal F}={\cal G} \circ \gamma $.
For the Keldysh component one can write \cite{eschrig09}
\begin{equation}
\label{ckelgf2}
\hat g^{\rm K} =
-2\pi i
\mat \plus {\cal G} & \plus {\cal F} \\ -\tilde{\cal F} &
-\tilde{\cal G} \matend^{\!\!\! \rm R } \circ
\mat x^{\rm K} & 0 \\ 0 & \tilde x^{\rm K} \matend \circ
\mat \plus {\cal G} & \plus {\cal F} \\ -\tilde{\cal F} &
-\tilde{\cal G} \matend^{\!\!\! \rm A } .
\end{equation}
Here, the $\circ $-symbol includes a time convolution as well as
matrix multiplication; the inversion is defined with respect
to the $\circ $-operation \cite{eschrig09}.

From the transport equation for the quasiclassical Green's functions
one obtains a set of matrix equations of motion for the 
six parameters above \cite{eschrig99,eschrig00}. For the coherence amplitudes this
leads to Riccati differential equations \cite{schopohl95}, hence the name Riccati parameterization.

\subsection{Transport equations }
The central equations that govern the transport phenomena have been derived
in Ref.~\cite{eschrig99,eschrig00}.
The transport equation for the coherence functions
$\gamma ({\vec p}_{F}, {\vec R},\epsilon,t)$ are given by
\bea
\label{cricc1}
&& 
(i \hbar \vec{v}_{\rm{F}} \cdot \nabla_{\vec{R}}
+2\epsilon )\gamma^{\rm R,A} = [ \gamma \circ \tilde \Delta \circ \gamma + 
 \Sigma \circ \gamma - \gamma \circ \tilde\Sigma - \Delta ]^{\rm R,A}. 
\eea
For the distribution functions
$x({\vec p}_{F}, {\vec R},\epsilon, t)$  the transport equations read
\bea
\label{keld1}
(i \hbar \vec{v}_{\rm{F}} \cdot \nabla_{\vec{R}}
+ i\hbar \, \partial_t ) x^{\rm K} &-&[\gamma \circ \tilde\Delta +\Sigma ]^{\rm R } 
\circ x^{\rm K}-
x^{\rm K} \circ [ \Delta \circ \tilde\gamma -\Sigma ]^{\rm A } 
\nonumber \\
&=&-\gamma^{\rm R} \circ \tilde\Sigma^{\rm K} \circ \tilde\gamma^{\rm A} 
+ \Delta^{\rm K} \circ \tilde\gamma^{\rm A} + \gamma^{\rm R} \circ \tilde\Delta^{\rm K}  -\Sigma^{\rm K} .
\eea
The equations for the remaining components are obtained by the symmetry 
relation Eq.~(\ref{tilde1}).

\subsection{Boundary conditions}

\begin{figure}[b]
\begin{center}
\scalebox{0.9}{
\includegraphics[width=10cm,clip]{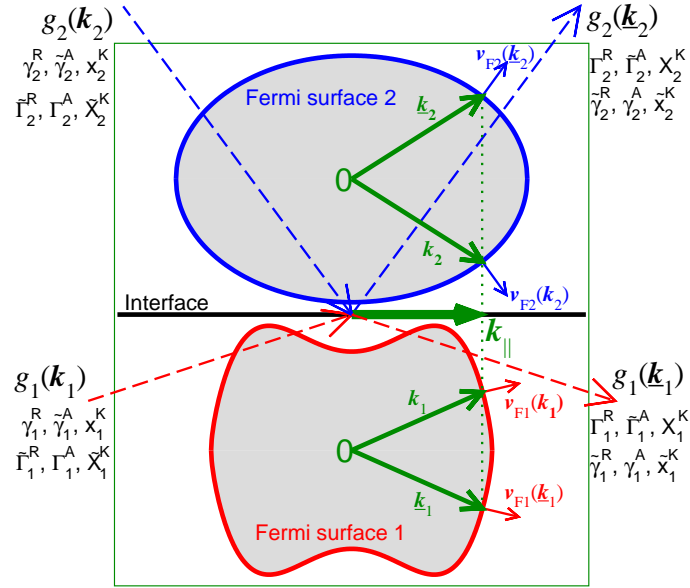} 
}
\end{center}
\caption{
Notation for the coherence amplitudes and distribution functions
at an interface. Indices 1 and 2 refer to the sides of the interface.
The arrows for the Fermi momenta are for particle like excitations.
The Fermi velocity directions are given by the directions
perpendicular to the Fermi surface at the corresponding Fermi momentum.
Quasiparticles move along the Fermi velocity directions (dashed lines).
The components of the Fermi momenta parallel to the surface are conserved
(indicated by the thin dotted line). 
For each trajectory, small case letters denote coherence functions and
distribution functions with initial conditions from the bulk, and
capital case letters denote functions with initial conditions at the interface.
The interface boundary conditions must express all capital case 
quantities in terms of the small case quantities. Here, the simplest case, that 
involves only one Fermi surface sheet on either side ('two-trajectory
scattering'), is shown.
After Ref.~\cite{eschrig00}.
}
\label{fig:E0}
\end{figure}

The transport equations must be complemented with boundary conditions for the
coherence amplitudes and distribution functions at interfaces and surfaces
\cite{shelankov84,zaitsev84}. For spin-active scattering such conditions
were obtained in Ref.~\cite{millis88}. Explicit formulations in terms
of special parameterizations were given in Refs.~\cite{yip97,eschrig00,fogelstrom00,ozana00,zhao04}. Further developments include strongly spin-polarized systems
\cite{eschrig03,kopu04,eschrig08,eschrig09,grein09}, diffusive interface
scattering \cite{luck03} or multi-band systems \cite{graser07}.
We adopt the notation \cite{eschrig00} that incoming amplitudes are denoted by small 
case letters and outgoing ones by capital case letters, see Fig.~\ref{fig:E0}. 
Note that the velocity direction of trajectories is opposite for holelike and 
particlelike amplitudes as well as advanced and retarded ones. 
The boundary conditions express outgoing amplitudes as a function of incoming ones and 
as a function of the parameters of the normal-state scattering matrix. 

\subsubsection{Coherence amplitudes}
The  boundary conditions for the coherence amplitudes
are formulated in terms of the solution of the equation \cite{eschrig09}
\bea
\label{gp}
[\gamma'_{kk'}]^{\rm R } &=& \sum_{p} S^{\rm R}_{kp} \circ \gamma^{\rm R}_{p} \circ \tilde S^{\rm R}_{pk'}
 \\
\label{GR}
\, [\Gamma_{k\leftarrow k'}]^{\rm R} &=& \big[ \gamma'_{kk'}
+\sum_{k_1\ne k}
\Gamma_{k\leftarrow k_1} \circ \tilde \gamma_{k_1} \circ \gamma'_{k_1k'}
\big]^{\rm R} , 
\eea
(the trajectory index $p$ runs over all incoming trajectories)
for $[\Gamma_{k\leftarrow k'}]^{\rm R}$, 
where the trajectory indices $k,k',k_1$ run over outgoing trajectories involved in the 
interface scattering process, and the scattering matrix parameters enter only via
the ``elementary scattering event'' $[\gamma'_{kk'}]^{\rm R,A}$.
The quasiclassical coherence amplitude is given by the
forward scattering contribution of $[\Gamma_{k\leftarrow k'}]^{\rm R}$,
\begin{equation}
\label{GRA}
\Gamma_{k}^{\rm R }= \Gamma_{k \leftarrow k}^{\rm R}.
\end{equation}
Analogous equations \cite{eschrig09} hold for the advanced and particle-hole conjugated components,
$[\tilde \Gamma_{p\leftarrow p'}]^{\rm R}$, $[\Gamma_{p'\rightarrow p}]^{\rm A}$,
and $[\tilde \Gamma_{k'\rightarrow k}]^{\rm A}$.

\subsubsection{Distribution functions }
For the Keldysh component not only the forward scattering contribution of
$[\Gamma_{k\leftarrow k'}]^{\rm R}$ is required, but also the 
off-scattering part
\begin{equation}
\label{GbarR}
\,[\overline \Gamma_{k \leftarrow k'}]^{\rm R } =
[\Gamma_{k \leftarrow k'} - \Gamma_{k} \delta_{kk'} ]^{\rm R}.
\end{equation}
The boundary conditions for the distribution functions read \cite{eschrig09}
\bea
\label{xp}
[x'_{kk'}]^{\rm K} &=& \sum_{p} S^{\rm R}_{kp}  \circ x^{\rm K}_{p} \circ  S^{\rm A}_{pk'}.
\\
\label{X}
X^{\rm K}_{k} &= & 
\sum_{k_1,k_2}
[\delta_{kk_1} + \overline \Gamma_{k \leftarrow k_1} \circ \tilde \gamma_{k_1}]^{\rm R}
\circ [x'_{k_1k_2}]^{\rm K} \circ
[\delta_{k_2k} + \gamma_{k_2} \circ \overline{\tilde \Gamma}_{k_2 \rightarrow k} ]^{\rm A}
\nonumber \\
&-&\sum_{k_1}
[\overline \Gamma_{k \leftarrow k_1}]^{\rm R} \circ \tilde x^{\rm K}_{k_1} \circ
[\overline{\tilde \Gamma}_{k_1 \rightarrow k} ]^{\rm A},
\eea
which depends on the scattering matrix parameters only via the elementary scattering event $[x'_{kk'}]^{\rm K}$.
Analogous relations hold for $\tilde X^{\rm K}_{p}$.

The transport equation for the distribution function is solved by any function of
energy in equilibrium. The correct boundary conditions in this case are
\be
x^{({\rm eq})} = (1-\gamma^{\rm R} \tilde \gamma^{\rm A} ) \tanh\left(\frac{\epsilon-q\Phi}{2k_{\rm B} T}\right),
\; \;
\tilde x^{({\rm eq})} = -(1-\tilde \gamma^{\rm R} \gamma^{\rm A} ) \tanh\left(\frac{\epsilon+q\Phi}{2k_{\rm B} T}\right)
\ee
for excitations of charge $q$ in an electrostatic potential $\Phi$.

\subsubsection{Case 1: one-trajectory scattering}

In this case only one incoming and one outgoing trajectory are coupled via the
boundary conditions. The corresponding normal state scattering matrix is  
denoted by $S$ and is a scalar in trajectory space. 
The boundary conditions read in this case simply
\bea
\label{surface}
[\gamma']^{\rm R } &=& S^{\rm R} \circ \gamma^{\rm R} \circ \tilde S^{\rm R} ,\qquad
\Gamma^{\rm R} = [\gamma']^{\rm R} ,
\eea
and
\bea
[x']^{\rm K} &=& S^{\rm R}  \circ x^{\rm K} \circ  S^{\rm A} , \qquad
X^{\rm K}_{k} =  [x']^{\rm K}  .
\eea

\subsubsection{Case 2: two-trajectory scattering}

This is the case of scattering from two incoming trajectories into two
outgoing trajectories. Examples are reflection and transmission at
an interface, or reflection from a surface in a two-band
system.
The scattering matrix and the elementary scattering events
have in this case the form
\begin{equation}
S= \left(\!\! \begin{array}{cc} S_{11} & S_{12}\\
S_{21} & S_{22} \end{array}\right), \quad
[\gamma'_{ij}]^{\rm R } = \sum_{l=1,2}
S^{\rm R}_{il} \circ \gamma^{\rm R}_{l} \circ \tilde S^{\rm R}_{lj},
\quad
\, [x'_{ij}]^{\rm K} = \sum_{l=1,2} S^{\rm R}_{il}  \circ x^{\rm K}_{l} \circ  S^{\rm A}_{lj}.
\end{equation}
We give the solutions for trajectory 1, the remaining solution can be obtained
by interchanging the indices 1 and 2.
The boundary conditions read for $i,j=1,2$
\bea
\Gamma_{1\leftarrow 1}^{\rm R} &=& \big[ \gamma'_{11} +
\Gamma_{1\leftarrow 2} \circ \tilde \gamma_{2} \circ \gamma'_{21} \big]^{\rm R} , \quad
\Gamma_{1\leftarrow 2}^{\rm R} = \big[ \gamma'_{12} +
\Gamma_{1\leftarrow 2} \circ \tilde \gamma_{2} \circ \gamma'_{22} \big]^{\rm R} .  
\eea
The equation for the $\Gamma_{1\leftarrow 2}$ 
can be solved by simple inversion, 
\bea
\label{G12}
\Gamma_{1\leftarrow 2}^{\rm R} = 
\big[ \gamma'_{12} \circ
(1-\tilde \gamma_{2} \circ \gamma'_{22})^{-1} \big]^{\rm R} ,  
\eea
and the solution introduced into the equation for
$\Gamma_{1\leftarrow 1}^{\rm R}=\Gamma_1^{\rm R}$,
\bea
\label{G11}
\Gamma_1^{\rm R} &=& \big[ \gamma'_{11} +
\gamma'_{12} \circ
(1-\tilde \gamma_{2} \circ \gamma'_{22})^{-1} 
\circ \tilde \gamma_{2} \circ \gamma'_{21} \big]^{\rm R} .
\eea
For the distribution function one needs the components
$\overline \Gamma_{1 \leftarrow 2}^{\rm R } =
\Gamma_{1 \leftarrow 2}^{\rm R}$
and obtains
\bea
\label{bX}
X^{\rm K}_{1} &= & 
[x'_{11}]^{\rm K} +
\Gamma_{1 \leftarrow 2}^{\rm R} \circ \tilde \gamma_{2}^{\rm R}
\circ [x'_{21}]^{\rm K} +
[x'_{12}]^{\rm K} \circ
\gamma_{2}^{\rm A} \circ {\tilde \Gamma}_{2 \rightarrow 1}^{\rm A}
\nonumber \\
&&+
\Gamma_{1 \leftarrow 2}^{\rm R} \circ \Big(\tilde \gamma_{2}^{\rm R}
\circ [x'_{22}]^{\rm K} \circ
\gamma_{2}^{\rm A} - \tilde x^{\rm K}_{2} 
\Big)
\circ {\tilde \Gamma}_{2 \rightarrow 1}^{\rm A}.
\eea

We present here formulas for the special case of the zero temperature conductance 
when a single band system is contacted by a normal metal. We assign the index 1 to the
normal metal side of the interface and the index 2 to the superconducting side.
The momentum for incoming trajectories on the superconducting 
side of the interface is
denoted by $\vk_2 $, and that for the outgoing trajectory on the superconducting side
by $\underline \vk_2$. For the normal side the corresponding momenta are
$\vk_1$ and $\underline \vk_1$ (see Fig.~\ref{fig:E0} for the scattering geometry).
The projection on the interface of all four momenta is equal. The corresponding
incoming coherence functions in the superconductor are
$\gamma_2(\epsilon )\equiv \gamma_2^{\rm R} (\vk_2,\epsilon )$ and 
$\tilde \gamma_2 (\epsilon ) \equiv \tilde \gamma^{\rm R}_2(\underline\vk_2,\epsilon )$.
Furthermore, $S_{12} \equiv S^{\rm R}_{12}(\underline \vk_1,\vk_2)$, 
$S_{22} \equiv S^{\rm R}_{22}(\underline \vk_2,\vk_2)$, and $\tilde S_{22} = \tilde S^{\rm R}_{22} (\vk_2,\underline \vk_2)$.
The Fermi velocity for outgoing directions
on the normal side will be denoted by $\vv_{\rm F1}\equiv \vv_{\rm F1}(\underline\vk_1)$.
Having thus specified all momentum dependencies, we will suppress in the formulas below
the momentum variables.
In the case under consideration, after introducing Eqs.~(\ref{G12}), (\ref{G11}), and (\ref{bX})
into Eq.~(\ref{curr}), we obtain after some algebra (we omit hereafter the
$\circ $ sign)
\bea
\label{cond}
\frac{G(eV)}{G_{\rm N}} &=& \Big\langle \hat \vn \vv_{\rm F1} \left\{
\Big|\Big| S_{12}\big[1+A_2(\epsilon ) S_{22}\big] {\Big|\Big|}^2 
-\Big|\Big| S_{12} A_2(\epsilon ) {\Big|\Big|}^2 \right\} \Big\rangle^+_{\epsilon = eV}
\nonumber \\
&& + \Big\langle \hat \vn \vv_{\rm F1}\Big|\Big| S_{12} \big[1+A_2(\epsilon ) S_{22}\big] \gamma_2(\epsilon ) \tilde S_{21} {\Big|\Big|}^2 \Big\rangle^+_{\epsilon=-eV}
\eea
where
\be
A_2(\epsilon )= \Big( 1-\gamma_2 (\epsilon ) \tilde S_{22} \tilde \gamma_2(\epsilon ) S_{22} \Big)^{-1}
\gamma_2 (\epsilon ) \tilde S_{22} \tilde \gamma_2 (\epsilon )
\ee
and we used the notation $||A{||}^2= \frac{1}{2}\mbox{Tr} (AA^\dagger )$ for any $2\times 2$ 
matrix $A$. The symbol $\langle \ldots \rangle^+_{\epsilon=eV}$ denotes Fermi surface
average only over outgoing directions, and the argument is to be taken at energy $eV$.
For $S_{22}=\tilde S_{22}=-\sqrt{R(\theta )}$, $S_{12}=\tilde S_{21}=\sqrt{D(\theta )}$ (with impact angle $\theta $), Eq.~(\ref{cond})
reduces to Eq.~(\ref{conductance}).
For the tunneling limit we can neglect the second line in Eq.~(\ref{cond}), and 
using the relation
\be
1+A_2(\epsilon ) S_{22} = \frac{1}{2} \left\{ {\cal N}_2(\epsilon )+1\right\}
\ee
with the complex quantity
\bea
{\cal N}_2(\epsilon )= \left\{
\Big( 1-\gamma_2 ( \epsilon ) \tilde S_{22} \tilde \gamma_2(\epsilon ) S_{22} \Big)^{-1}
\Big( 1+\gamma_2 (\epsilon ) \tilde S_{22} \tilde \gamma_2(\epsilon ) S_{22} \Big)
\right\}
\nonumber
\eea
the conductance simplifies after some re-arrangements to
\be
\label{tunnel}
\frac{G(eV)}{G_{\rm N}} = 
\frac{1}{2}
\mbox{Re} 
\mbox{Tr}
\Big\langle \hat \vn \vv_{\rm F1} \left\{
S_{12} {\cal N}_2 (eV) (S_{12})^\dagger \right\} \Big\rangle^+ .
\ee
For the tunneling limit, in ${\cal N}_2$ the surface scattering matrix (i.e. for $S_{12}=S_{21}=0$) 
can be used, for which the
local density of states at an impenetrable surface is
\be
\frac{N_2(\epsilon )}{ N_{2,{\rm F}}}= 
\frac{1}{2}
\mbox{Re} 
\mbox{Tr}
\Big\langle {\cal N}_2(\epsilon ) + S_{22} {\cal N}_2 (\epsilon) S_{22}^\dagger \Big\rangle^{+} .
\ee

\subsubsection{Scattering matrix for non-centrosymmetric/normal-metal junction}

For the case that a non-centrosymmetric material with small spin-orbit
splitting is brought in contact with a normal metal, we can use
the formulas of the last subsection. 
The scattering matrix for scattering between the two helicity bands in
the non-centrosymmetric metal (index 2) and the two spin bands in the 
normal metal (index 1)
can be expressed in terms of the scattering matrix for scattering between
spin states on both sides of the interface. The corresponding
transformation is
\bea
\left(\!\! \begin{array}{cc} S'_{11} & S'_{12}\\
S'_{21} & S'_{22} \end{array}\right) &=&
\left(\!\! \begin{array}{cc} 1 & 0\\
0 & U_{\underline \vk} \end{array}\right) \cdot
\left(\!\! \begin{array}{cc} S_{11} & S_{12}\\
S_{21} & S_{22} \end{array}\right)  \cdot
\left(\!\! \begin{array}{cc} 1 & 0\\
0 & U^\dagger_{\vk} \end{array}\right)
, \\
\left(\!\! \begin{array}{cc} \tilde S'_{11} & \tilde S'_{12}\\
\tilde S'_{21} & \tilde S'_{22} \end{array}\right) &=&
\left(\!\! \begin{array}{cc} 1 & 0\\
0 & U^\ast_{-\vk} \end{array}\right) \cdot
\left(\!\! \begin{array}{cc} \tilde S_{11} & \tilde S_{12}\\
\tilde S_{21} & \tilde S_{22} \end{array}\right)  \cdot
\left(\!\! \begin{array}{cc} 1 & 0\\
0 & U^T_{-\underline \vk} \end{array}\right)
.
\eea
For the simple case of a spin-conserving scattering in the spin/spin
representation, the spin/helicity representation of the scattering
matrix takes the form
\bea
\label{scatt}
\left(\!\! \begin{array}{cc} S'_{11} & S'_{12}\\
S'_{21} & S'_{22} \end{array}\right) &=&
\left(\!\! \begin{array}{cc} r & t\; U_{\vk}^\dagger \\
t^\ast \; U_{\underline \vk}  & -r\; U_{\underline \vk} 
U_{\vk}^\dagger
\end{array}\right)
, \quad
\left(\!\! \begin{array}{cc} \tilde S'_{11} & \tilde S'_{12}\\
\tilde S'_{21} & \tilde S'_{22} \end{array}\right) =
\left(\!\! \begin{array}{cc} r & t^\ast \; U^T_{-\underline\vk}\\
t\; U^\ast_{-\vk} & - r\; U^\ast_{-\vk} 
U^T_{-\underline\vk} \end{array}\right)
.
\eea
where $r\equiv r_{\underline \vk \vk}$ and $t=t_{\underline \vk \vk }$ with
$r^2+|t|^2=1$
are reflection and transmission coefficients that depend on the 
(conserved) momentum projection on the interface.
We have chosen $r$ real, as in quasiclassical approximation possible reflection
phases do not affect the results.
The case $t=0$ can be used to describe scattering at a surface.

In the case of a contact with a non-centrosymmetric metal with strong spin-orbit
split bands the scattering matrix has a more complicated structure. 
It connects in this case three incoming with three outgoing trajectories,
and the scattering at the interface will not be spin-conserving. 
For this case,
it does then not make sense anymore do use a spin/spin representation, 
but a spin/helicity representation must be used consistently.
The scattering matrix must be obtained in agreement with the symmetry group of
the interface, and it  cannot in general be related anymore to the
$U_\vk$ matrices in a simple way.

\subsection{Superconducting order parameter}

For the case of weak spin-orbit splitting one expects that to leading order in
the small expansion parameters either a singlet or a triplet component
nucleates. 
On the other hand, any finite spin-orbit interaction leads to 
a mixture of spin singlet ($\Delta_s$) and triplet ($\Delta_t$) components
\cite{ser04,fri06}. Consequently, the singlet or triplet states a never
pure, but they are mixed. This mixing becomes in particular prominent
when the spin-orbit interaction is strong.
In this case, it does not make sense anymore to speak about singlet or
triplet components, but it is necessary to start from the helicity basis.

It is interesting to consider what happens in the weak case first.
In this case the triplet component is expected to be induced directly
by the structure of the spin-orbit interaction, and
the spin triplet component aligns with $\vg(\vk )$.
The gap function is in this case in spin representation given by, 
\be 
\label{gapspin}
\Delta^\sm{s}  = (\Delta_\vk   + D_\vk \; \vg (\vk) \cdot \vsigma ) \I \sigma^{(2)} 
\ee
which transforms in helicity basis into
\bea
\Delta  &=& U^{\,}_\vk (\Delta_\vk   + D_\vk \; \vg (\vk) \cdot \vsigma ) 
\I \sigma^{(2)}  U^T_{-\vk}\nonumber \\&= &
U^{\,}_\vk (\Delta_\vk   + D_\vk \; \vg (\vk) \cdot \vsigma ) U_\vk^\dagger 
U^{\, }_\vk \I \sigma^{(2)} U^T_{-\vk} \nonumber \\&=&
(\Delta_\vk   + D_\vk |\vg (\vk)| \sigma^{(3)} )
U^{\, }_\vk U^\dagger_{-\vk} \I \sigma^{(2)}  .
\eea
We introduce the notation
\be
(U^{\, }_{\vk} U_{-\vk}^\dag )_{\lambda \lambda'}= 
\left(
\begin{array}{cc}
0 &  \E^{-i \phi_\vg} \\
-\E^{i \phi_\vg} & 0 \end{array} \right) 
\equiv
-i \sigma^{(\vg)}_{\lambda \lambda' }.
\label{eq:ng}
\ee
Note that the identities $ (\sigma^{(\vg)})^2 =1$,
$ \sigma^{(-\vg)} = -\sigma^{(\vg)} $,
and 
$\sigma^{(2)} \sigma^{(\vg)}\sigma^{(2)} = -\sigma^{(\vg)\ast } $,
hold.
With this, we can obtain the
Nambu-Gor'kov space structure of the order parameter
\begin{eqnarray}
\hat \Delta&=&
\left(\begin{array}{cc} 0& \Delta \\ \tilde \Delta &0
\end{array}\right) =
\left( \begin{array}{cc} 0& 
(\Delta_\vk   + D_\vk |\vg| \sigma^{{3}} ) \sigma^{(\vg )}
\sigma^{(2)}\\
(\Delta_{-\vk}+D_{-\vk} |\vg| \sigma^{(3)})^\ast
\sigma^{(\vg)\ast} \sigma^{(2)} 
&0 \end{array} \right)  \nonumber \\
&=&\left( \begin{array}{cc} 0& (\Delta_\vk+D_\vk |\vg| \sigma^{(3)})\I \sigma^{(2)} \\
(\Delta_{\vk}^\ast+D_{\vk}^\ast |\vg| \sigma^{(3)}) \I \sigma^{(2)}
&0 \end{array} \right)
\left( \begin{array}{cc} 
\I \sigma^{(\vg)} &0\\0& -\I \sigma^{(-\vg)\ast}
\end{array} \right),
\end{eqnarray}
where $\Delta_{-\vk} =\Delta_{\vk}$, and $D_{-\vk}=D_{\vk} $, and we have used
$\sigma^{(\vg )\ast} \sigma^{(3)} \sigma^{(\vg )\ast}= -\sigma^{(3)}$.
With $\Delta_\pm (\vk) = \Delta_\vk \pm D_\vk |\vg|$ the order parameter
can be cast in the form
\bea
\label{OP}
\Delta (\vk )&=&
\left(\begin{array}{cc} 
\Delta_{+}(\vk )\; t_+(\vk) &0\\ 0&\Delta_{-}(\vk ) \; t_-(\vk)
\end{array}\right)
\\
\tilde \Delta (\vk )&=&
\left(\begin{array}{cc} 
\Delta_{+}(\vk )^\ast \; t_+(-\vk)^\ast&0 \\ 0& \Delta_{-}(\vk )^\ast \; t_-(-\vk)^\ast
\end{array}\right)
.
\eea
with phase factors $t_\lambda(\vk)=-\E^{-i\lambda \phi_\vg}$. Note that 
$t_\lambda(-\vk)=-t_\lambda(\vk)$, and $|t_\lambda (\vk)|=1$, and
$\Delta_\pm (-\vk) =\Delta_\pm (\vk) $.

\begin{petit}
We note that other possibilities to define the canonical transformation
that diagonalizes the kinetic part of the Hamiltonian exist, which differ
by the relation between particle and hole components. Using these alternative
definitions (e.g. in Refs. \cite{Frigeri,Vorontsov}), the order parameter
is purely off-diagonal instead of diagonal in the band representation,
and the symmetry relation
Eq.~(\ref{tilde1}) becomes non-trivial (see e.g. Ref.~\cite{Vorontsov}).
Here, we prefer a transformation that preserves the symmetry (\ref{tilde1}),
and renders the order parameter above diagonal. This is a natural
choice when treating strongly spin-orbit split systems, where the order
parameter should be band diagonal.
\end{petit}

The coherence amplitudes in a bulk system with 
order parameter Eq.~(\ref{OP}) are of a similar form,
\bea
\label{cohbulk1}
\gamma (\vk, \epsilon )&=&
\left(\begin{array}{cc} 
\gamma_{+}(\vk, \epsilon )\; t_+(\vk) &0\\ 0&\gamma_{-}(\vk,\epsilon )\; t_-(\vk) 
\end{array}\right) 
\\
\label{cohbulk2}
\tilde \gamma (\vk, \epsilon )&=&
\left(\begin{array}{cc} 
\tilde \gamma_{+}(\vk, \epsilon ) \; t_+(-\vk)^\ast &0\\ 0& \tilde \gamma_{-}(\vk,\epsilon ) \; t_-(-\vk)^\ast \end{array}\right) 
\eea
with $\tilde \gamma_{\pm}(\vk, \epsilon) =
\gamma_{\pm}(-\vk,-\epsilon )^\ast $.
In inhomogeneous systems helicity-mixing can take place. If this
happens, the form of the coherence functions is the same band-diagonal form
as above for the case of strong spin-orbit splitting, however 
has the full matrix structure for the case of weak spin-orbit splitting.

\subsection{Results}
\subsubsection{Andreev bound states near the surface}

The surface bound states are determined by the poles of the
Green's function. Following Refs.~\cite{Iniotakis07,Vorontsov},
we consider specular
reflection, whereby the component of $\vk$ normal to surface
changes sign, $\vk\to\underline\vk$, whereas the component parallel
to the surface is conserved. We
find the amplitudes $\gamma (\vk,\epsilon )$ 
by integrating forward 
along the incoming, $\vk$,
trajectory starting from the values in the
bulk, and the amplitudes
$\tilde \gamma({\ul\vk},\epsilon )$ by
integrating backward along the outgoing, $\ul\vk$,
trajectory, again starting from the values in the bulk
\cite{eschrig00}.
For the homogeneous solutions one obtains
\be
\label{gammabulk}
\gamma^0_\pm(\vk,\epsilon )=-\frac{\Delta_\pm(\vk )}
{\epsilon+\I \sqrt{|\Delta_\pm(\vk)|^2-\epsilon^2}}, \;
\tilde\gamma^0_\pm(\underline\vk ,\epsilon )= \frac{\Delta_\pm(\underline\vk)^\ast }
{\epsilon+\I  \sqrt{|\Delta_\pm(\underline\vk )|^2-\epsilon^2}},
\ee
\begin{petit}
Note that the spin-orbit interaction in the helicity basis enters as a term
proportional to $\sigma^{(3)}$, see Eq.~(\ref{eilen2}). Consequently,
this term commutes with any term diagonal in the helicity basis, and thus drops
out of the homogeneous solutions in Eq.~(\ref{gammabulk})
(see Ref.~\cite{Hayashi06} for the case of a Rashba-type spin-orbit coupling). 
Note, however, that this is not in general the case for non-homogeneous solutions:
when helicity mixing takes place due to impurities or surfaces and
interfaces, and a fully self-consistent solution is obtained,
then the spin-orbit coupling term in Eq.~(\ref{eilen2}) enters through
the transport equation. 
\end{petit}

The amplitudes $\Gamma_{\ul\vk}$ and
$\tilde\Gamma_\vk$, are determined from the boundary conditions at the surface.
We consider here a simple model of a non-magnetic surface, that
conserves the spin under reflection (this assumption only holds
for a {\it small} spin-orbit interaction in the bulk material).
In this case the components of $\hg$ in the spin basis,
\be
\hg^\sm{s} (\vk ,\epsilon ) =\hat{U}^\dagger_{\vk} \hg (\vk , \epsilon) \hat{U}^{\; }_{\vk},
\ee
are continuous at the surface.
This leads to a surface induced
mixing of the helicity bands according to 
\bea
\label{Gam1}
U^\dag_{\underline\vk}
\Gamma (\underline\vk ,\epsilon ) U^\ast_{-\underline\vk} =
\Gamma^\sm{s} ( \underline{\vk},\epsilon ) = \gamma^\sm{s} (\vk ,\epsilon ) = U^\dag_{\vk}
\gamma ( \vk,\epsilon ) U^\ast_{-\vk}, \\
\label{Gam2}
U^T_{-\vk}
\tilde\Gamma (\vk , \epsilon ) U^{\; }_{\vk}= 
{\tilde\Gamma}^\sm{s} ( \vk ,\epsilon )=
{\tilde\gamma}^\sm{s} (\underline{\vk} , \epsilon ) = U^T_{-\underline\vk}
\tilde\gamma (\underline{\vk} , \epsilon )  U^{\; }_{\underline\vk} . 
\eea
Note that these boundary conditions correspond to Eq.~(\ref{surface}) with
$S^{\rm R}$ and $\tilde S^{\rm R}$ given by the (22)-components of 
Eq.~(\ref{scatt}).

We proceed with discussing the local density of states at the surface,
$N(\epsilon )$, that is
defined in terms of the momentum resolved density of states, $N(\vk, \epsilon )$
by
\be
\label{LDOS1}
N(\vk ,\epsilon )/N_{\rm F}= -(2\pi )^{-1}
\mbox{Im} \mbox{Tr}_{\lambda }\left\{ g (\vk ,\epsilon )\right\} ,
\quad
N(\epsilon ) = \langle N(\vk ,\epsilon ) \rangle ,
\ee
which can be expressed in terms of the coherence amplitudes in the following way
(here $\vk $ points towards the surface and $\underline\vk $ away from it),
\bea
N(\vk ,\epsilon )/N_{\rm F}
&=& 
\mbox{Re} \mbox{Tr}_\lambda 
\left\{ 
\left[ 1-\gamma (\vk , \epsilon ) \tilde \Gamma (\vk ,\epsilon ) \right]^{-1} -1/2 \right\}
\nonumber \\
N(\underline \vk , \epsilon )/N_{\rm F}&=& 
\mbox{Re} \mbox{Tr}_\lambda 
\left\{ 
\left[1-\Gamma (\underline \vk , \epsilon ) \tilde \gamma (\underline \vk ,\epsilon ) \right]^{-1} -1/2 \right\} .
\label{LDOS2}
\eea
We obtain $\Gamma (\underline \vk, \epsilon )$ and $\tilde \Gamma ( \vk , \epsilon )$ from 
Eqs.~(\ref{Gam1}) and (\ref{Gam2}), with $\gamma (\vk, \epsilon )$ and
$\tilde \gamma (\underline \vk , \epsilon )$ from Eqs. (\ref{cohbulk1}), (\ref{cohbulk2}),
(\ref{gammabulk}), and (\ref{phasefactor}).

\begin{petit}
The bound states in the surface density of states
correspond to the zero eigenvalues of the matrix
\be
 1 - \gamma(\vk, \epsilon ) \tilde \Gamma(\vk ,\epsilon)= 1 -
\gamma (\vk,  \epsilon )
(U^\ast_{-\vk } U^T_{-\underline\vk})
\tilde \gamma(\ul\vk,\epsilon )
(U^{\; }_{\underline\vk} U^\dag_{\vk} )
\ee
at the surface. An explicite calculation results in an equation for
the Andreev bound states energy in terms of the
surface coherence amplitudes in the helicity basis \cite{Vorontsov},
\be 
\frac{
\left\{1 + \gamma_+ \tilde \gamma_+  \right\}
\left\{1 + \gamma_- \tilde \gamma_-  \right\} }
{
\left\{1 + \gamma_+ \tilde\gamma_- \right\}
\left\{1 + \gamma_- \tilde \gamma_+ \right\} }
= 
- \cM \,,
\label{eq:BS} 
\ee 
where we used the abbreviations 
$\gamma_\pm \equiv \gamma_\pm(\vk, \epsilon) $ and
$\tilde\gamma_\pm \equiv \tilde \gamma_\pm(\underline \vk, \epsilon) $.
The ``mixing'' factor $\cM$ is determined by the
change of $\vg(\vk ) \to \vg(\underline\vk)$ 
under reflection $\vk \to \underline\vk $ at
the surface,
 \be
 \cM =
\frac{ \sin^2{\theta_\vg-\theta_{\ul\vg} \over 2} +
\sin^2{\theta_\vg+\theta_{\ul\vg} \over 2} \tan^2{\phi_\vg -\phi_{\ul\vg} \over 2} }
{ \cos^2{\theta_\vg-\theta_{\ul \vg} \over 2} +
\cos^2{\theta_\vg+\theta_{\ul\vg} \over 2} \tan^2{\phi_\vg -\phi_{\ul\vg} \over 2} } \,,
    \ee
where $\theta_\vg,\phi_\vg $ and $\theta_{\ul\vg},\phi_{\ul\vg}$
are the polar and azimuthal angles of $\vg(\vk )$ and $\vg({\ul\vk})$,
respectively.
\end{petit}

In general, the order parameter must be obtained self-consistently
at the surface. Helicity mixing at the surface will lead necessarily
to a suppression of the order parameter. To gain insight in the role
of the order parameter suppression it is useful to model it by
a normal layer of width $W$ next to the interface.
Trajectories incident at an
angle $\alpha_\vk $ from the surface normal 
travel through a normal region of an effective width
$2W_\vk =2W/\cos(\alpha_\vk )$. 
Thus, the surface
coherence amplitudes gain a phase factor, 
\be
\label{phasefactor}
\gamma_\pm (\vk ,\epsilon )=
\gamma^\sm{0}_\pm (\vk ,\epsilon )\, 
\E^{2i \epsilon W/ v_{\rm F}\cos(\alpha_\vk)}, 
\quad 
\tilde\gamma_\pm (\underline \vk,\epsilon )=
\tilde\gamma^\sm{0}_\pm (\underline \vk , \epsilon )\, 
\E^{2i \epsilon W/ v_{\rm F}\cos(\alpha_\vk)}.
\ee

Similarly like for $\vg $ we will use in the following
polar and azimuthal angles for the vector
$\vk$, defined by $\{k_x,k_y,k_z\}=|\vk| \{
\sin(\theta_\vk) \cos(\phi_\vk),
\sin(\theta_\vk) \sin(\phi_\vk),
\cos(\theta_\vk) \}$ (where $0\le \theta_\vk \le \pi $).
We also introduce the notation $\hat \vg =\vg /\mbox{max}(|\vg |)$.
For the order parameter, we assume isotropic $\Delta_\vk =\Delta $ and $D_\vk =D$ in
Eq.~(\ref{gapspin}), and 
introduce the parameter
$q= \Delta/D' $ where $D'=D\cdot \mbox{max}(|\vg |)$
\cite{Iniotakis07}. In this case $\Delta_\pm = D'(q\pm |\hat \vg |)$
with maximal gap amplitudes $\Delta_0=D'(q+1)$.

\begin{figure}[t]
\begin{center}
\scalebox{0.57}{
\includegraphics[width=10cm,clip]{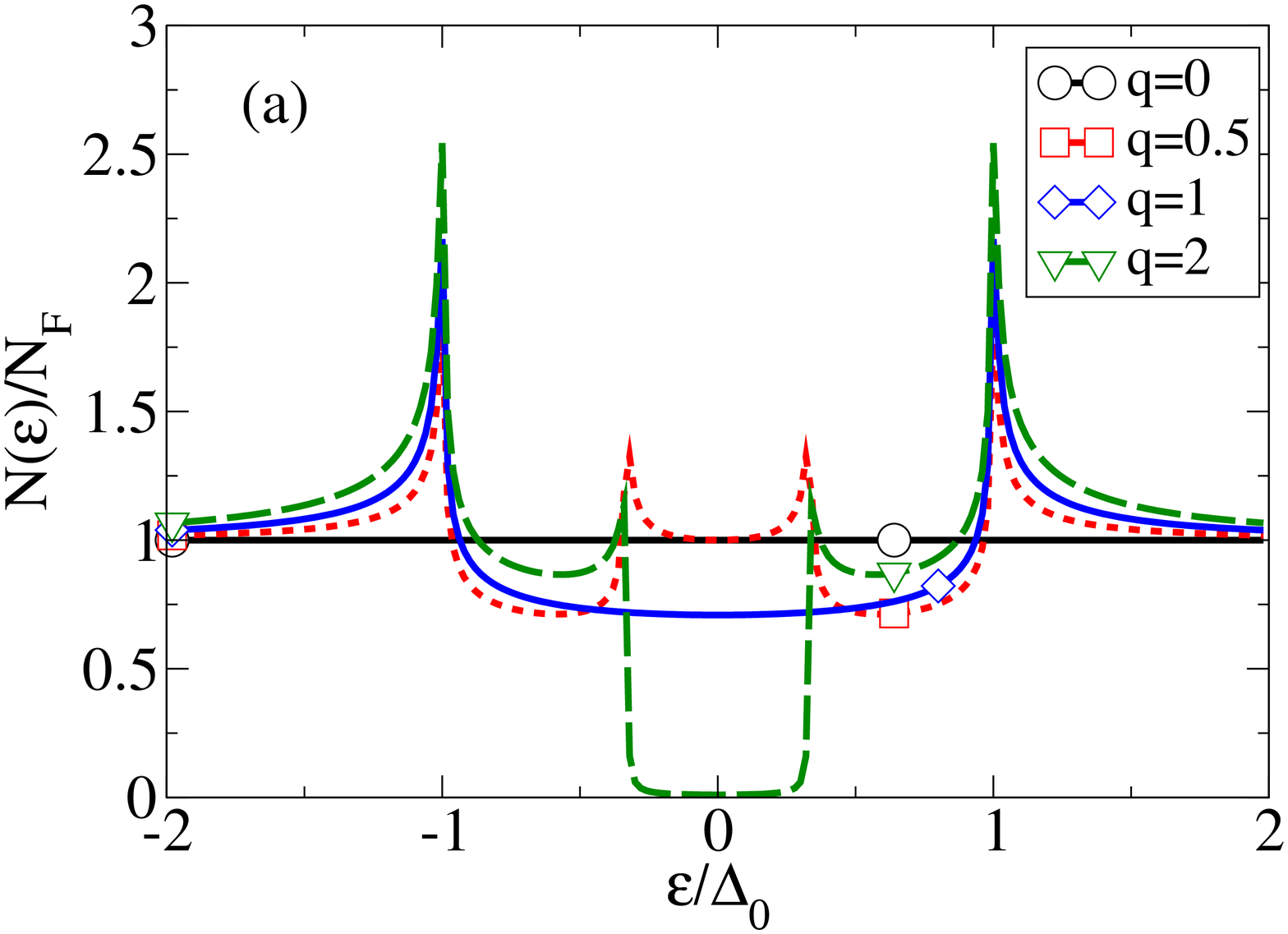} 
\includegraphics[width=10cm,clip]{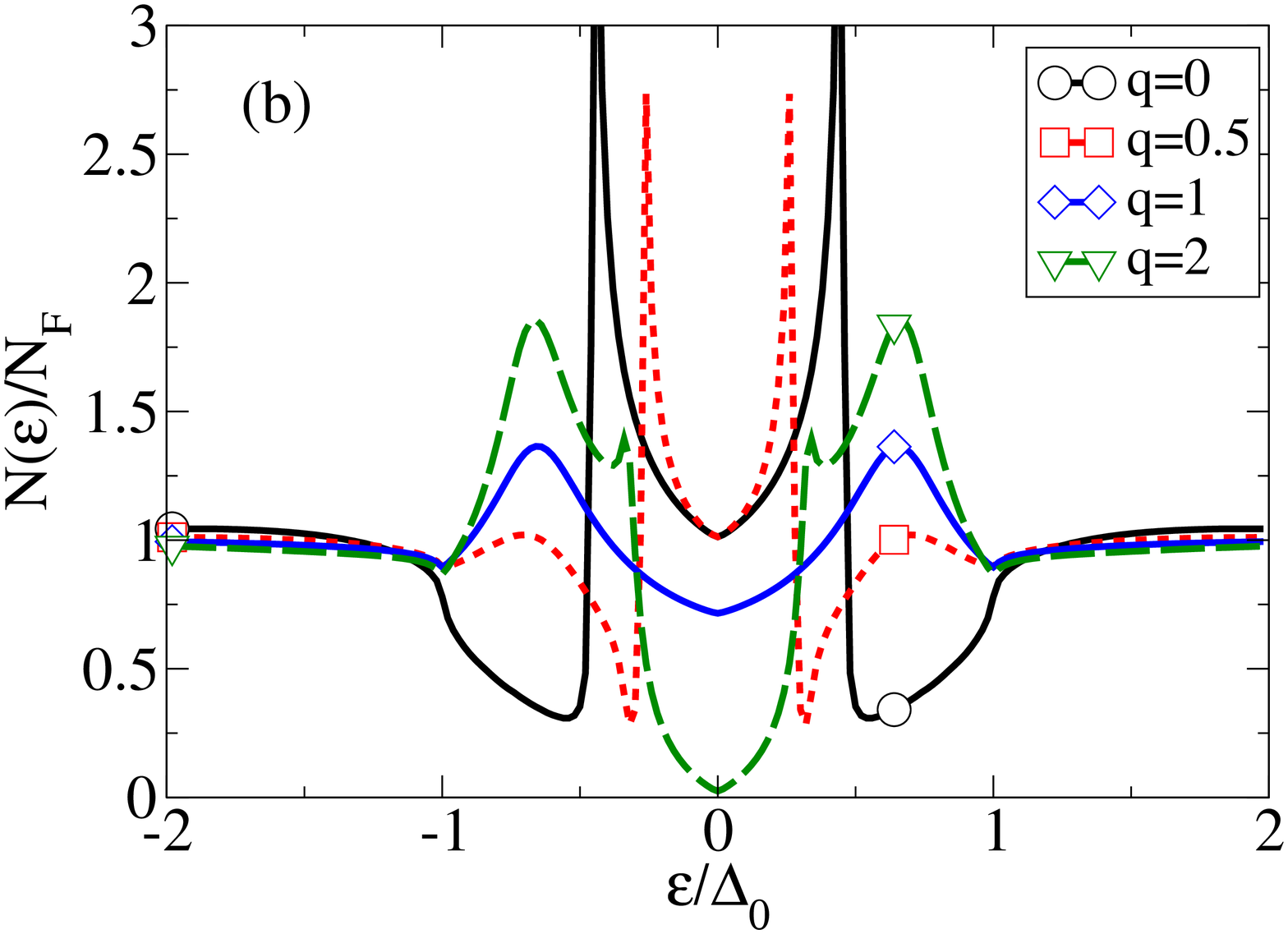} 
}
\scalebox{0.57}{
\includegraphics[width=10cm,clip]{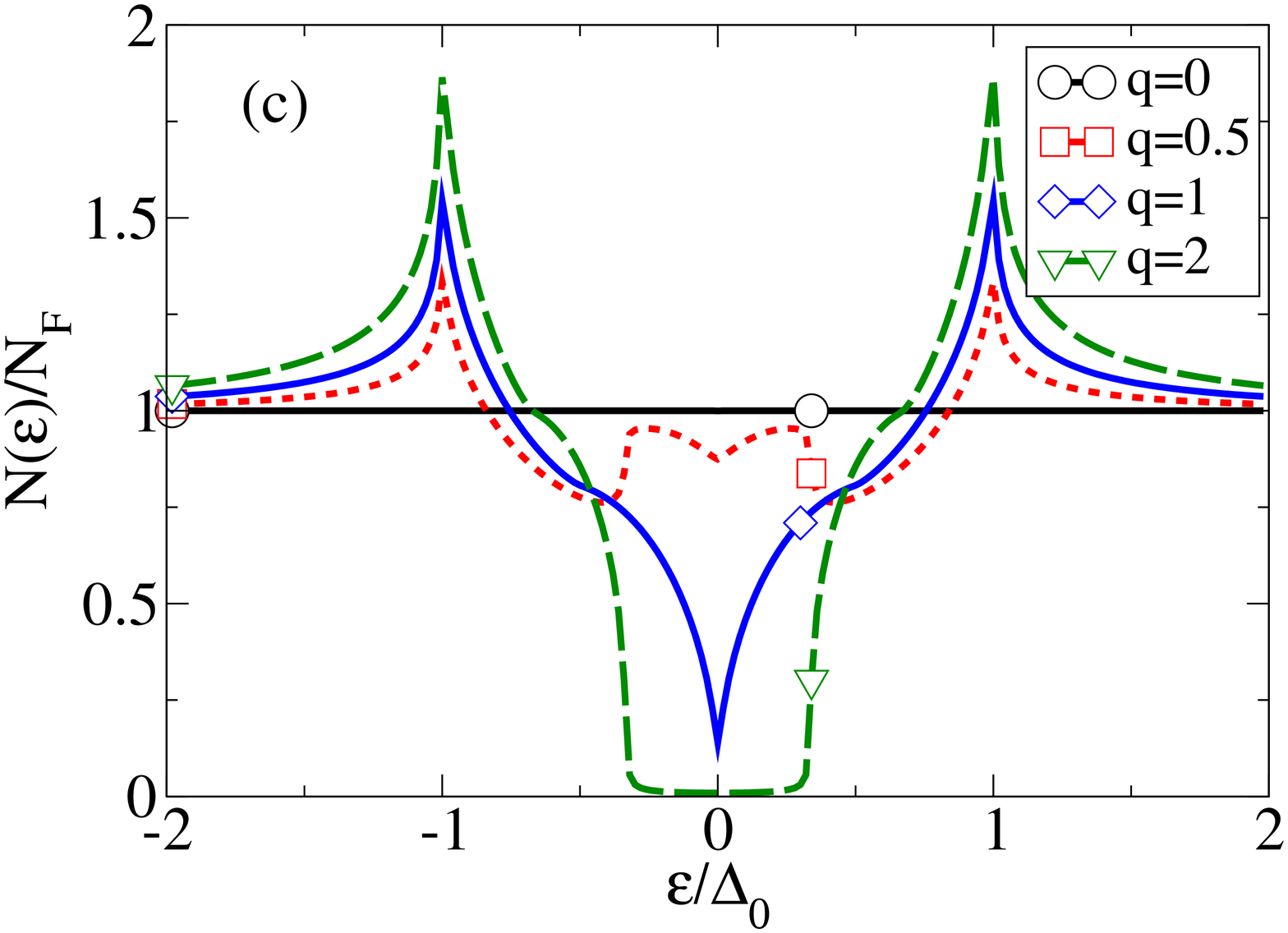} 
\includegraphics[width=10cm,clip]{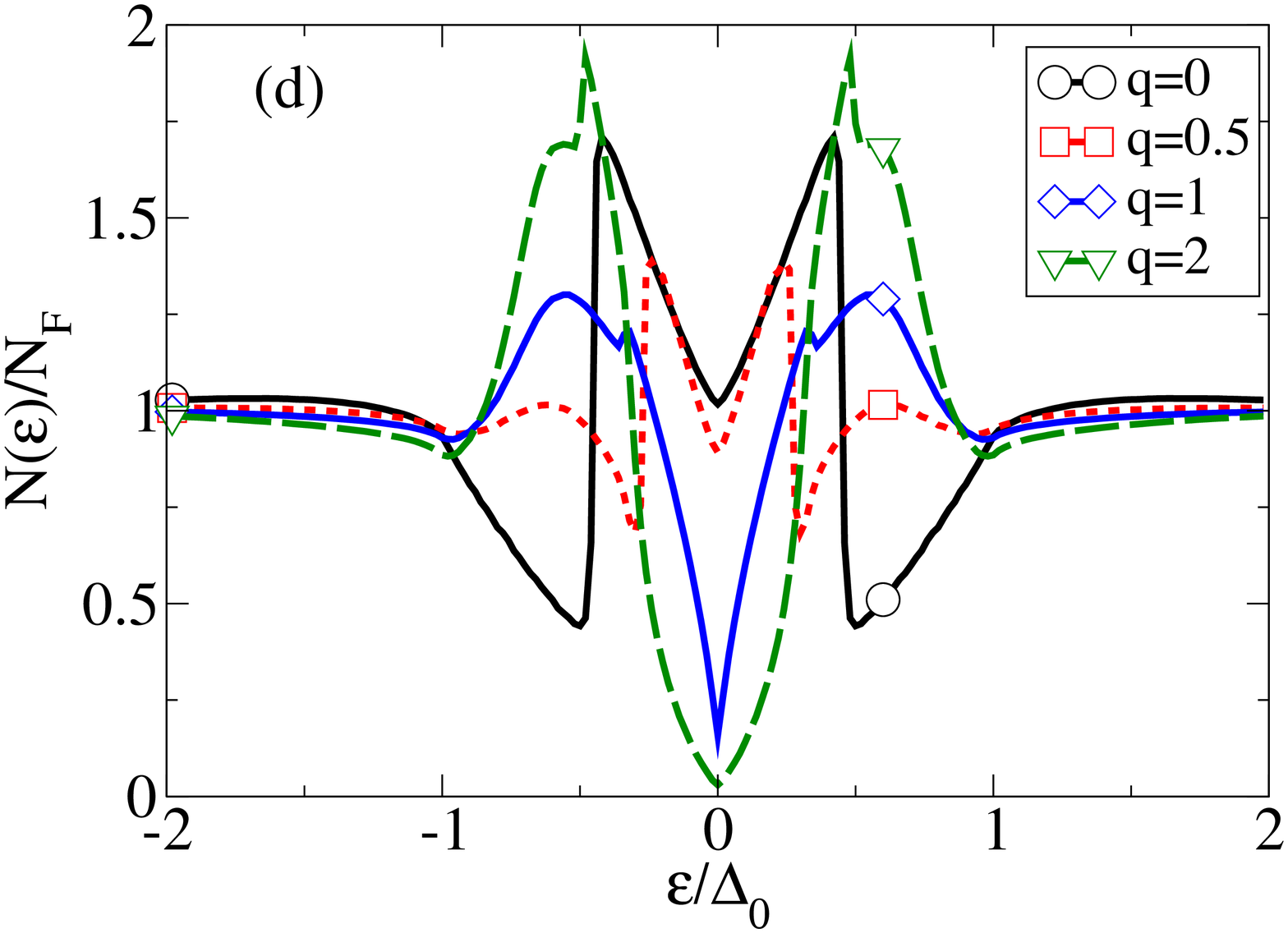} 
}
\end{center}
\caption{(Color online) 
Local surface density of states $N(\epsilon )/N_{\rm F}$ for 
a Rashba superconductor, $\vg(\vk )= \alpha_R \vk \times \hat \vz $. 
The surface is parallel to $\hat \vz $. The curves are for
$\Delta_\pm =\Delta_0(q \pm |\hat \vg (\vk )|) /(q+1)$,
with $q $ ranging from 0 to 2.
In (a) and (c)
the order parameter is assumed constant up to the surface, and in (b) and (d)
a suppression of the order parameter to zero in a surface
layer of thickness $W=2\xi_0$ with $\xi_0=\hbar v_{\rm F}/2\pi k_{\rm B}T_c$ is
assumed. (a) and (b) is for a cylindrical Fermi surface,
$\vv_{\rm F} = (v_x,v_y,0)$, and
(c) and (d) is for a spherical Fermi surface.
(The symbols are labels for the curves only).
}
\label{fig:E1}
\end{figure}
In Fig.~\ref{fig:E1} we show results for a Rashba-type spin-orbit coupling,
$\vg (\vk)=\alpha_R \vk \times \hat \vz $. 
The surface is aligned with the $\hat z$ direction.
In (a) and (b) we use a cylindrical Fermi surface, for which $|\hat \vg (\vk )|=1$.
In (c) and (d) the results for a spherical Fermi surface are shown,
for which $|\hat \vg (\vk )|=\sin (\theta_\vk )$.
The effect of a surface layer with suppressed order parameter is 
illustrated in Fig.~\ref{fig:E1} (b) and (d), where Eq.~(\ref{phasefactor})
with $W=2\xi_0$ is used, where $\xi_0$ is the coherence length
$\xi_0=\hbar v_{\rm F}/2\pi k_{\rm B}T_c$.

\begin{petit}
For the special case $q=0$ we have 
$\Delta_+=-\Delta_-=\Delta_0\sin (\theta_\vk )$ (we use a real gauge).
For this case, $\theta_\vg=\theta_{\underline \vg} =\pi/2$, and consequently,
$\cM = \tan^2 \phi_\vg=\cot^2 \phi_\vk $. 
The bound states are then given by \cite{Iniotakis07,Vorontsov}
    \be \frac{\vare}{\Delta_0 } = 
    - \sin \left({2 W \, \vare \over v_{\rm F} \cos\phi_\vk } \pm \phi_\vk \right)
\sin (\theta_\vk ) .
    \label{eq:ABS-W}
    \ee
\end{petit}
Numerical solution of the problem shows that the
``principal'' bound state branches 
$\varepsilon (\phi_\vk )$ with energies away from the continuum edge
contribute the most to the subgap DOS. 
For $W\ne0 $ the main branch
$\varepsilon_{bs}(\phi_\vk )$ develops a maximum at
$\varepsilon^\star<\Delta_0$, which gives rise to a
peak in the surface DOS near $\varepsilon^\star$, see Fig.~\ref{fig:E1} (b) and (d).
Fully self-consistent solution confirms this \cite{Vorontsov}. 
For $q\to \infty$ 
the order parameter becomes insensitive to helicity mixing, i.e. the effective
$W$ decreases for increasing $q$.

Andreev bound states in non-centrosymmetric superconductors have unusual spin structure \cite{Vorontsov,Rapid}.
It is found, that the
states corresponding to different branches of
Eq.~(\ref{eq:ABS-W}) have opposite spin polarization. Since the
spin polarization changes sign for reversed trajectories, the
Andreev states carry spin current along the interface.
Such spin currents exist in
NCS materials because the spin is not conserved, 
and consequently precession terms enter the continuity equation
\cite{EIRashba:2003}.
There are spin currents both in the normal state and in the superconducting
state.
As was found in Ref.~\cite{Vorontsov},
the most prominent feature is a large surface current with
out of plane spin polarization (reminiscent to that in spin Hall bars
\cite{EGMishchenko:2004}) that flows
along the surface, and decays rapidly into the bulk on a Fermi wavelength
scale.  In addition, there is also a surface induced
superconducting spin current with out of plane spin polarization, 
that adds to the background microscopic spin currents
and greatly exceeds them in the limit of small spin-orbit band splitting.
This effect is in this case solely determined by the structure of the superconducting gap.
Superconducting spin currents decay into the bulk on the scale
of the coherence length and show
oscillations determined by the spin-orbit strength due to Faraday-like
rotations of the spin coherence functions along quasiparticle
trajectories \cite{Vorontsov}.

\subsubsection{Tunneling conductance}

For a three-dimensional model, which for the Rashba-type spin-orbit coupling was
discussed in Ref.~\cite{Iniotakis07}, we present in the following tunneling conductances
in various geometries. 
We will discuss several types of spin-orbit interaction:
\bea
\tens{C}_{4v}:&& \quad \vg= 
\eta \left(\begin{array}{c} 
\hat k_y \\ -\hat k_x \\ 0
\end{array}\right)
+\eta'
\left(\begin{array}{c} 
\hat 0\\ 0 \\ \hat k_x \hat k_y\hat k_z(\hat k_x^2-\hat k_y^2) 
\end{array}\right),
\nonumber
\\
\tens{T}_d:&& \quad
\vg= \eta
\left(\begin{array}{c} 
\hat k_x(\hat k_y^2-\hat k_z^2) \\\hat k_y(\hat k_z^2-\hat k_x^2) \\\hat k_z(\hat k_x^2-\hat k_y^2)
\end{array}\right), 
\qquad
\tens{O}: \quad
\vg = \eta
\left(\begin{array}{c} 
\hat k_x\\  \hat k_y \\ \hat k_z 
\end{array}\right),
\label{gvectors}
\eea
For the symmetry ${\tens{C}_{4v}}$, corresponding to the tetragonal point group, 
the two parameters $\eta $ and $\eta'$ can
both be non-zero. We will discuss below the special cases $\eta=0$ and $\eta'=0$.
The case $\eta'=0$ corresponds to a Rashba spin-orbit coupling.
The type of spin-orbit coupling we consider for the full tetrahedral point group, 
${\tens{T}_d}$, is also known as Dresselhaus coupling.
Finally, for the cubic point group, $\tens{O}$, the simples form for $\vg$ 
is considered here, which is fully isotropic.
All the cases above are relevant for non-centrosymmetric superconductors:
$\tens{C}_{4v}$ for CePt$_3$Si, CeRhSi$_3$, and CeIrSi$_3$, $\tens{T}_d$ for
Y$_2$C$_3$ and possibly KOs$_2$O$_6$, and $\tens{O}$ for Li$_2$(Pd$_{1-x}$Pt$_x$)$_3$B.

The zero temperature tunneling conductance is obtained 
according to the formula Eq.~(\ref{tunnel}), which leads for a spin-inactive
$\delta $-function barrier to
\be
\frac{G(eV)}{G_{\rm N}}=\frac{\langle \cos (\alpha_\vk ) \; D(\alpha_\vk ) \; N(\vk ,eV) \rangle}{
\langle \cos (\alpha_\vk ) \; D(\alpha_\vk ) \rangle},\quad
D(\alpha_\vk )=\frac{D_0 \cos^2(\alpha_\vk )}{1-D_0 \sin^2(\alpha_\vk) }
\label{Dfactor}
\ee
where $\alpha_\vk $ is the angle between the surface normal and $\vk $.
A remark is in place here. In principle, the interface barrier will be spin-dependent
once a spin-orbit split material is brought in contact with a normal metal. However,
for the limit of small spin-orbit splitting we can neglect the spin-dependence of
the interface potential consistent with the quasiclassical approximation. The corrections
are of the same order as the corrections for the quasiparticle velocity in this case, 
and are of higher order in the parameter {\sc small}.

\begin{figure}[t]
\begin{center}
\scalebox{0.475}{
\includegraphics[width=10cm,clip]{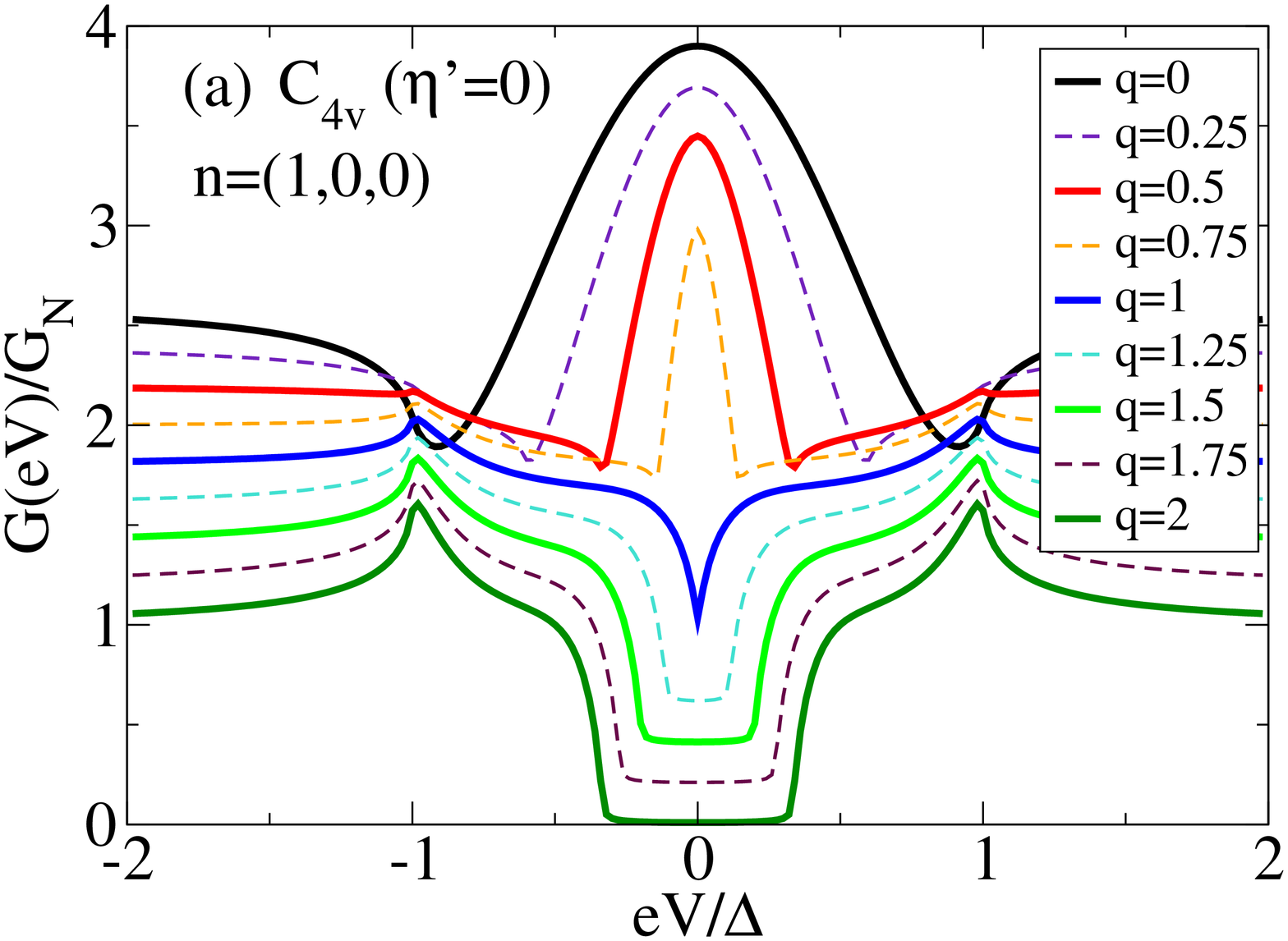} 
\hspace{1cm}
\includegraphics[width=10cm,clip]{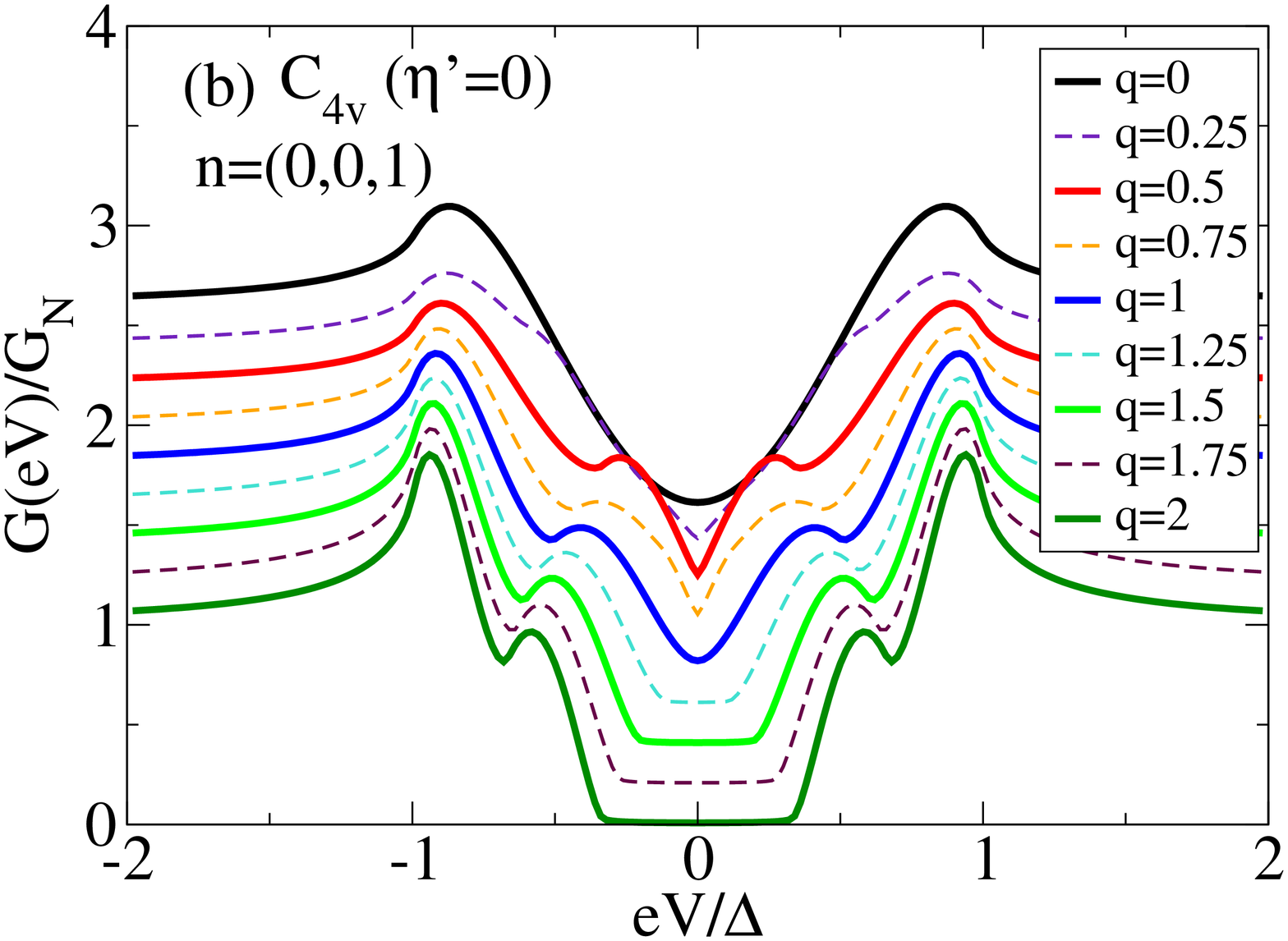} 
}
\scalebox{0.475}{
\includegraphics[width=10cm,clip]{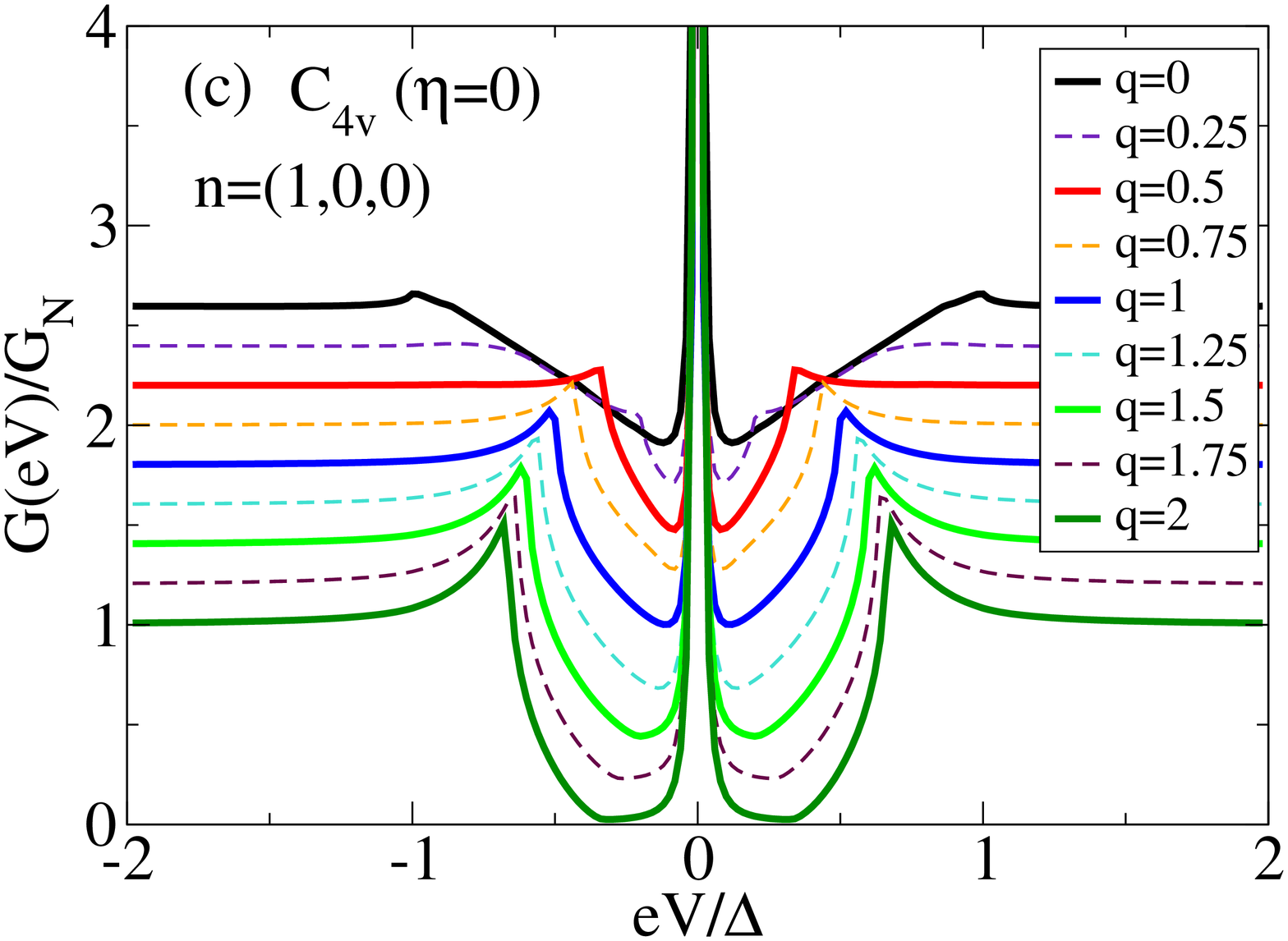} 
\hspace{1cm}
\includegraphics[width=10cm,clip]{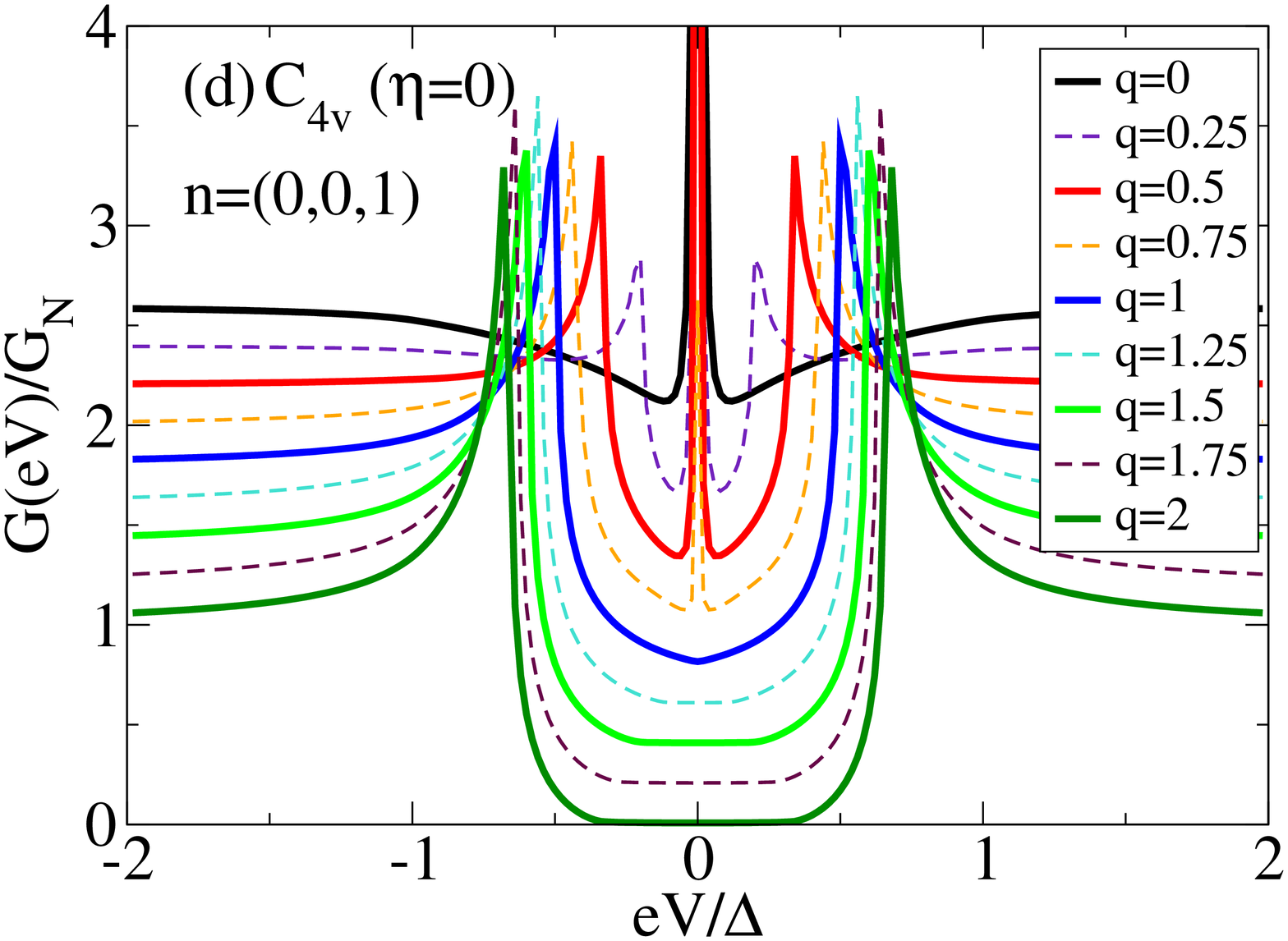} 
}
\scalebox{0.475}{
\includegraphics[width=10cm,clip]{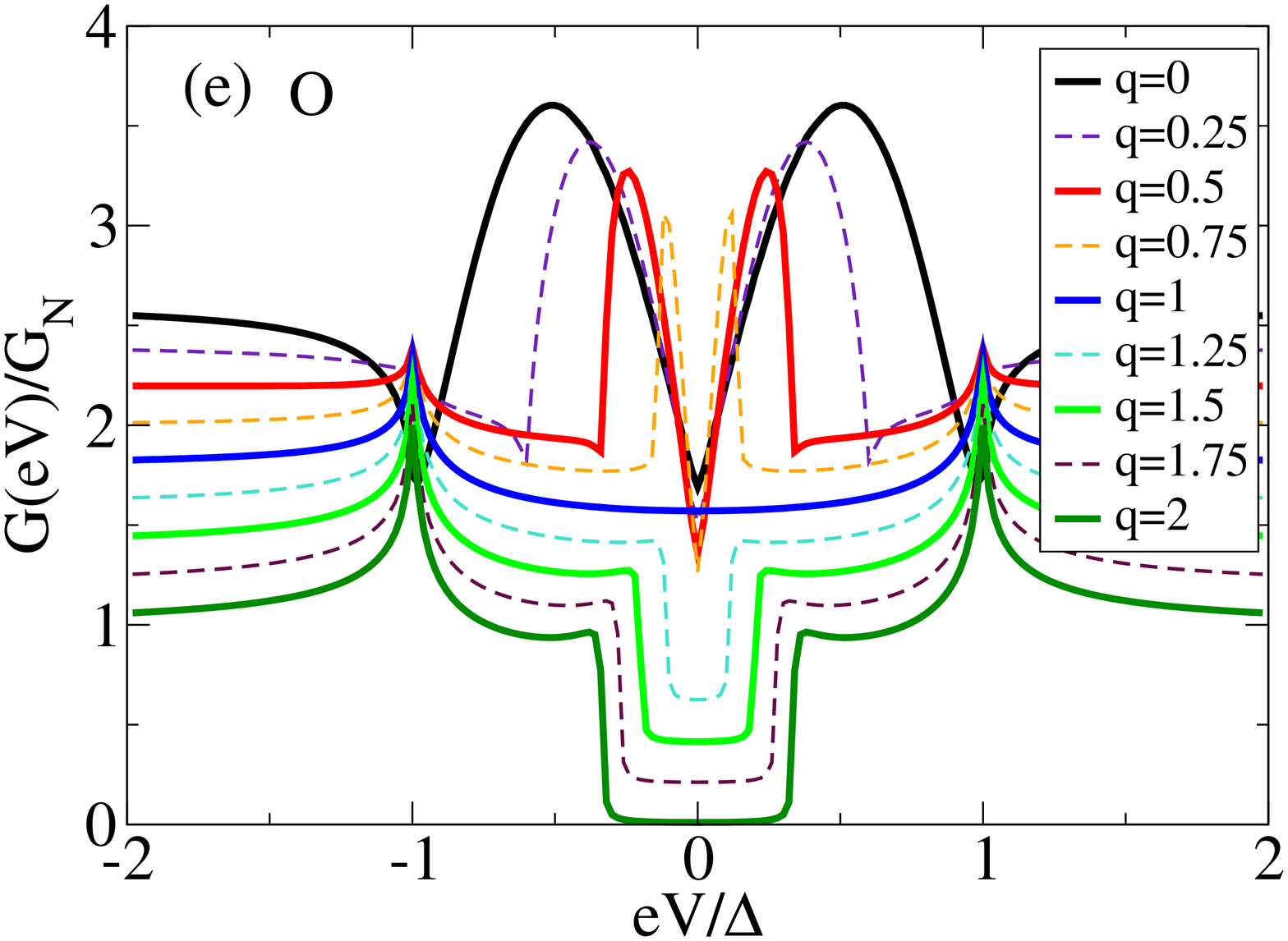}
\hspace{1cm}
\includegraphics[width=10cm,clip]{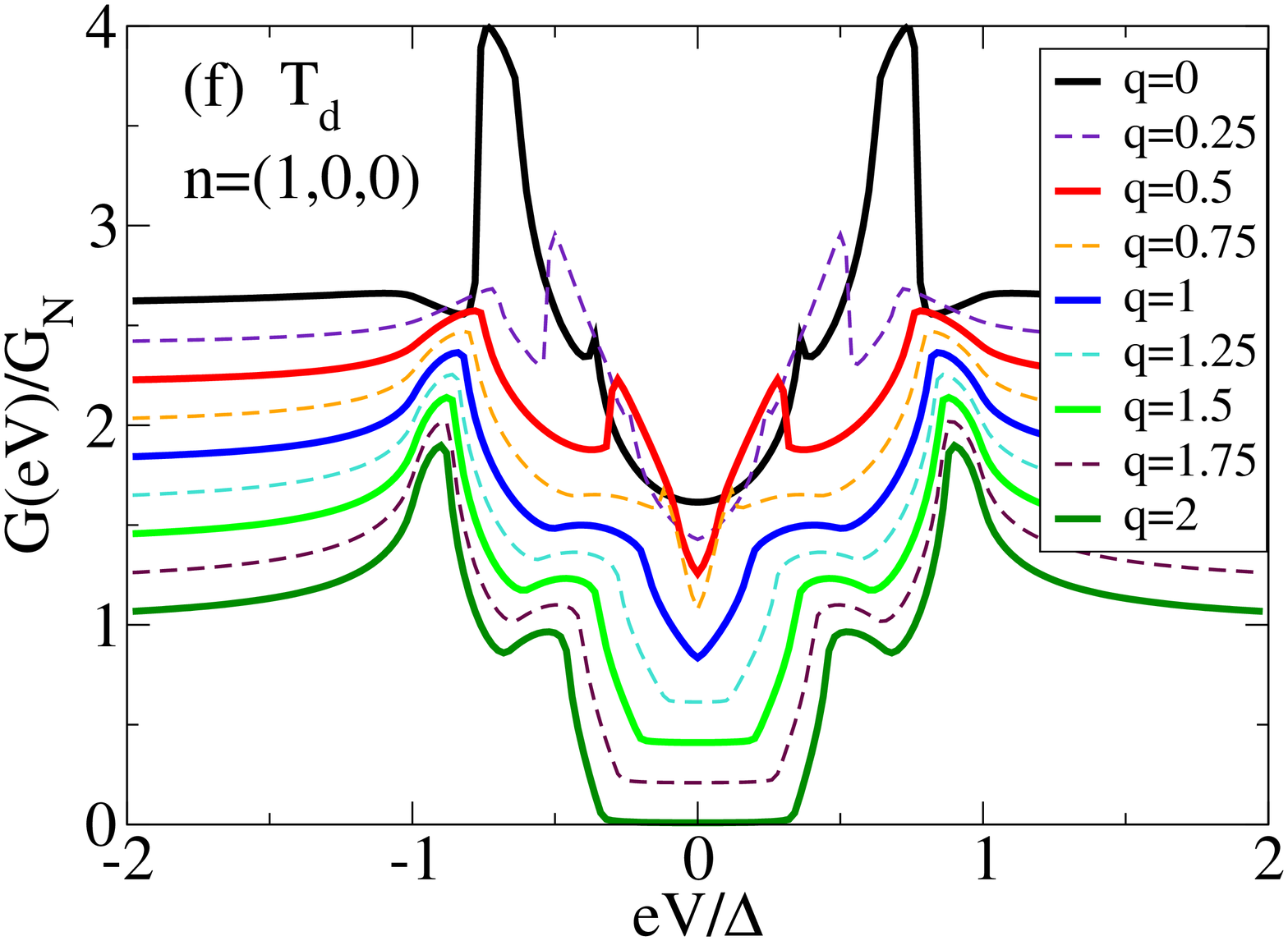} 
}
\scalebox{0.475}{
\includegraphics[width=10cm,clip]{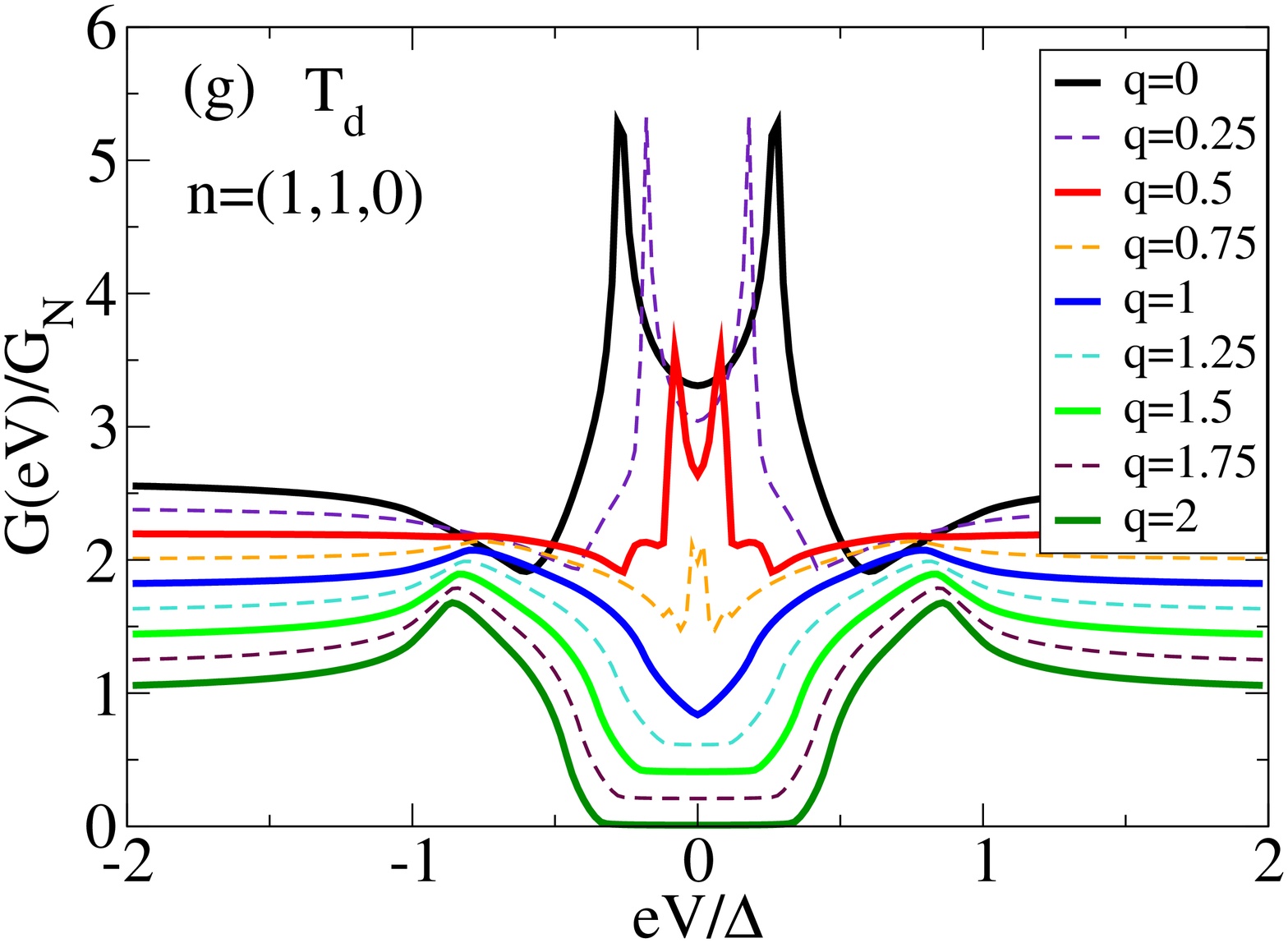} 
\hspace{1cm}
\includegraphics[width=10cm,clip]{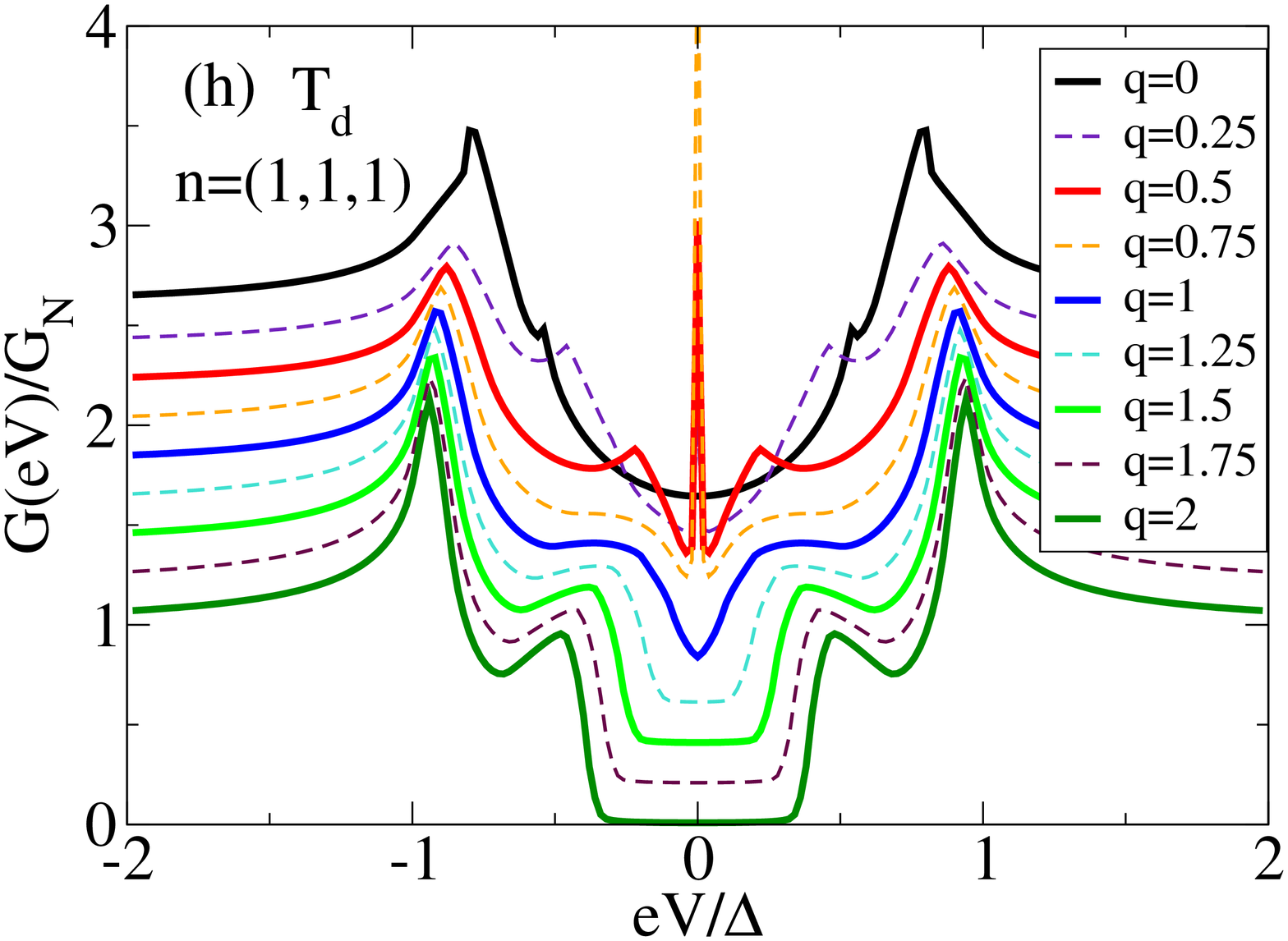} 
}
\end{center}
\caption{(Color online) 
Tunneling conductance $G(eV )/G_{\rm N}$ for various
types of spin-orbit coupling corresponding to the indicated symmetry groups,
and for various alignments of the surface normal $\hat \vn$ as indicated.
The spin-orbit vector is of the form $\tens{C}_{4v}$:
$\vg= \eta [\hat k_y, -\hat k_x, 0 ]+
\eta'[0, 0, \hat k_x\hat k_y\hat k_z (\hat k_x^2-\hat k_y^2) ]$;
${\tens{O}}$:
$\hat \vg = [\hat k_x,\hat k_y,\hat k_z]$ (this case is fully isotropic);
${\tens{T}}_d$:
$\hat \vg= 2[\hat k_x(\hat k_y^2-\hat k_z^2),\hat k_y(\hat k_z^2-\hat k_x^2), \hat k_z(\hat k_x^2-\hat k_y^2)]$.
The curves are for $\Delta_\pm =\Delta_0(q \pm |\hat \vg (\vk )|) /(q+1)$,
with $q $ ranging from 0 to 2.
The order parameter is assumed constant up to the surface, and a
spherical Fermi surface with isotropic Fermi velocity is assumed. 
The tunneling parameter is $D_0=0.1$. Curves are vertically 
shifted by multiples of 0.2.
}
\label{fig:E2}
\end{figure}

In Figure. ~\ref{fig:E2} we show the
tunneling conductance $G(eV )/G_{\rm N}$ obtained from Eq.~(\ref{Dfactor}) with
Eqs.~(\ref{LDOS1})-(\ref{LDOS2})
for various types of spin-orbit coupling corresponding to the spin-orbit couplings 
in Eq.~(\ref{gvectors}), 
and for various alignments of the surface normal with respect to the
crystal symmetry directions.
The tunneling parameter in this figure is $D_0=0.1$.
We show curves for an order parameter
$\Delta_\pm =\Delta_0(q \pm |\hat \vg (\vk )|) /(q+1)$,
with $q $ ranging from 0 to 2. For simplicity, we concentrate here on the
assumption that the order parameter is constant up to the surface,
i.e. we use the bulk solutions Eqs.~(\ref{cohbulk1})-(\ref{gammabulk}).
For a detailed quantitative description, a self-consistent determination of
the order parameter suppression near the surface must be obtained. 
We also use the simplifying assumptions of a
spherical Fermi surface with isotropic Fermi velocity.

As seen from Fig.~\ref{fig:E2}, a rich structure of Andreev bound states 
below the bulk gap energy develops, that depends strongly on the 
alignment of the surface with the crystal symmetry axes. In (a) and (b) a
pure Rashba spin-orbit coupling $\vk=[\hat k_y, -\hat k_x,0]$
on a Fermi sphere is assumed. Below the critical value $q=1$, a zero bias
peak appears for tunneling in the direction perpendicular to the $\hat z$ direction,
i.e. the (1,0,0) or (0,1,0) direction,
however a dependence quadratic in energy appears for tunneling parallel to
the $\hat z$ direction, i.e. the (0,0,1) direction \cite{Iniotakis07}. 
For $q>1$ the tunneling density of 
states acquires a gap, as then the singlet character of the order parameter
dominates. 

In Fig.~\ref{fig:E2} (b) and (c), we show results for a hypothetical spin-orbit coupling
of the form $\vg=\eta'[0, 0, \hat k_x\hat k_y\hat k_z (\hat k_x^2-\hat k_y^2) ]$,
that is consistent with the same point group symmetry $\tens{D}_{4v}$ as the Rashba
spin-orbit coupling. For this case, a sharp zero bias conductance peak exists for
$q<1$ in all tunneling directions. In contrast, for $q>1$, the zero bias conductance
peak only exists when tunneling perpendicular to the $z$-direction, however, not
when tunneling parallel to the $z$-direction.

In Fig.~\ref{fig:E2} (d) we consider the cubic point group symmetry, and assume
the simplest form of a fully isotropic spin-orbit interaction of the form
$\hat \vg = [\hat k_x,\hat k_y,\hat k_z]$. Here, for $q<1$ the tunneling conductance
is zero at zero bias, but raises sharply away from zero bias, showing side
peaks due to Andreev bound states. At $q=1$ this structure disappears with only
a pseudogap remaining. For $q>1$ a gap opens.

Finally, in Fig.~\ref{fig:E2} (e)-(g) we show results for 
the full tetrahedral point group $\tens{T}_d$. 
We compare tunneling in (1,0,0), (1,1,0), and (1,1,1) directions.
Note that in this case, the relation (\ref{gvectors}) between $\vg $ and $\vk$ 
is not invariant under a rotation of both vectors by 90 degree around the 
$\hat k_x$-, $\hat k_y$-, or $\hat k_z$-axis, but an overall sign
change appears; however, the conductance spectra are insensitive to this
sign change.
For $q<1$ there is a vanishing zero bias conductance for tunneling in (1,0,0)
direction, and a low-energy dispersive Andreev bound state branch
for tunneling in (1,1,0) direction.
For tunneling in (1,1,1) direction, the zero bias conductance vanishes for
$q=0$, and shows 
a sharp zero bias peak for $0<q<1$.
For $q>1$ the tunneling conductance becomes gapped for all directions.

As can be seen from these results, studying directional resolved tunneling
in non-centrosymmetric superconductors gives important clues about the 
order parameter symmetry and the type of spin-orbit interaction.

\subsubsection{Andreev point contact spectra}

Here we present results for the case of a point contact between a normal metal
and a non-centrosymmetric superconductor. We use Eq.~(\ref{cond}) to calculate the
spectra, with a scattering matrix that has the form shown in Eq.~(\ref{scatt}).
We assume isotropic Fermi surfaces in the materials on both sides of the
interface, and for simplicity use equal magnitudes for Fermi momenta and velocities.
The transmission amplitude is modeled by that for a $\delta $-function barrier,
\be
t(\alpha_\vk )=\frac{t_0 \cos(\alpha_\vk )}{\sqrt{1-t_0^2 \sin^2(\alpha_\vk)} } ,
\ee
and the component of the Fermi velocity along the interface normal
in direction of current transport is $\hat \vn \vv_{{\rm F}1}=v_{\rm F} \cos(\alpha_\vk)$.

\begin{figure}[t]
\begin{center}
\scalebox{0.46}{
\includegraphics[width=10cm,clip]{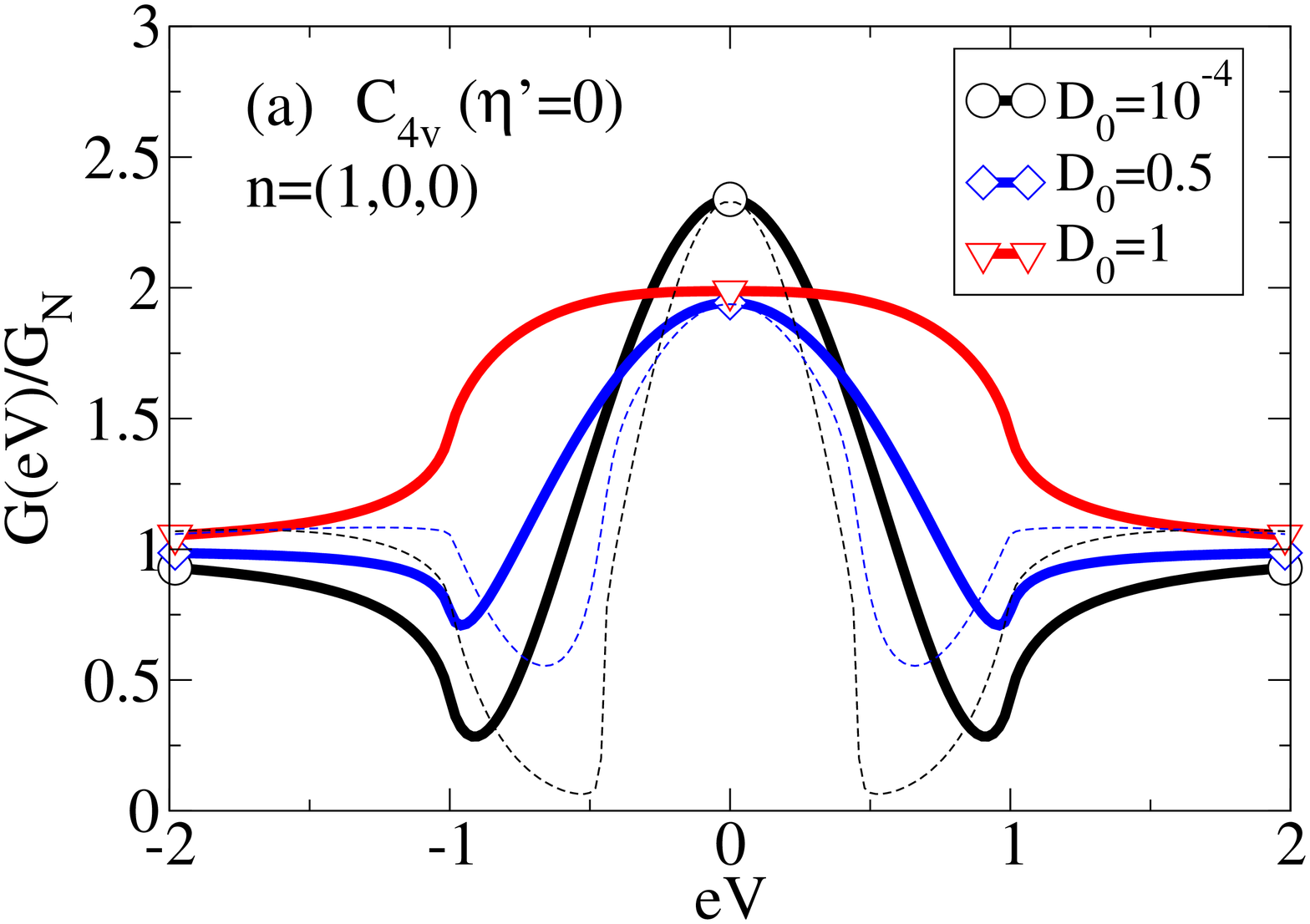}
\hspace{1cm}
\includegraphics[width=10cm,clip]{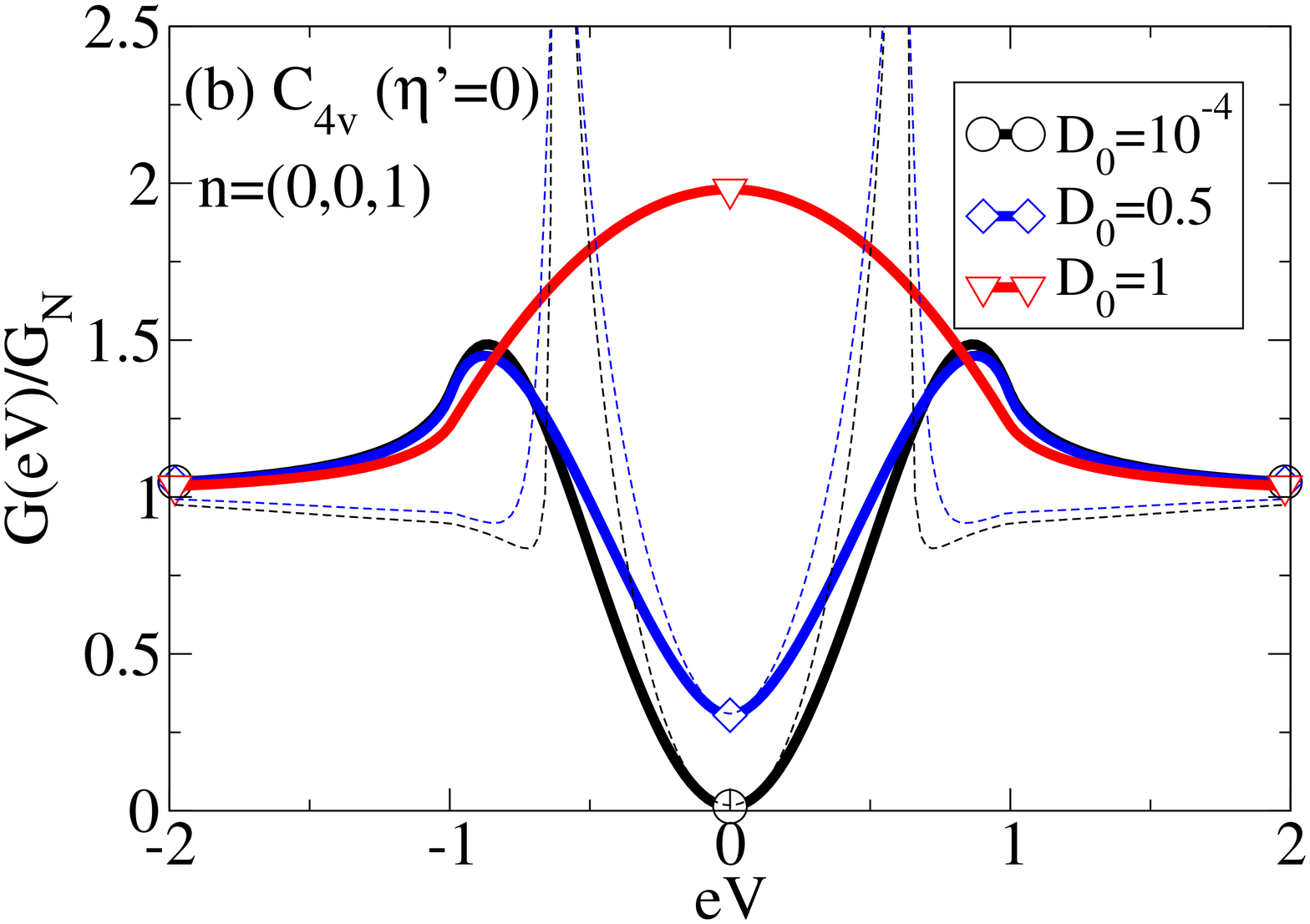} 
}
\scalebox{0.46}{
\includegraphics[width=10cm,clip]{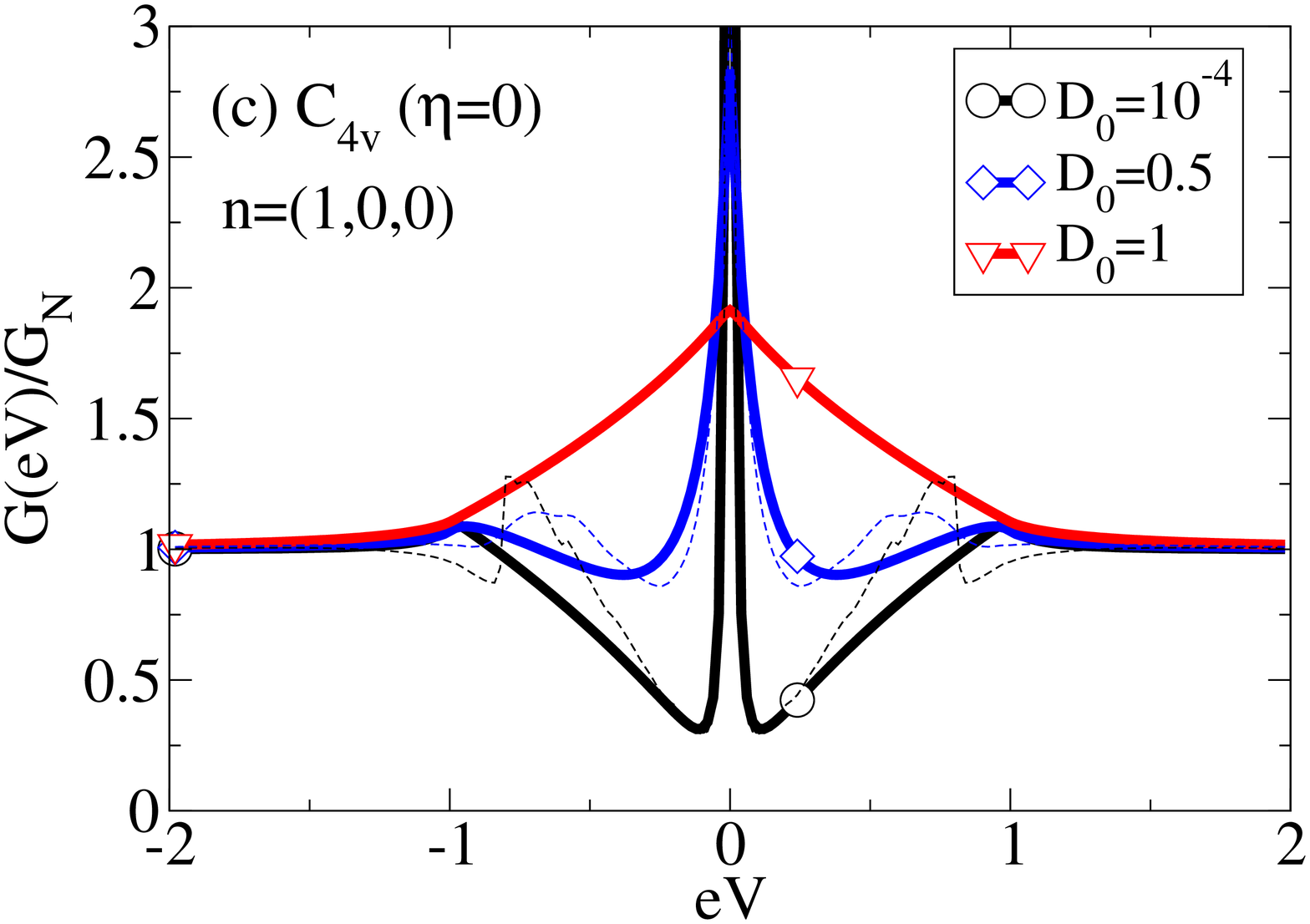}
\hspace{1cm}
\includegraphics[width=10cm,clip]{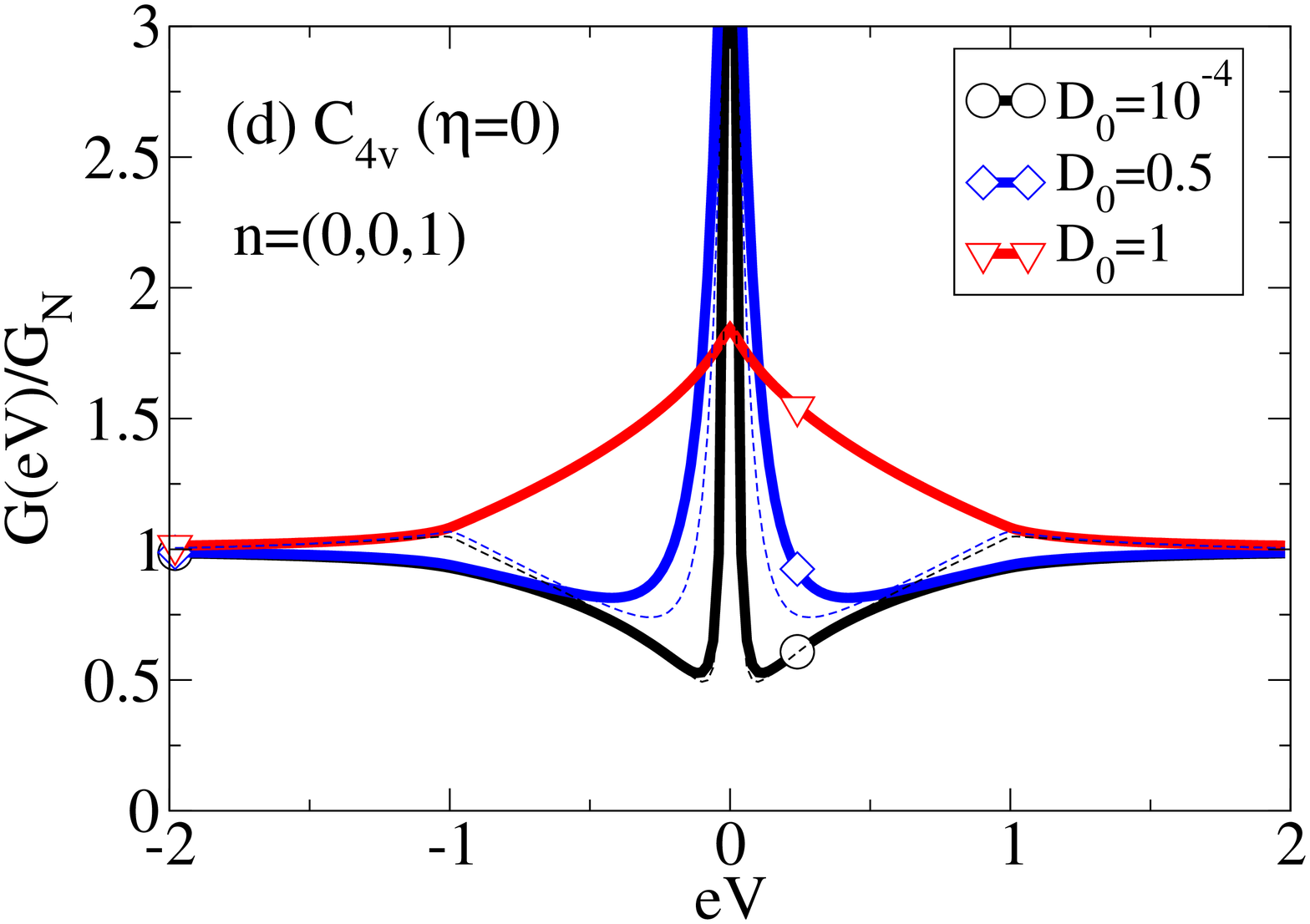} 
}
\scalebox{0.46}{
\includegraphics[width=10cm,clip]{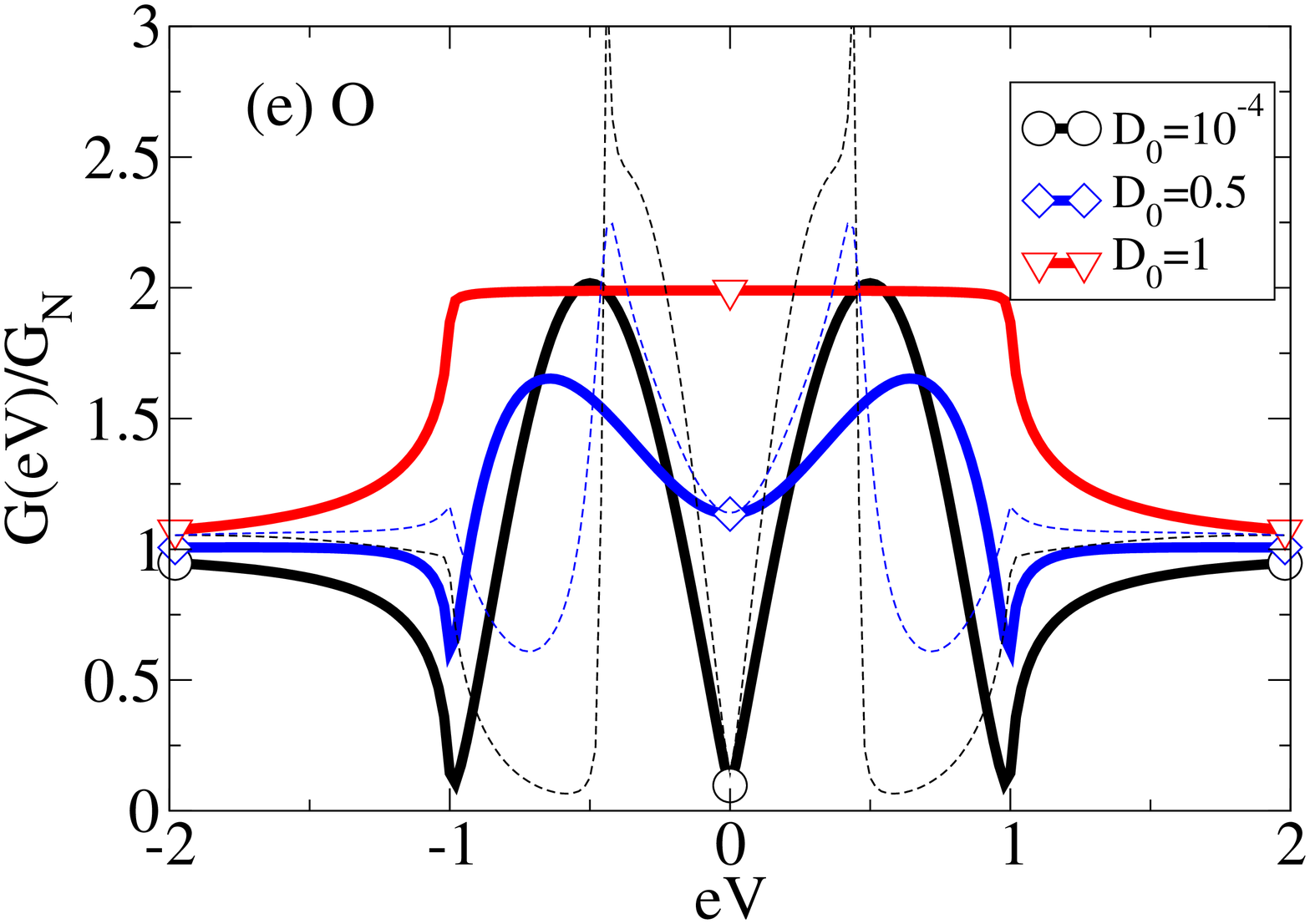}
\hspace{1cm}
\includegraphics[width=10cm,clip]{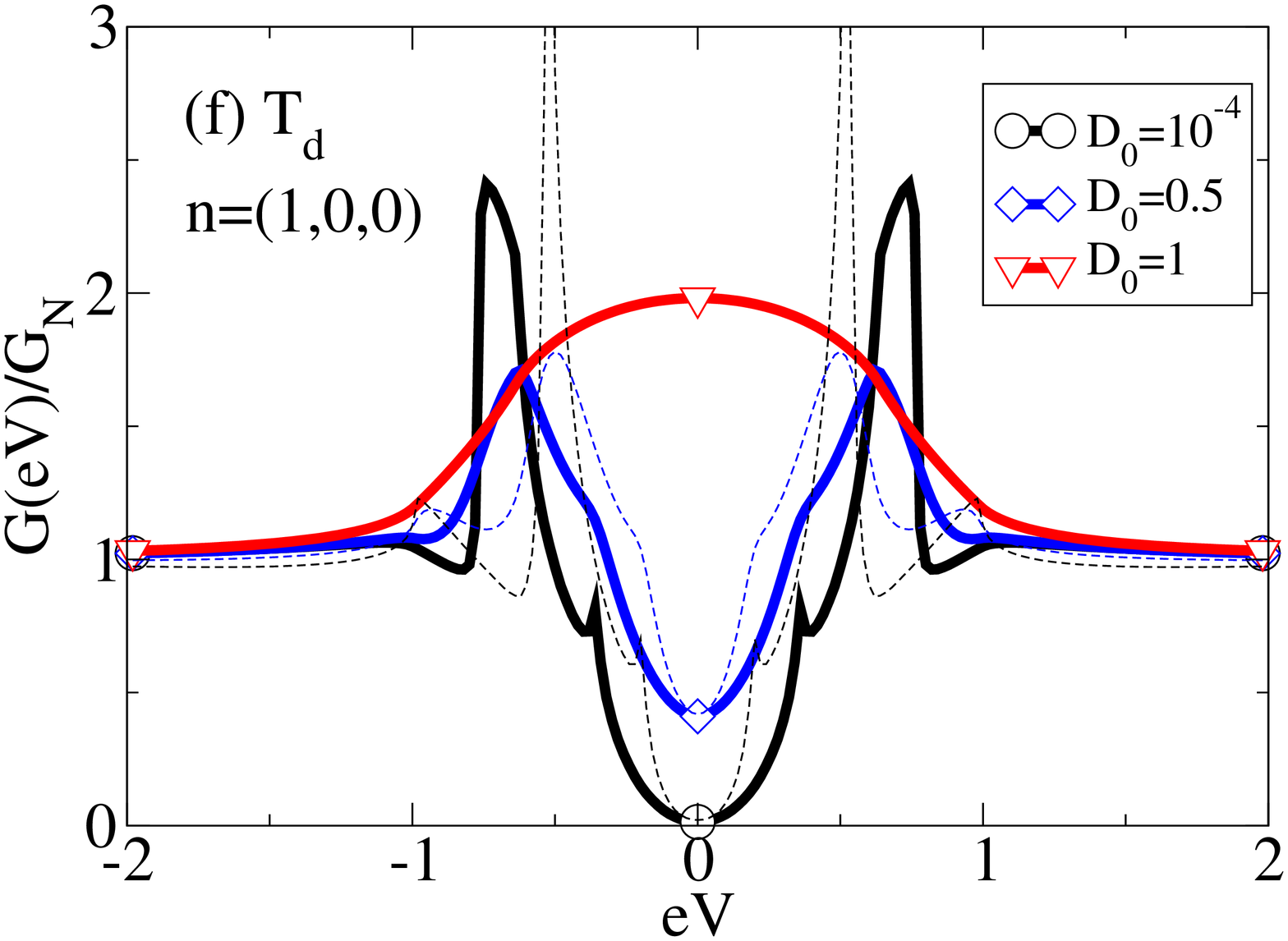} 
}
\scalebox{0.46}{
\includegraphics[width=10cm,clip]{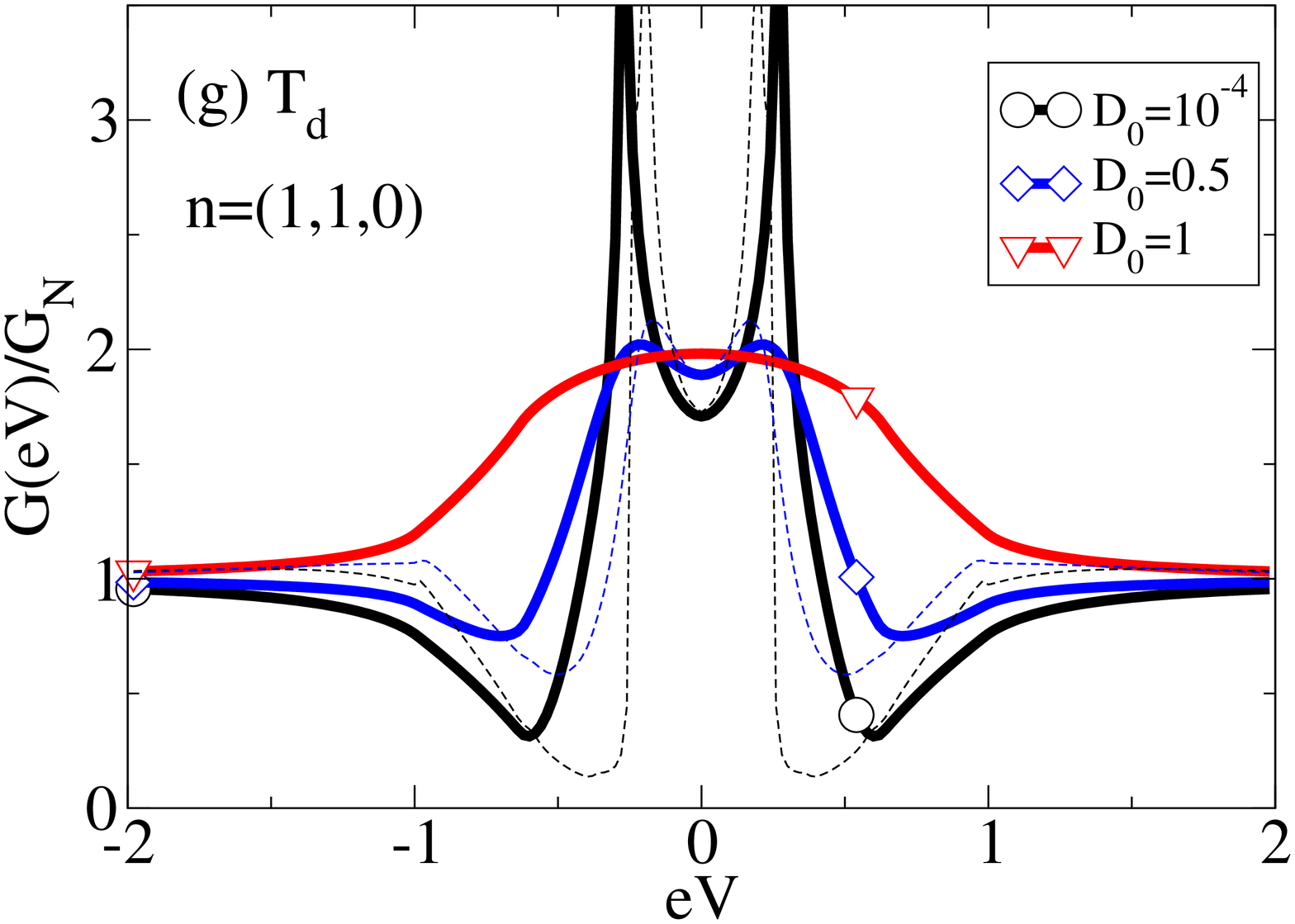} 
\hspace{1cm}
\includegraphics[width=10cm,clip]{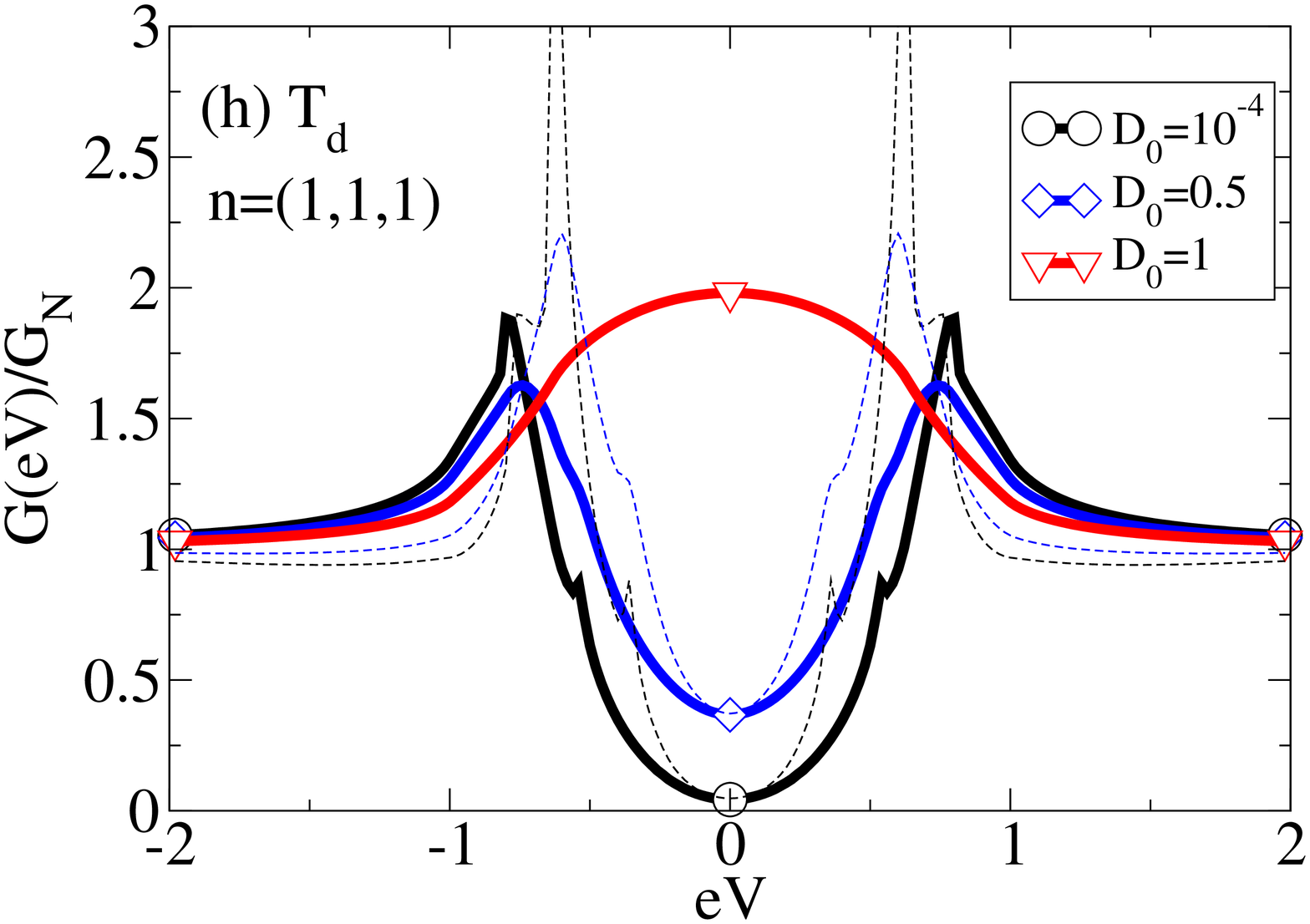} 
}
\end{center}
\caption{(Color online) 
Andreev conductance $G(eV )/G_{\rm N}$ for various
types of spin-orbit coupling corresponding to the indicated symmetry groups,
and for the indicated alignments of the surface normal $\hat \vn$.
The transmission probability $D_0=t_0^2$ is varied.
The spin-orbit vector is of the form $\tens{C}_{4v}$:
$\vg= \eta [\hat k_y, -\hat k_x, 0 ]+
\eta'[0, 0, \hat k_x\hat k_y\hat k_z (\hat k_x^2-\hat k_y^2) ]$;
${\tens{O}}$:
$\hat \vg = [\hat k_x,\hat k_y,\hat k_z]$ (this case is fully isotropic);
${\tens{T}}_d$:
$\hat \vg= 2[\hat k_x(\hat k_y^2-\hat k_z^2),\hat k_y(\hat k_z^2-\hat k_x^2), \hat k_z(\hat k_x^2-\hat k_y^2)]$.
The curves are for $\Delta_\pm =\pm \Delta_0|\hat \vg (\vk )|)$.
The order parameter is assumed constant up to the surface (full lines)
or suppressed to zero in a surface layer
of thickness $W=2\xi_0$ with $\xi_0=\hbar v_{\rm F}/2\pi k_{\rm B}T_c$ (dotted lines).
For $D_0=1$ these two cases give identical results.
A spherical Fermi surface with isotropic Fermi velocity is assumed. 
}
\label{fig:E3}
\end{figure}

In Fig.~\ref{fig:E3}, the
Andreev conductance $G(eV )/G_{\rm N}$ for various
types of spin-orbit coupling 
and for various alignments of the surface normal $\hat \vn$ are shown.
Here, the transmission probability $D_0=t_0^2$ is varied from zero to one.
We restrict here to the case $q=0$, i.e. an order parameter of the form
$\Delta_\pm =\pm \Delta_0|\hat \vg (\vk )|)$.
Again, a spherical Fermi surface with isotropic Fermi velocity is assumed. 
We also compare the case of a surface layer with suppressed order parameter
(dashed lines) with that of an order parameter constant
constant up to the surface (full lines).
To model the order parameter suppression, we assume a layer of
thickness $W=2\xi_0$ with $\xi_0=\hbar v_{\rm F}/2\pi k_{\rm B}T_c$ (dotted lines) in
which the order parameter vanishes. Thus, we use Eq.~(\ref{phasefactor}) as
incoming solutions for the coherence amplitudes.
Note that for $D_0=1$ the surface layer with zero order parameter does not
affect the Andreev conductance. This is due to the fact that for perfect transmission
the normal region simply extends slightly further towards the superconductor, and
within our approximation we neglect the spin-orbit effects in interface potential. 
\begin{petit}
For a larger spin-orbit coupling the interface between a normal metal and a 
normal conducting non-centrosymmetric metal with strong spin-orbit interaction
becomes necessarily spin-active, as the interface potential term
in the Hamiltonian must be hermitian. Thus, a perfect transmission is not realistic
in such a case. For weak spin-orbit splitting these effects are also present, 
however modify the results only to order $\hat v_{\rm SO}/E_{\rm F}$, or
on energy scales $\hat v_{\rm SO}^2/E_{\rm F}$.
\end{petit}
For lower transmission, we remark as an overall observation that the suppression of the
order parameter does not affect the value of the Andreev conductance at
zero bias. This is simply due to the fact that in the clean limit
the coherence amplitudes become effectively spatially constant for $\epsilon =0$.
For higher bias, deviations can be observed, that in general lead to a
shift of Andreev bound states to lower bias.

We turn now to the Andreev point contact spectra for $t_0=1$.
As can be seen, the form of the spectrum is sensitive to the type of spin 
orbit coupling, and the associated order parameter symmetry.
For a Rashba spin-orbit coupling, Fig.~\ref{fig:E3} (a) and (b), the
Andreev conductance is enhanced to twice the normal conductance at zero bias,
however to a smaller value for finite bias. There is a pronounced anisotropy
in the shape of the Andreev conductance spectra.
In (c) and (d) the Andreev conductance
shows a sharp kink feature at zero bias, associated with the complex nodal structure
of the spin-orbit vector. For cubic symmetry, (e), we observe an Andreev conductance
resembling that of an $s$-wave spin singlet superconductor. And, finally, for
a Dresselhaus spin-orbit coupling, (f)-(h), the Andreev conductance shows a 
behavior similar to the case of a Rashba spin-orbit interaction, however with a
not so pronounced anisotropy.

\section{Conclusions}

We have given an overview over the current status of the theoretical 
understanding of Andreev bound states at the surface of a non-centrosymmetric
superconducting material, and have presented results for tunneling conductance,
point contact spectra, and spin polarized Andreev bound state spectra.

The new feature in non-centrosymmetric superconductors is the possible
appearance of spin polarized Andreev states, that carry a spin-current 
along the interface or surface. The presence of such
Andreev bound states that cross the chemical potential as a function of
incident angle to the surface,
is a topologically stable superconducting property.
Such bound states exist as long as triplet order parameter components 
(in spin representation) dominate singlet components of the
order parameter. When both components are equal, the bound states
at the chemical potential disappears, and a topologially new ground
state establishes. The transition between the two states is 
a quantum phase transision.

The spectrum of Andreev states at the surface provides valuable information
about both the structure of the superconducting order parameter and the
vector field of spin-orbit vectors on the Fermi surface. 
In this chapter we have concentrated on the rich structure that 
appears for the limiting case of a small spin-orbit splitting of the
energy bands in the non-centrosymmetric material. In this limit,
the spin quantum number is approximately conserved during scattering from
surfaces and interfaces with normal metals, which leads to strong mixing 
between the helicity bands in the non-centrosymmetric material.
The opposite limit of strong spin-orbit splitting is still largely 
unexplored. We have provided a theoretical basis in this chapter 
that allows to treat this case as well.

Finally, we would like to mention that
interesting effects, like e.g. effects related to the spin Hall effect, or to
Berry phases associated with the change of the spin-orbit vector along 
closed paths, are interesting subjects left for future studies.

\begin{acknowledgement}
The authors would like to thank 
A. Balatsky,
W. Belzig,
J. Inoue,
S. Kashiwaya,
K. Kuroki,
N. Nagaosa, 
J.A. Sauls,
G. Sch\"on,
M. Sigrist,
Y. Tanuma,
I. Vekhter, 
A. Voron\-tsov, and
T. Yokoyama
for valuable discussions or contributions in connection
with the topic of this chapter.
\end{acknowledgement}


\begin{thebibliography}{99}
\bibitem{Giaever} I. Giaever, Phys. Rev. Lett. {\bf 5}, 147 (1960).
\bibitem{Tanaka95} Y. Tanaka and S. Kashiwaya, Phys. Rev. Lett. {\bf 74}, 3451 (1995).
\bibitem{KashiwayaReport} S. Kashiwaya and Y. Tanaka, Rep. Prog. Phys. {\bf 63}, 1641 (2000).

\bibitem{Bruder} C. Bruder, Phys. Rev. B {\bf 41}, 4017 (1990).
\bibitem{Hu} C.R. Hu, Phys. Rev. Lett. {\bf 72}, 1526 (1994).
\bibitem{Buchholtz} L.J. Buchholtz, M. Palumbo, D. Rainer, and J.A. Sauls, J. Low Temp. Phys. {\bf 101}, 1099 (1995).

\bibitem{MABS}
 L.J. Buchholtz and G. Zwicknagl, Phys. Rev. B \textbf{23}
5788 (1981); J. Hara and K. Nagai, Prog. Theor. Phys. {\bf 74} (1986) 1237.

\bibitem{Honerkamp} C. Honerkamp and M. Sigrist, J. Low Temp. Phys. \textbf{111}, 895 (1998); M. Yamashiro, Y. Tanaka, and S. Kashiwaya,
Phys. Rev. B \textbf{56}, 7847 (1997).

\bibitem{Matsumoto} M. Matsumoto and M. Sigrist, J. Phys. Soc. Jpn. {\bf 68}, 994 (1999).

\bibitem{Kuroki}
Y. Tanaka, T. Tanuma, K. Kuroki and S. Kashiwaya,
J. Phys. Soc Jpn. \textbf{71} 2102 (2002).

\bibitem{Experiments}
J. Geerk, X.X. Xi, and G. Linker: Z. Phys. B. \textbf{73}, (1988), 
329; S. Kashiwaya, 
Y. Tanaka, M. Koyanagi, H. Takashima, and K.  Kajimura, 
Phys. Rev. B \textbf{51}  (1995) 1350; 
L. Alff, 
H. Takashima, S. Kashiwaya, N. Terada, H. Ihara, Y.
Tanaka, M. Koyanagi, and K. Kajimura, Phys. Rev. B {55},  (1997) R14757; 
M. Covington, 
M. Aprili, E. Paraoanu, L.H. Greene, F. Xu, J.
Zhu, and C.A. Mirkin, 
Phys. Rev. Lett. \textbf{79},  (1997) 277; 
J. Y. T. Wei, 
N.-C. Yeh, D. F. Garrigus and M. Strasik, 
Phys. Rev. Lett. \textbf{81},  (1998) 2542. 

\bibitem{Laube} F. Laube, G. Goll, H. v. L\"ohneysen, M. Fogelstr\"om, and F. Lichtenberg, 
Phys. Rev. Lett. {\bf 84}, 1595 (2000).

\bibitem{Mao} Z.Q. Mao, K.D. Nelson, R. Jin, Y. Liu, and Y. Maeno, Phys. Rev. Lett. {\bf 87}, 037003 (2001); 
M. Kawamura, H. Yaguchi, N. Kikugawa,
Y. Maeno, H. Takayanagi, J. Phys. Soc. Jpn. \textbf{74}, (2005) 
531.
 


\bibitem{Ott}Ch. W\"{a}lti, H.R. Ott, Z. Fisk, and J.L. Smith,
Phys. Rev. Lett. {\bf 84}, (2000) 5616.

\bibitem{Wei}
P. M. C. Rourke, M. A. Tanatar, C. S. Turel, 
J. Berdeklis, C. Petrovic, and J. Y. T. Wei, Phys. Rev. Lett. 94, (2005) 107005.
\bibitem{Ichimura}
K. Ichimura, S. Higashi, K. Nomura and A. Kawamoto, 
Synthetic Metals Vol. 153 (2005) 409. 

\bibitem{Turel}C.S. Turel, J.Y.T. Wei, W.M. Yuhasz and M.B. Maple, Physica C, 463-465 32 (2007). 

\bibitem{Aoki}
Y. Aoki, Y. Wada, M. Saitoh, R. Nomura, Y. Okuda, Y. Nagato, M. Yamamoto, S. Higashitani, and K. Nagai, Phys. Rev. Lett. {\bf 95}, (2005) 075301. 




\bibitem{Blonder} G. E. Blonder, M. Tinkham, and T. M. Klapwijk, Phys. Rev. B
\textbf{25}, 4515 (1982).

\bibitem{eschrig00} 
M. Eschrig, Phys. Rev. B {\bf 61}, 9061 (2000).




\bibitem{Maeno}
A. P. Mackenzie and Y. Maeno, Rev. Mod. Phys. {\bf 75}, (2003) 657;  
Y. Maeno et al., Nature \textbf{394}, 532 (1994)


\bibitem{Maeno1}
Y. Maeno, H. Hashimoto, K. Yoshida, S. Nishizaki, T.
Fujita, J. G. Bednorz, and F. Lichtenberg, Nature (London)
372, 532 (1994).

\bibitem{Maeno2}
K. Ishida, H. Mukuda, Y. Kitaoka, K. Asayama, Z. Q.
Mao, Y. Mori, and Y. Maeno, Nature (London) 396, 658
(1998).

\bibitem{Maeno3}
G. M. Luke, Y. Fudamoto, K. M. Kojima, M. I. Larkin, J.
Merrin, B. Nachumi, Y. J. Uemura, Y. Maeno, Z. Q. Mao,
Y. Mori, H. Nakamura, and M. Sigrist, Nature (London)
394, 558 (1998).

\bibitem{Maeno4}
A. P. Mackenzie and Y. Maeno, Rev. Mod. Phys. 75, 657
(2003).

\bibitem{Maeno5}
K. D. Nelson, Z. Q. Mao, Y. Maeno, and Y. Liu, Science
306, 1151 (2004); Y. Asano, Y. Tanaka, M. Sigrist, and S.
Kashiwaya, Phys. Rev. B 67, 184505 (2003); Phys. Rev. B
71, 214501 (2005).


\bibitem{Kambara}
H. Kambara, 
S. Kashiwaya, H. Yaguchi, Y. Asano, Y. Tanaka 
and Y. Maeno, Phys. Rev. Lett. \textbf{101}, 267003 (2008). 

\bibitem{Tanuma}
Y. Tanuma, N. Hayashi, Y. Tanaka, and A. A. Golubov,  
Phys. Rev. Lett. 102, 117003 (2009); 
T. Yokoyama, C. Iniotakis, Y. Tanaka, and M. Sigrist,  
Phys. Rev. Lett. \textbf{100}, 177002 (2008). 

\bibitem{Doppler}
M. Fogelstr\"{o}m, D. Rainer and J. A. Sauls, Phys. Rev. Lett.
\textbf{79} 281
(1997).

\bibitem{Laube04}
F.~Laube, G.~Goll, M.~Eschrig, M.~Fogelstr{\"o}m, and R.~Werner,
Phys. Rev. B {\bf 69}, 014516 (2004). 

\bibitem{Yamashiro99}
M. Yamashiro, Y. Tanaka, N. Yoshida, and S. Kashiwaya,
J. Phys. Soc. Jpn. {\bf 68}, 2019 (1999).

\bibitem{Girvin}
See for e.g., {\it The Quantum Hall effect}, edited by R.E. Prange and 
S.M. Girvin, (Springer-Verlag, 1987), and references therein.

\bibitem{Thouless}
D. J. Thouless, 
M. Kohmoto, M. P. Nightingale, and M. den Nijs,
Phys. Rev. Lett. \textbf{49}, 405 (1982).

\bibitem{Furusaki} J. Goryo and K. Ishikawa, J. Phys. Soc. Jpn.
\textbf{67}, 3006 (1998); 
A. Furusaki, M. Matsumoto, and M. Sigrist, Phys. Rev. B \textbf{64}
054514 (2001).

\bibitem{Bauer} E. Bauer, G. Hilscher, H. Michor, Ch. Paul, E.W. Scheidt, A. Gribanov, Yu. Seropegin,
H. No\"el, M. Sigrist, and P. Rogl, Phys. Rev. Lett. {\bf 92}, 027003 (2004).

\bibitem{Frigeri} P. A. Frigeri, D. F. Agterberg , A. Koga and M. 
Sigrist, 
Phys. Rev. Lett. \textbf{92}, 097001 (2004).

\bibitem{Interface} N. Reyren et al., Science \textbf{317}, 1196 (2007).

\bibitem{Yada}
K. Yada, S. Onari, Y. Tanaka, and J. Inoue,  
Phys. Rev. B \textbf{80}, 140509 (2009). 

\bibitem{Qi}
X.L. Qi, T. L. Hughes, S. Raghu and S.C. Zhang, 
Phys. Rev. Lett. 102, 187001 (2009); 
M. Sato and S. Fujimoto, Phys. Rev. B \textbf{79}, 094504 (2009); 
R. Roy, arXiv:cond-mat/0608064; 
C. K. Lu and S.-K. Yip Phys. Rev. B 78, 132502 (2008); 
S.-K. Yip, arXiv:0910.0696.

\bibitem{Rapid}
Y. Tanaka, T. Yokoyama, A. V. Balatsky and N. Nagaosa, 
Phys. Rev. B \textbf{79}, 060505(R) (2009).  

\bibitem{Andreev} A. F. Andreev, Sov. Phys. JETP \textbf{19}, 1228 (1964).

\bibitem{Iniotakis07}
C. Iniotakis, N. Hayashi, Y. Sawa, T. Yokoyama, U. May, 
Y. Tanaka, and M. Sigrist, Phys. Rev. B {\bf 76}, 012501 (2007).

\bibitem{Yokoyama} T. Yokoyama, Y. Tanaka and J. Inoue,
Phys. Rev. B \textbf{72} 220504(R) (2005).

\bibitem{Vorontsov}
A.B. Vorontsov, I. Vekhter, M. Eschrig, Phys. Rev. Lett. 
\textbf{101}, 127003 (2008). 

\bibitem{Linder}
J. Linder and A. Sudb{\o}, Phys. Rev. B \textbf{76}, 054511 (2007).

\bibitem{Borkje}
K. B{\o}rkje and A. Sudb{\o}, Phys. Rev. B \textbf{74}, 054506 (2006);  
K. B{\o}rkje, Phys. Rev. B {\bf 76}, 184513 (2007). 

\bibitem{Kashiwaya99}
S. Kashiwaya, Y. Tanaka, N. Yoshida, and M.R. Beasley, 
Phys. Rev. B, \textbf{60} 3572 (1999).

\bibitem{Konig}
M. K\"{o}nig,
S. Wiedmann, C. Br\"{u}ne, A. Roth, H. Buhmann,
L. Molenkamp, X.-L. Qi, and S.-C. Zhang, 
Science, \textbf{318}, 766 (2007).

\bibitem{Mele}
C. L. Kane and E. J. Mele, Phys. Rev. Lett. \textbf {95}, 146802 (2005);
C. L. Kane and E. J. Mele, Phys. Rev. Lett. \textbf {95}, 226801 (2005).

\bibitem{Bernevig}
B. A. Bernevig, and S. C. Zhang, Phys. Rev. Lett. \textbf{96}, 106802 
(2006).

\bibitem{Fu}
L. Fu and C. L. Kane, Phys. Rev. B \textbf{74}, 195312 (2006);
L. Fu and C. L. Kane, Phys. Rev. B \textbf{76}, 045302 (2007).



\bibitem{landau57}
L.~D. Landau, Zh. Eksp. Teor. Fiz. {\bf 32},  59  (1957), [Sov.\ Phys.\ JETP
  {\bf 5}, 101 (1957)].

\bibitem{landau59}
L. D. Landau, Sov. Phys. JETP {\bf 8},  70 (1959).

\bibitem{Serene}
J.~W. Serene and D. Rainer, Phys. Rep. {\bf 101},  221  (1983).

\bibitem{rai86}
D. Rainer, in {\it Progress in Low Temperature Physics X}, p.\ 371, 
edited by D. F. Brewer (Elsevier Science Publishers, Amsterdam, 1986).

\bibitem{eschrig94}
M. Eschrig, J. Heym, and D. Rainer, J. Low Temp. Phys. {\bf 95}, 323 (1994).

\bibitem{rai95}
D. Rainer and J.~A. Sauls,  in {\em Superconductivity: From Basic Physics to
New Developments}, edited by P.~N. Butcher and Y. Lu (World Scientific,
Singapore, 1995), pp.\ 45--78.

\bibitem{esc99fluk}
M. Eschrig, D. Rainer, and J.A. Sauls, Phys. Rev. B {\bf 59}, 12095 (1999);
Appencix C.

\bibitem{larkin68}
A.~I. Larkin and Y.~N. Ovchinnikov, Zh. Eksp. Teor. Fiz. {\bf 55},  2262
  (1968), [Sov. Phys. JETP {\bf28}, 1200 (1969)].

\bibitem{eilen}
G. Eilenberger, Z. Phys. {\bf 214},  195  (1968).

\bibitem{schmid75}
A. Schmid and G. Sch\"on, J. Low. Temp. Phys. {\bf 20}, 207 (1975).

\bibitem{schmid81}
A. Schmid, 
in {\it Nonequilibrium Superconductivity, Phonons and Kapitza Boundaries}, 
Proceedings of NATO 
Advanced Study Institute,
edited by K. E. Gray (Plenum Press, New York, 1981), Chapter 14.

\bibitem{rammer86}
J. Rammer and H. Smith, Rev. Mod. Phys. {\bf 58}, 323 (1986).

\bibitem{Larkin86}
A. I. Larkin and  Y. N. Ovchinnikov,
 in {\em Nonequilibrium Superconductivity},
edited by D. N. Langenberg and A. I. Larkin (Elsevier
Science Publishers, 1986),  p. 493.

\bibitem{FLT}
M. Eschrig, J. A. Sauls, H. Burkhardt, and D. Rainer,
in {\em High-$T_{\rm c}$ Superconductors and Related Materials, Fundamental
Properties, and Some Future Electronic Applications}, Proceedings
of the NATO Advanced Study Institute,
edited by S.-L. Drechsler and T. Mishonov,
pp. 413-446 (Kluwer Academic, Norwell, MA, 2001).

\bibitem{gorkov58}
L.~P. Gor'kov, Zh. Eksp. Teor. Fiz. {\bf 34},  735  (1958), [Sov.\ Phys.\ JETP
  {\bf7}, 505 (1958)],
L.~P. Gor'kov, Zh. Eksp. Teor. Fiz. {\bf 36},  1918  (1959), [Sov.\ Phys.\ JETP
  {\bf9}, 1364 (1959)].

\bibitem{keldysh64}
L.~V. Keldysh, Zh. Eksp. Teor. Fiz. {\bf 47},  1515  (1964), [Sov. Phys. JETP
  {\bf20}, 1018 (1965)].

\bibitem{shelankov85}
A.~L. Shelankov, J. Low. Temp. Phys. {\bf 60}, 29 (1985).

\bibitem{alexander85}
J. A. Alexander, T. P. Orlando, D. Rainer, and P. M. Tedrow,
Phys. Rev. B {\bf 31}, 5811 (1985).

\bibitem{eschrig99}
M. Eschrig, J. A. Sauls, and D. Rainer, Phys. Rev. B {\bf 60}, 10447 (1999). 

\bibitem{nagato93}
Y. Nagato, K. Nagai, and J. Hara, J. Low Temp. Phys. {\bf 93},  33  (1993),
S. Higashitani and K. Nagai, J. Phys. Soc. Jpn. {\bf 64},  549  (1995),
Y. Nagato, S. Higashitani, K. Yamada, and K. Nagai, J. Low Temp. Phys. {\bf
  103},  1  (1996).

\bibitem{schopohl95}
N. Schopohl and K. Maki, Phys. Rev. B {\bf 52},  490  (1995),
N. Schopohl, cond-mat/9804064 (unpublished, 1998).

\bibitem{eschrig04} 
M. Eschrig, J. Kopu, A. Konstandin, J. C. Cuevas, M. Fogelstr\"om, and G. 
Sch\"on, Adv. in Sol. State Phys. {\bf 44}, pp. 533-546, (Springer Verlag, Heidelberg, 2004).

\bibitem{cuevas06}
J. C. Cuevas, J. Hammer, J. Kopu, J. K. Viljas, and M. Eschrig,
Phys. Rev. B {\bf 73}, 184505 (2006).

\bibitem{shelankov84}
A. L. Shelankov, Sov. Phys. Solid State {\bf 26}, 981 (1984) [Fiz. Tved. Tela
{\bf 26}, 1615 (1984)].

\bibitem{zaitsev84}
A.~V. Zaitsev, Zh. Eksp. Teor. Fiz. {\bf 86}, 1742 (1984)
[Sov. Phys. JETP {\bf 59}, 1015 (1984)].

\bibitem{millis88}
A. Millis, D. Rainer, and J. A. Sauls, Phys. Rev. B {\bf 38}, 4504 (1988).

\bibitem{yip97}
S.-K. Yip, J. Low Temp. Phys. {\bf 109},  547  (1997);
C.-K. Lu and S.-K. Yip, Phys. Rev. B {\bf 80}, 024504 (2009).


\bibitem{fogelstrom00} 
M. Fogelstr\"om, Phys. Rev. B {\bf 62}, 11812 (2000).

\bibitem{ozana00}
A. Shelankov and M. Ozana, Phys. Rev. B {\bf 61}, 7077 (2000);
A. Shelankov and M. Ozana, J. Low Temp. Phys. {\bf 124}, 223 (2001).

\bibitem{zhao04} 
E. Zhao, T. L\"ofwander, and J.A. Sauls, Phys. Rev. B {\bf 70}, 134510 (2004).

\bibitem{eschrig09}
M. Eschrig, Phys. Rev. B {\bf 80}, 134511 (2009).

\bibitem{grein09}
R. Grein, M. Eschrig, G. Metalidis, and G. Sch\"on, 
Phys. Rev. Lett. {\bf 102}, 227005 (2009).

\bibitem{eschrig03}
M. Eschrig, J. Kopu, J. C Cuevas, and G. Sch\"on, Phys. Rev. Lett. {\bf 90},
137003 (2003).

\bibitem{kopu04}
J. Kopu, M. Eschrig, J. C. Cuevas, and M. Fogelstr\"om, Phys. Rev. B {\bf 69}, 094501 (2004).

\bibitem{eschrig08}
M. Eschrig and T. L\"ofwander, Nat. Phys. {\bf 4}, 138 (2008).

\bibitem{luck03}
T. L\"uck, P. Schwab, U. Eckern, and A. Shelankov,
Phys. Rev. B {\bf 68}, 174524 (2003).

\bibitem{graser07}
S. Graser and T. Dahm, Phys. Rev. B {\bf 75}, 014507 (2007).

\bibitem{fri06}
P.A.Frigeri {\it et al.}, Europ. Phys. Journ. B {\bf 54},
435 (2006).

\bibitem{ser04}
I. A. Sergienko and S. H. Curnoe,
Phys. Rev. B {\bf 70}, 213410 (2004).

\bibitem{Hayashi06} N. Hayashi, K. Wakabayashi, P.A. Frigeri, and M. Sigrist,
Phys. Rev. B {\bf 73}, 024504 (2006).

\bibitem{EIRashba:2003}
E. I. Rashba, Phys. Rev. B {\bf 68}, 241315 (2003).

\bibitem{EGMishchenko:2004}
E. G. Mishchenko, A. V. Shytov, and B. I. Halperin,
Phys. Rev. Lett. {bf 93}, 226602 (2004).

\end{thebibliography}
\end{document}